\documentclass[traditabstract]{aa}
\usepackage{graphicx}
\usepackage{txfonts}

\begin{document}
                
\title{The mixing and transport properties of the intra cluster medium: a numerical study using tracers particles}

\author{F. Vazza\inst{1}, C.Gheller \inst{2}, G. Brunetti\inst{1}}

\offprints{Franco Vazza \\ \email{vazza@ira.inaf.it}}

\institute{INAF/Istituto di Radioastronomia, via Gobetti 101, I-40129 Bologna,
Italy
\and CINECA, High Performance System Division, Casalecchio di
Reno--Bologna, Italy}

\date{Received / Accepted}

\authorrunning{F. Vazza, C.Gheller, G.Brunetti}
\titlerunning{Mixing in the ICM with Tracers}

\abstract
{We present a study of the mixing properties
of the simulated intra cluster medium, using 
tracers particles that are advected by the gas flow
during the evolution of cosmic structures.
Using a sample of seven galaxy clusters (with masses in the range of $M \sim 2 - 3 \cdot 10^{14}M_{\odot}/h$) 
simulated with a peak resolution of $25kpc/h$ up to the distance
of two virial radii from their centers, 
we investigate the application of tracers to some important
problems concerning the mixing of the ICM. 
The transport properties of the evolving ICM are studied through
the analysis of pair dispersion statistics and mixing distributions.
As an application, we focus on the transport of metals in the ICM.
We adopt simple scenarios for the injection of metal tracers
in the ICM, and find remarkable differences of metallicity
profiles in relaxed and merger systems also through
the analysis of simulated emission from Doppler-shifted
Fe XXIII lines.}

\maketitle


\section{Introduction}
\label{sec:intro}

The mixing properties of the intra cluster medium (ICM) 
are still poorly understood
and a number of open fields of research are tightly
connected to this important topic:
the 'heating cooling flows' problem (e.g. Bruggen \& Kaiser 2002; Omma et al. 2004),
 the spreading of metals in the ICM (e.g. Schindler
\& Diaferio 2008; Borgani et al.2008), the stability of thermal profiles
in the innermost region of galaxy clusters (e.g. Sharma et al.2009), the
efficiency in the ram pressure stripping of accreting satellites and 
the emergence of cold fronts in galaxy clusters (e.g. Ascasibar \&
Markevitch 2006; Markevitch \& Vikhlinin 2007) and many others.

From the theoretical point of view, in the last few years meaningful
 progress has been made in understanding the convective
properties of the ICM. Focusing on the role played by
instabilities in a magnetized, low-collisional plasma (e.g. Parrish 
\& Stone 2005; Quataert 2008; Ruszkowski \& Oh 2009), a new picture
of the ICM stability was presented, which drastically alters the ``classic''
picture provided by the Schwarzschild criterion, in which stability is ensured 
by $dlog(S)/dr>0$ (Schwarzschild 1959), where $S$ is the specific entropy.

These approaches cannot take into account all sources of mixing motions
in galaxy clusters though, which are due to large scale accretions of
matter and turbulent motions. 
Indeed there is clear evidence nowadays that
a sizable amount of turbulent motions may be present in the ICM.
Observations suggest large scale subsonic turbulent motions 
in the range of $\sim 100\rm kpc-1 \rm Mpc$  
(e.g. Schuecker et al.2004; Henry et al.2004; Churazov et al.2004). 
In addition,
also studies of Faraday Rotation allow a complementary approach and suggest
that the ICM magnetic field is  tangled over a broad range of 
scales (e.g. Murgia et al.2004; Vogt \& Ensslin 2005; Kuchar \& Ensslin 2009; Bonafede et al.2010); also, the diffuse 
radio emission detected in a fraction of merging clusters may provide indirect
evidence for turbulence in the ICM (e.g. Brunetti et al. 2008).

Obtaining a self-consistent picture of the evolving ICM, where large scale and
small scale mixing properties of galaxy clusters are followed during 
the whole cluster evolution is challenging, and in this respect
cosmological high resolution numerical simulations may provide additional and
valuable
information.

At present, two main numerical methods are massively applied to cosmological
numerical simulations: Lagrangian methods, which sample both the
Dark Matter and the gas properties using unstructured point-like fluid elements, usually
regarded as particles (e.g.
smoothed particles hydrodynamic codes such as GADGET, Springel 2005, or HYDRA,
Couchman et al.1995) and Eulerian methods, which reconstruct the
gas properties with a discrete sampling of the space using meshes of various
resolution (e.g. cosmological TVD code, Ryu et al.1993; ENZO, Norman et al.2007;
FLASH, Fryxell et al.2000; RAMSES, Teyssier 2002 etc) and model the Dark Matter 
properties with a particle mesh approach.

The typical advantages and drawbacks of the two approaches 
have been extensively investigated.
Lagrangian methods allow one to obtain a high spatial resolution where matter 
concentrates, but in low density regions the sampling is rather poor.
 By construction, this class of methods allows one to follow in a natural way
the advection of a single parcel of (gas or DM) matter during the whole
simulation and its dynamical evolution.

Eulerian methods have a spatial resolution 
fixed to the size of the cell in the computational mesh, and this often 
is inadequate to  
follow the details of cosmological and physical processes of interest
in the innermost region of collapsed objects.
The adaptive mesh refinement technique (AMR)
overcomes this problem by increasing the spatial resolution of the
Eulerian mesh in regions of particular interest (e.g. Berger \& Oliger 1984;
Berger \& Colella 1989;
Kravtsov et al.1997). 
A further important feature of Eulerian methods is that they are strictly
conservative for the total fluid energy, and therefore they accurately follow
the Rankine-Hugoniot relations for shocks (a property shared by shock
capturing methods, also known as Godunov schemes, Godunov 1959).
This is particular important in cosmological simulations, where cosmic shocks
play an important role in setting both thermal and non thermal properties
of the ICM (e.g. Ryu et al.2003; Pfrommer et al.2006; Vazza et al.2009).

In the last few years a number of works highlighted and investigated the causes 
which lead to the main  
differences between these numerical approaches (Agertz
et al.2006; Tasker et al.2008; Wadsley et al. 2008; Mitchell et al.2009; Robertson {\it et
al} 2009; Springel 2009). 
The adoption of artificial viscosity by SPH and the limited spatial 
resolution for AMR, were the reasons for the more significant differences 
between the two approaches; if however both sources of error are tuned 
appropriately (e.g. by reducing the 
action of artificial viscosity outside shocks in SPH, or by increasing
the mesh resolution in AMR codes), the two approaches produce consistent results.

\bigskip

High resolution AMR simulations (such as the ENZO simulations
presented in this paper) can provide an accurate representation of the
cosmic gas dynamics in galaxy clusters, achieving a very broad dynamical
range. Recent works showed that opportune refinements
(e.g. Iapichino \& Niemeyer 2008, Vazza et al. 2009;
Maier et al.2009; Zhu, Feng \& Fang 2010) allow for 
efficiently studying the onset and evolution of chaotic
motions in the ICM.

Yet, some interesting information of Lagrangian nature cannot be followed by
these methods. One example is the distribution of metals in the ICM
due to diffusion and turbulent mixing which cannot be followed, 
unless the conservation equations are
self-consistently implemented for every metal species, with a sizable increase in algorithmic complexity 
and computational expense.
However, it is possible to treat metals in a
post-processing stage by
following their spatial evolution through that of
mass-less particles (tracers), which
move along with the baryon gas.

An other interesting case is the advection of cosmic rays (CR) particles 
in the ICM. Shocks, turbulence, collisions between high energy hadrons
and AGN/supernovae activities are expected to inject relativistic particles
in the ICM over cosmological time-scales (e.g. Brunetti \& Lazarian 2007), whose
interplay with diffuse $\sim \mu G$ ICM magnetic fields is
responsible for the diffuse non-thermal radio emissions
observed in galaxy clusters (e.g. Ferrari et al.2008 for a recent review).
If the energy stored in CR is much smaller than the thermal ICM 
(as shown by recent observations of the central regions
of clusters, e.g. Reimer 2004; Aharonian et al.2008; Brunetti et al.2007; THE MAGIC
collaboration et al.2009), 
the spatial evolution of CR can be studied through that of 
passive tracers. This applies as long as the effect of CR diffusion is negligible and the
evolution of CR particles can be regarded as the simple advection problem 
of tracers in the evolving ICM.

\bigskip

The objective of the present work is to investigate the mixing
properties of the ICM with an appropriate numerical recipe. 
For this purpose high algorithmic accuracy and high spatial and mass 
resolution are needed.
These requirements are well provided by ENZO, which is a high-order and 
cosmological AMR Eulerian code (
e.g. Norman
et al. 2007).
Here we present ENZO simulations with two important 
customizations: first, an additional refinement criterion is added to increase 
the spatial resolution on propagating shocks (Vazza et al.2009);
second, since a pure Eulerian method cannot provide 
the details on the behavior of each 
specific fluid element (e.g. its trajectory and velocity),
we explicitly follow the 
the advection of a large number of Lagrangian passive (mass-less)
{\it tracer particles}, which move according to the 3--D velocity field of
the gas.

The paper is organized as follows: in Sect.\ref{sec:methods} we present the details
of the simulation runs for this project, and the method to inject and follow tracers.
In Sect.\ref{sec:convergence} we discuss the result of several convergence tests to assess the 
reliability of tracers simulation with different possible setups. In Sect.\ref{sec:results}
we present astrophysical results of tracers, studying in particular
the average transport properties of tracers in all simulated clusters by discussing
the evolution of the pair dispersion statistics (Sect.\ref{subsec:dispersion}); the morphology and the
evolution of radial mixing in merging and non merging clusters through the
evolution of tracers initially located in spherical shells (Sect.\ref{subsec:mix});
the spreading of metal tracers in the ICM, studying toy models of metal
injections from galaxies (Sect.\ref{subsec:metals});
the simulated emissivity from the broadened Fe XXIII line from
metal tracers and its dependence on the the dynamical
history of clusters (Sect.\ref{subsec:iron_line}). Our conclusions are given
in Sect.\ref{sec:conclusions}.

\bigskip

\begin{table}
\label{tab:clusters}
\caption{Main characteristics of the simulated clusters. 
Column 1: identification number. Columns 2:
total mass ($M_{\rm DM}+M_{gas}$) inside the virial radius.  Columns 3: virial radius. Column 4: 
dynamical state at $z=0$ (qualitative classification); MM=major merger for $z<1$; mm=only minor mergers for $z<1$, rr=visually relaxed at $z \approx 0$. Cluster H1d is the same as H1, but
re-simulated with mesh refinement on gas/DM density only.}
\begin{center}
\begin{tabular}{c|c|c|c|c}
ID & $M_{vir}[10^{14} M_{\odot}/h]$ & $R_{\rm v}[kpc/h]$ & dynamical state\\
\hline

H1 & 3.10 & 1890 & mm    \\  
H1d & 3.10 & 1890 & mm  \\
H2 & 3.05 & 1810 & MM   \\   
H3 & 2.95 & 1710 & rr   \\  
H4 & 2.63 & 1738 & rr    \\ 
H5 & 2.41 & 1703 & MM    \\  
H6 & 2.32 & 1593 & MM    \\  
H7 & 2.14 & 1410 & mm   \\   

\end{tabular}
\label{tab:char}
\end{center}
\end{table}

\begin{figure*} 
\includegraphics[width=0.3\textwidth]{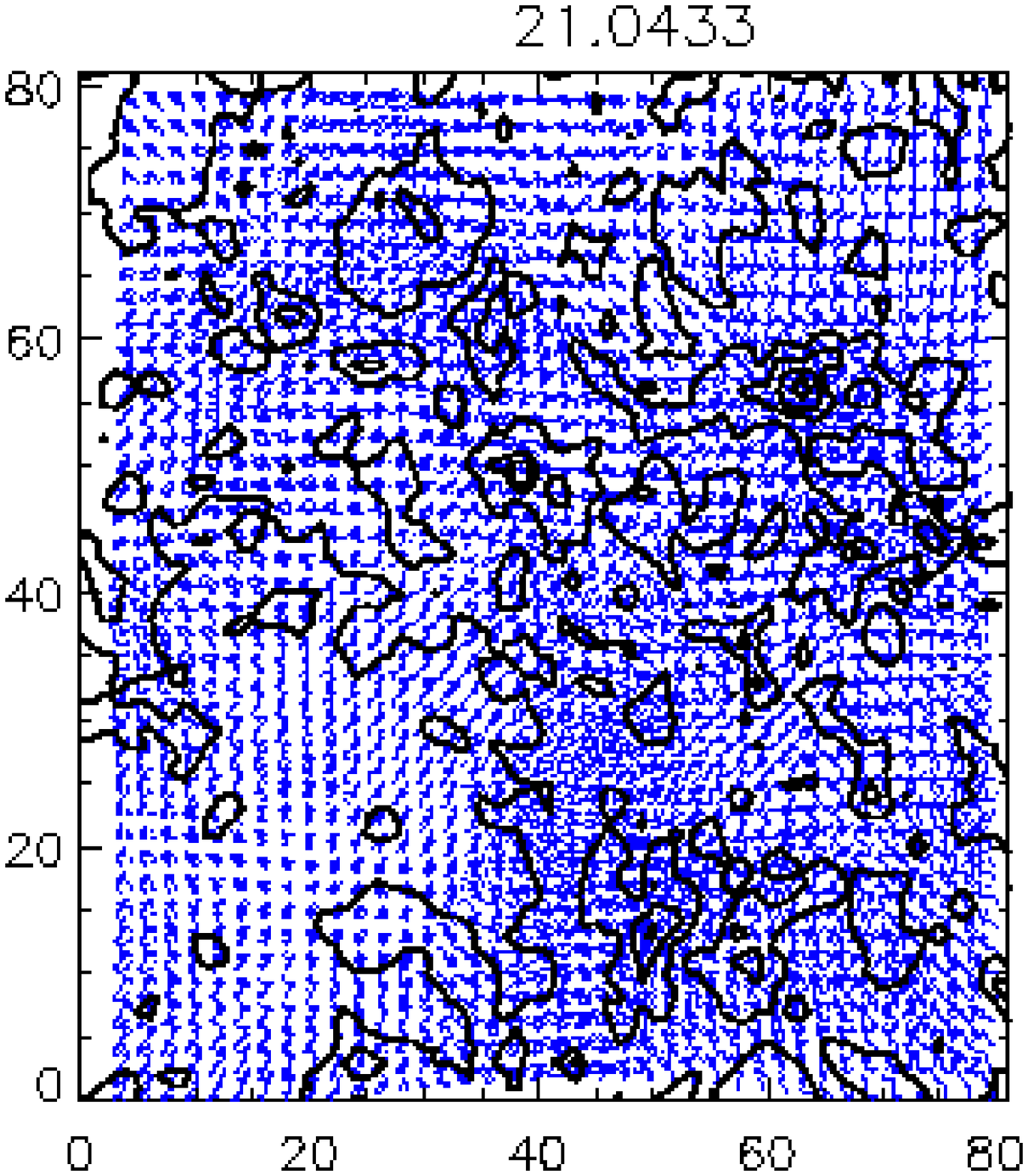}
\includegraphics[width=0.3\textwidth]{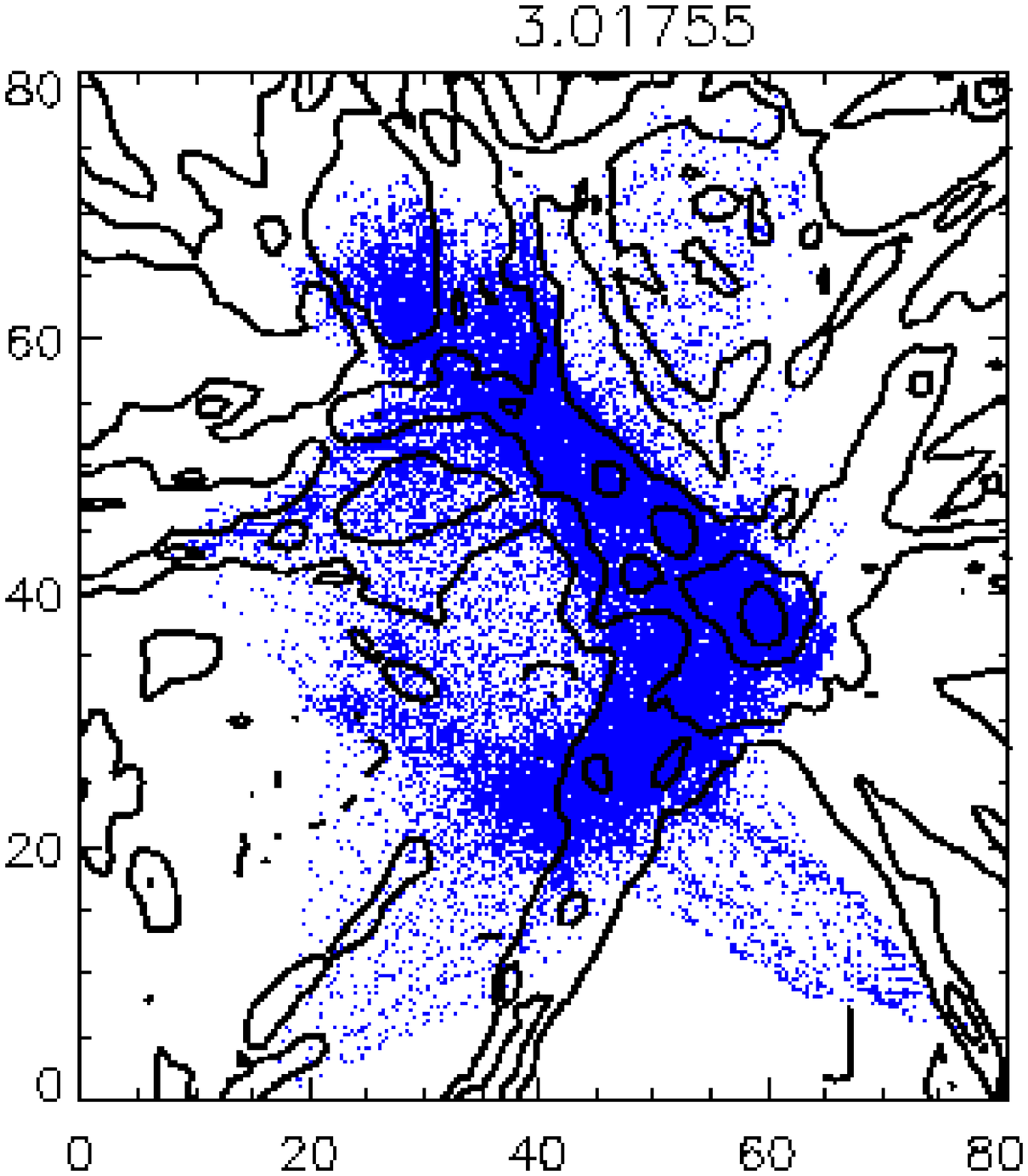}
\includegraphics[width=0.3\textwidth]{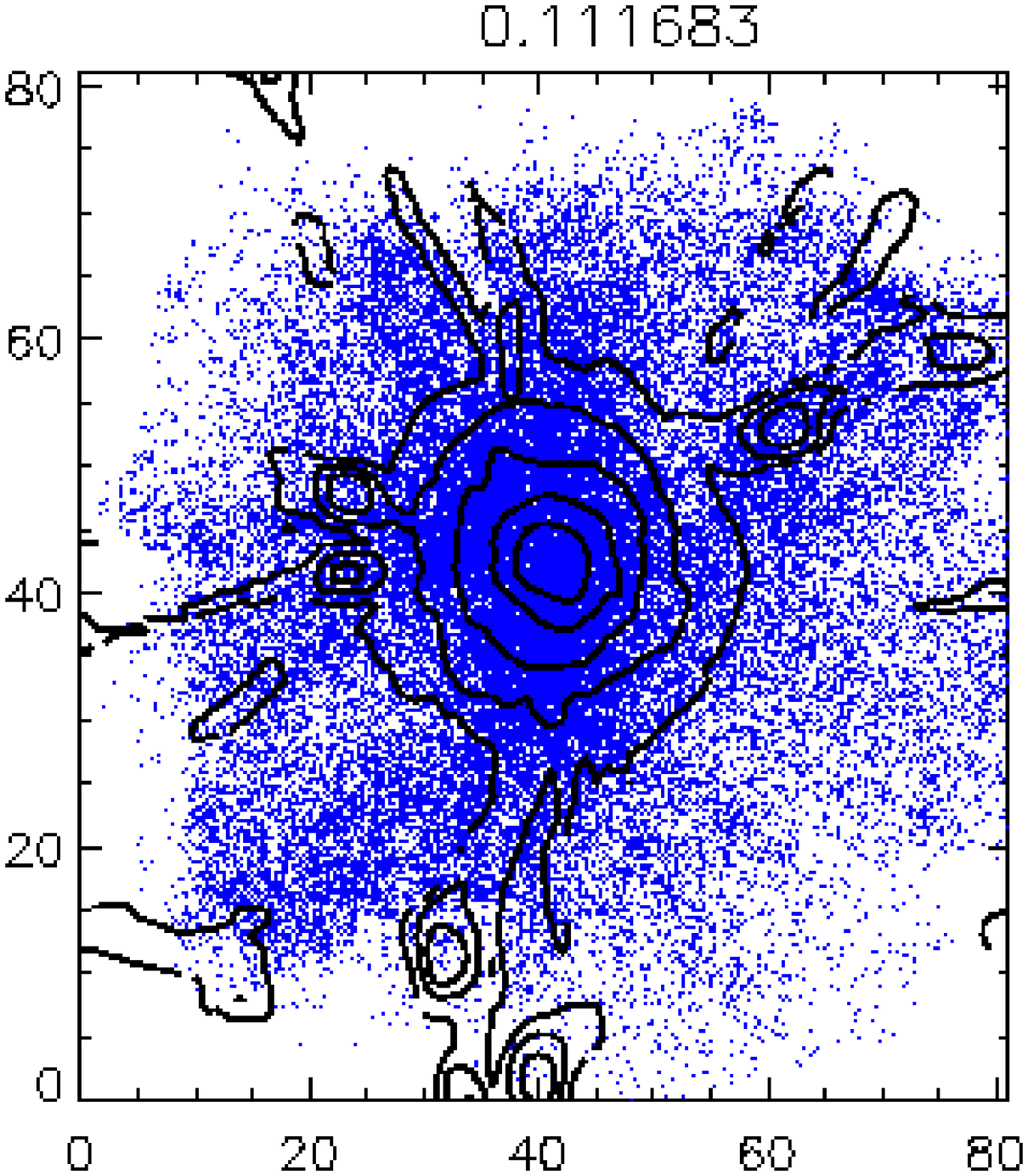}
\caption{An example of the advection of tracers within a forming galaxy cluster.
Contours: projected gas density across a volume of side 10 Mpc/h. Blue points: projected positions of $N \approx 10^{6}$ at $z=21$, $z=3$ and $z=0$.}
\label{fig:map1}
\end{figure*}

\section{Numerical methods}
\label{sec:methods}

\subsection{The ENZO code}

ENZO is an AMR cosmological hybrid code 
originally written by G. Bryan and M. Norman
(Bryan \& Norman 1997; Norman et al.2007). 

It uses a particle-mesh N-body method (PM) to follow
the dynamics of the collision-less Dark Matter (DM) component (Hockney \& Eastwood 1981). The DM component is coupled to
the baryonic matter (gas), via gravitational forces, calculated from the 
total mass distribution (DM+gas) solving the Poisson equation with 
a FFT based approach.

The gas component is described as a perfect fluid, and its dynamics
are calculated by solving the conservation equations of mass, energy and momentum over a computational mesh with
an Eulerian solver based on the 
Piecewise Parabolic 
Method (PPM, Woodward \& Colella, 1984). 
The PPM algorithm belongs to a class
of schemata in which shock waves are accurately described 
by building into the numerical method the calculation of the
propagation and interaction of non--linear waves with a directionally split
approach. 
It is a higher order extension of Godunov's shock capturing
method (Godunov 1959). It is at least second--order accurate in space (up
to the fourth--order in 1--D, for smooth flows and small time-steps) and
second--order accurate in time. 
The PPM method requires 
no artificial viscosity, which leads to an 
optimal treatment of energy
conversion processes, to the
minimization of errors due to the finite size of the cells of the grid and
to a spatial resolution close to the nominal one. 
The basic PPM technique has been modified to include the
gravitational interaction and the expansion of the Universe (e.g. Bryan
et al.1995).  In ENZO cosmological simulations, several comparison tests (e.g. O'Shea et al. 2005;
Tasker et al. 2008)) provided evidence of the better performance of the PPM
method implemented, compared
to the alternative method of ZEUS artificial viscosity (which is also implemented in ENZO).

\begin{figure} 
\includegraphics[width=0.48\textwidth]{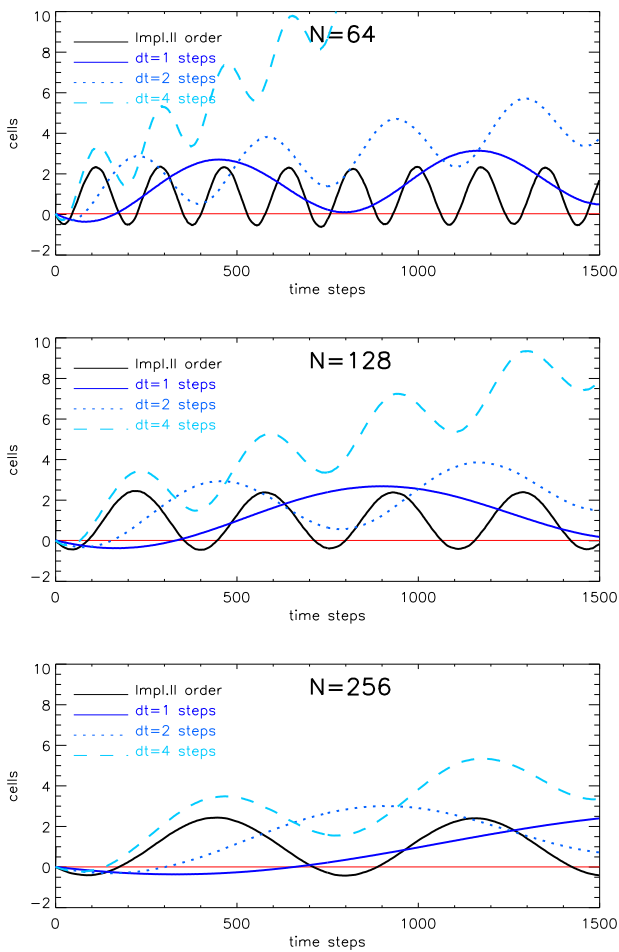}
\caption{Radial position of tracers in a simulated 2-D rotating disk for
different time interpolation schemes: NGP with updates every time step
of the simulation ({\it solid lines}), every two time steps ({\it dotted}),
every 4 time steps ({\it dashed}) or with an implicit second order scheme ({\it solid black}). From top to bottom, results are shown for increasing mesh
resolution; the horizontal red line shows the ideal results.}
\label{fig:vortex}
\end{figure}

\begin{figure*} 
\includegraphics[width=0.95\textwidth]{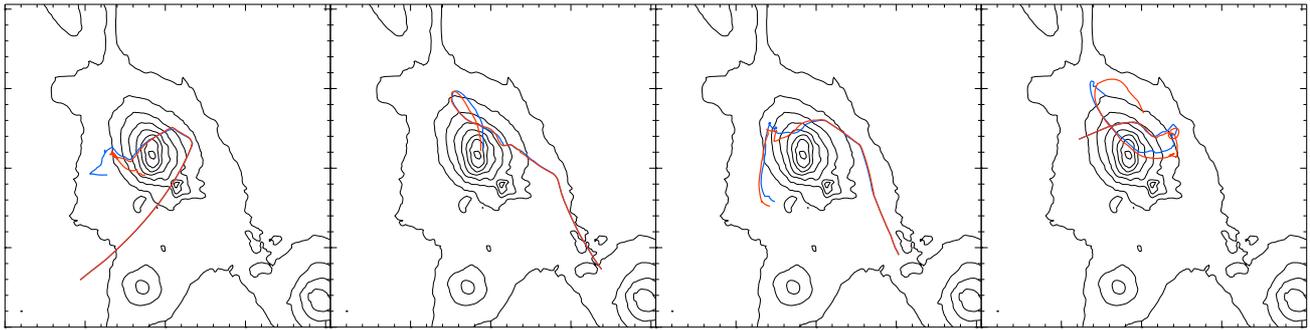}
\caption{Projected paths of four random tracers, evolved from $z=30$
to $z=0$ with
the NGP method (red lines) and the CIC method (blue lines); 
the isocontours show the projected gas density at $z=0$. The side of the image
is 10Mpc.}
\label{fig:map_path}
\end{figure*}

\subsection{Cluster simulations}
\label{subsec:amr}
For the simulations presented here, we assumed a ``concordance'' $\Lambda$CDM cosmology with
the parameters $\Omega_0 = 1.0$, $\Omega_{BM} = 0.0441$, $\Omega_{DM} =
0.2139$, $\Omega_{\Lambda} = 0.742$, the Hubble parameter $h = 0.72$ and
a 
normalization for the primordial density power
spectrum $\sigma_{8} = 0.8$.

The clusters considered in this paper were extracted
from independent cosmic volumes, each with the size of 
$109Mpc/h$, obtained with root grids of
$256^{3}$ cells and with DM mass resolution of $m_{dm} \sim 5.4 \cdot 10^{9} M_{\odot}/h$.
We additionally sub-sampled 
cubic volumes of side $54.5Mpc/h$ with 
a second $256^{3}$ grid around
the region of the formation of our clusters,
achieving an effective root grid resolution
of $\sim 210kpc$ and an effective DM mass
resolution of $m_{dm} \sim 6.7 \cdot 10^{8}M_{\odot}/h$. 
For every cluster run, we identified cubic regions
of the size of $\sim 4 R_{vir}$ (where
 $R_{vir}$ is the virial
radius of clusters at z=0), and allowed for three additional levels of mesh refinement, 
achieving a peak spatial resolution of $\Delta \approx 25kpc/h$ {\footnote{Below we will refer to this sub-volume as to the 'AMR region'.}}.

The mesh refinement was triggered by gas or DM over-density criteria from the
beginning of all simulations ($z=30$). From $z=2$ an additional refinement 
criterion based on 1--D velocity
jumps was switched on (Vazza et al.2009). This
second AMR criterion is
designed to  
capture shocks and intense turbulent motions in the 
ICM out to the external cluster regions.
The grid was refined of a factor two whenever one
of the following conditions was matched:

\begin{itemize}

\item either the gas or the DM density at a given cell exceeded $\Delta \rho/\rho_{mean}>\delta_{\rho}$,
where $\rho$ is the gas (DM) density within the cell at the l-AMR level, and $\rho_{mean}$
is the mean gas (DM) density at the root grid level (in the AMR region);

\item the 1--D velocity jump across a patch of three cells (normalized to the minimum velocity modulus in the
same patch) was larger than a threshold $|\delta v_{i}|/min(v_{i})> \delta_{v}$,
where $i=x,y$ or $z$.

\end{itemize}

The adopted threshold values to trigger AMR were $\delta_{\rho} = 3$ {\footnote {We note that
this threshold values are slightly smaller than in other AMR simulations, e.g.
Nagai et al.2007; Iapichino \& Niemeyer 2008.}}  and $\delta_{v} = 3$. The 
second AMR criterion was switched on only from $z=2$, since this produced only
a negligible difference in the cluster simulations compared to its application
starting from larger redshifts (Vazza et al.2009).

At $z=0$, the number of cells refined up to the peak resolution in our runs
corresponded to a $\sim 20-40$ per cent of the total volume of the AMR region
($N \sim 10^{8}-10^{9}$  cells). Compared to standard cluster simulations, where
mesh refinement is triggered by  gas/DM over-density, the dynamical range 
available for chaotic motions within $R_{vir}$ is much larger, since turbulent
motions with significant 1--D jumps in velocity are not artificially damped by
the under-sampling effects due to poor resolution, which would otherwise 
arise if they are not following
large enough over-densities (of gas or DM) in the cluster volume.

Table \ref{tab:clusters} summarizes the main properties of the galaxy clusters simulated in this work.
The last column in the Table reports a broad classification of their dynamical
state, based on the presence of merger events in the range $0\leq z \leq 1$.
All runs presented here are non-radiative and no treatment of the reheating background
due to AGN and/or stars is considered.

In order to understand the effect played by our refinement strategy on the 
properties of simulated tracers, we additionally re-simulated the most massive cluster
of our sample using the standard refinement criterion (triggered by gas/DM over-density)
starting from $z=30$ (run H1d). A more general comparison on the effect played by the refinement
criteria on the Eulerian properties  of the simulated clusters can be found in
Vazza et al. (2009).

\begin{figure} 
\includegraphics[width=0.48\textwidth]{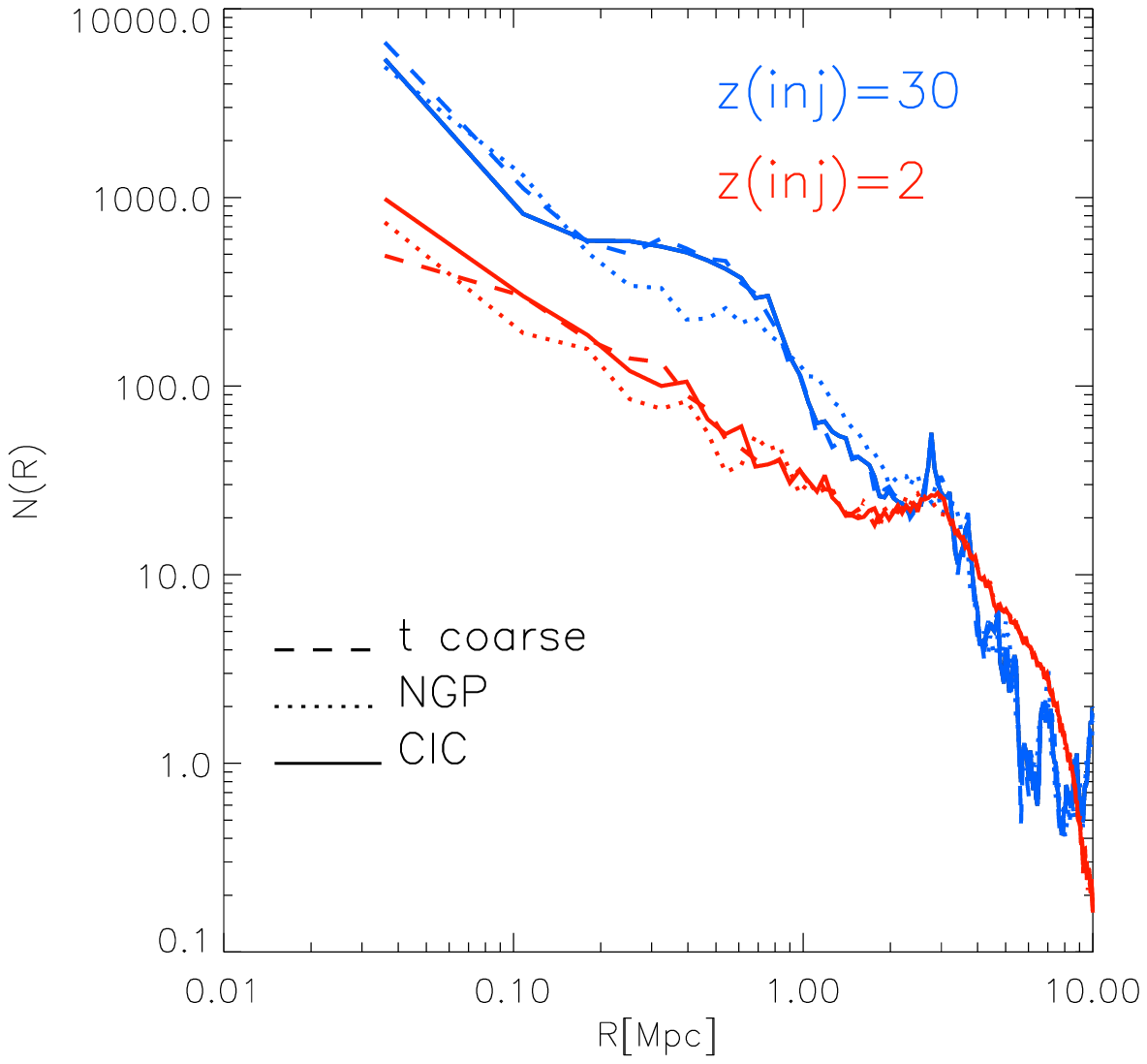}
\caption{Profiles of number distribution of tracers in run H5 at
z=0, for tracers injected at $z=30$ (blue) and tracers generated at $z=2$ (red) . The continuous lines are for the CIC interpolation, the dotted lines
are for the NGP interpolation and the dashed lines are for the CIC interpolation with time integration with coarse time steps.}
\label{fig:pro1}
\end{figure}

\begin{figure} 
\includegraphics[width=0.48\textwidth]{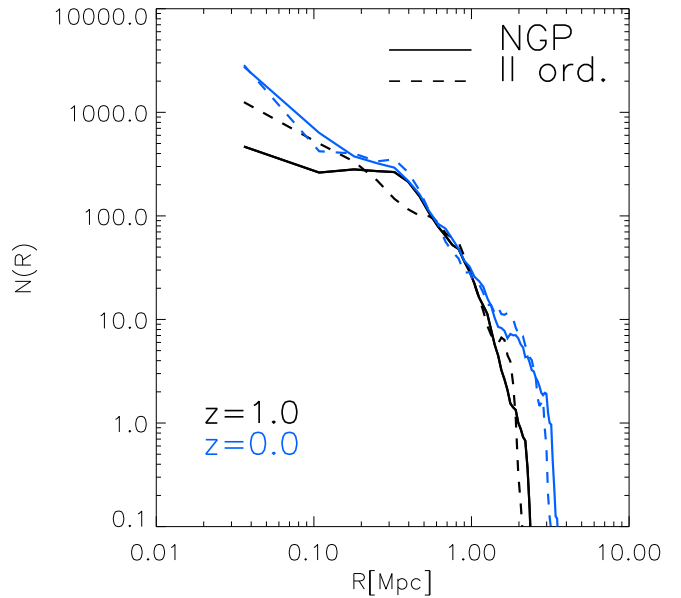}
\caption{Profiles of number distribution of tracers in run H7 at
z=0 ({\it blue lines}) and at z=1.0 ({\it black lines}). Tracers have
been evolved using with NGP method with updates every two time steps (in
solid) and with an implicit second order scheme (dashed).}
\label{fig:pro_IIorder}
\end{figure}

\begin{figure*} 
\includegraphics[width=0.95\textwidth]{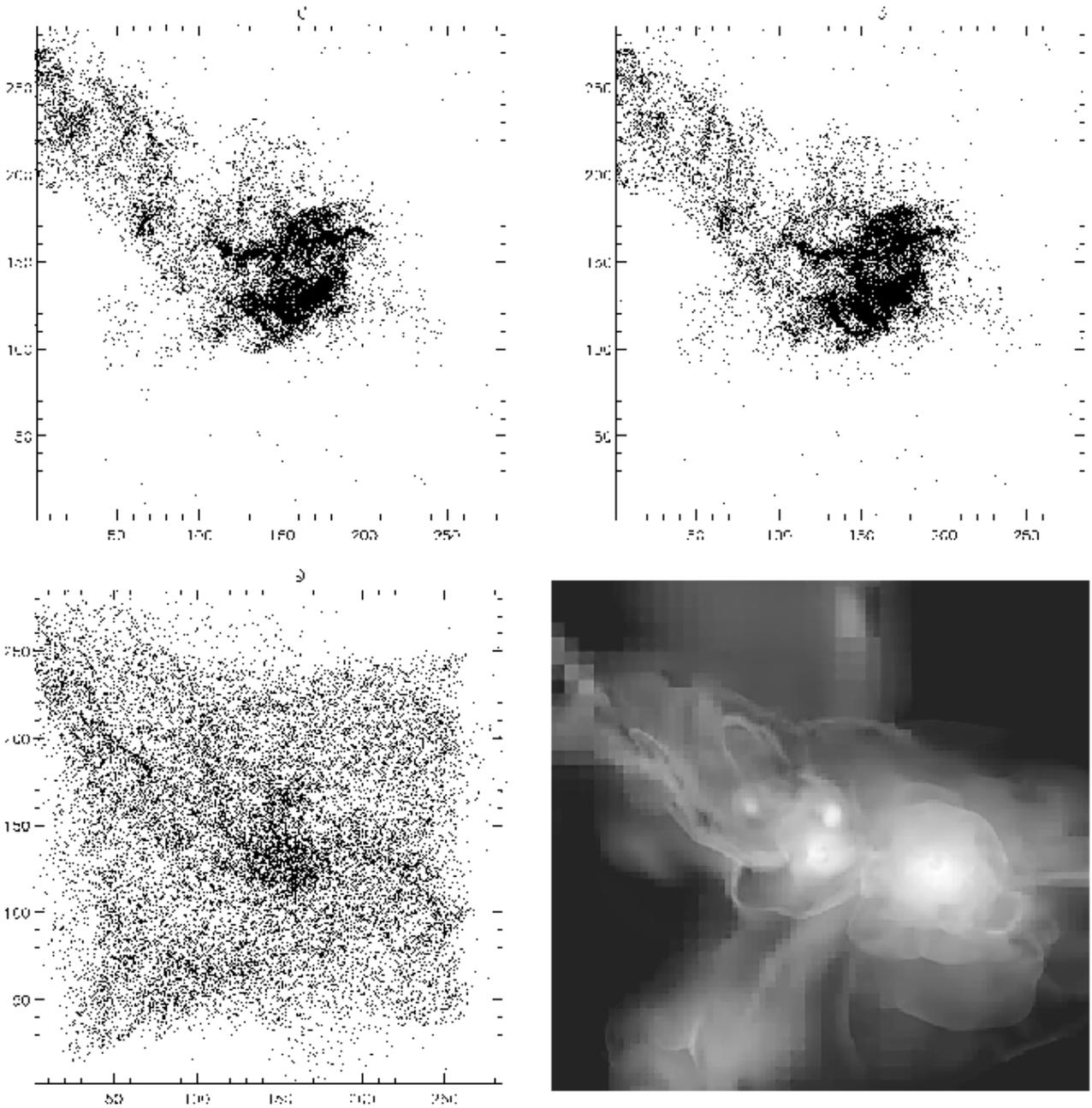}
\caption{Black points: projected position for three different generation
of tracers (generation '0' is generated at $z=30$, gen. '3' at $z \approx 6$,
and generation '9' at $z \approx 0.1$) for a cluster run at $z=0$.
Color maps: slice through the center of the cluster, showing gas density (top)
and gas temperature (bottom). The side of the images is 10Mpc.}
\label{fig:map2}
\end{figure*}

\subsection{Tracer simulations}
\label{subsec:trac_intro}

The transport and diffusion properties of turbulent media are
naturally addressed in the Lagrangian viewpoint, and many 
highly sophisticated numerical techniques haven been developed
and tested over years (e.g. Toschi
\& Bodenschatz 2009 and references therein).
Usually, the statistical properties of fluid particles transported by
a 3--D fully developed turbulent flow are investigated by means
of direct numerical hydro simulations and by following
the advection of mass-less (or inertial) particles, whose 3--D 
position is timely updated by assuming the motion of 
particle tracked by the flow (e.g. Maxey \& Riley,1983). 
In the last few years, a number of interesting works have investigated 
in detail the behavior of tracers in astrophysical turbulent flows by
taking into account the effects played by inertia effects (e.g. Bec et al. 2006; Calzavarini et al. 2008), intermittency (e.g. Biferale et al. 2004), gravity (e.g. Bosse et al. 2006) and magnetic fields (e.g. Arzner et al. 2006).

In the framework of galaxy clusters study with cosmological numerical
simulations, to our knowledge no study of 
the Lagrangian properties using tracers particles have been 
performed so far. In this work, we present a method to inject
and follow tracer particles in cosmological adaptive mesh refinement
simulations with the ENZO code.
A qualitative similar analysis was recently presented by 
Federrath et al. (2008), who used ENZO PPM simulations to study the 
turbulent transport in the Interstellar Medium.

We followed the kinematics of passive tracers by injecting and evolving 
Lagrangian mass-less points in the AMR region of each cluster
and by updating positions according to the underlying gas velocity field.
Since the volume considered for tracer studies is always centered on
a forming massive galaxy cluster, the bulk of tracers is progressively
advected towards the center of the computational domain; in the (rare) 
case of tracers exiting from the considered volume, we simply removed them
from the computation.

In order to better sample the entire cluster volume
during time, we injected different populations of tracers in the AMR region.
For the remainder of the paper, we will refer to tracer ``generation'' as to 
the injection of mass-less tracer particles in the AMR region
at a given cosmic epoch. 
When new tracers were generated, they were 
placed at the center of cells 
and were uniformly distributed in the grid.
If different mesh resolutions were present, as in the case we 
analyze here, the initial sampling was always taken from 
the most refined mesh.

In our fiducial setup,
tracers were moved 
by updating their comoving coordinates through:

\begin{equation}
\vec{x}(t+\Delta t)=\vec{x}(t)+{\vec{u}} \Delta t.
\label{eq:one}
\end{equation}

The time step used to evolve the tracer positions is the
time step of the simulation. 
In non-radiative simulations
the time step in ENZO is computed as the minimum
between three time steps related to three typical velocities:
the maximum velocity module among Dark Matter particles, $v_{DM}$, the maximum gas velocity in the grid, $v_{grid}$, and the the expansion velocity of the Universe at each time, $v_{H}$.
The actual time step in ENZO is thus: 

\begin{equation}
\Delta t \equiv n_{C}/max|v_{DM},v_{grid},v_{H}|;
\end{equation}

where $n_{C}$ (the Courant safety number) is customary set to 0.4.
The above choice ensures that no signal can travel for more than a half-cell during
a time step (at all AMR level) and it also ensures that 
updating the tracer velocities every two time steps is accurate enough.

In principle, more accurate time integration schemes can be adopted 
to better preserve stability, like the Runge-Kutta or the leapfrog scheme
(e.g. Hockney \& Eastwood, 1981); in Sect.\ref{subsec:interp} we specifically
present a comparison between different strategies for the time-integration
of tracer positions.

\bigskip

The tracer behavior was calculated through a post-processing procedure 
using the outputs of ENZO simulations as input for evolving the
tracers positions in time.
In principle the whole procedure can be incorporated in 
ENZO as an additional run-time 
routine, but we found it much more convenient
to follow a post-processing strategy, even if this led to a general overhead 
in terms of data storage and management.
Indeed, the post-processing approach
first allowed us to perform a large number of numerical
tests to find the best setup for the generation of
tracers and for the computation of their evolution. 
Second, most of the physical mechanisms
that we planned to simulate with tracers were expected to have negligible
dynamical feedback on the surrounding baryonic matter (e.g.
metal enrichment and spreading in the ICM, cosmic rays) and thus
the same high-resolution
cluster simulation can be used for a large number of studies.
Finally, some of the physical mechanisms that can be studied
with tracers, like the injection of CR at shocks,
are still poorly understood theoretically
have a still incomplete theoretical
understanding and the iterative application of tracers may help
to explore a wide number of different assumptions.
 
As an example of a simple tracer run, we show in Fig.\ref{fig:map1} the projected positions 
for a single injection (generated at $z=30$) of $N \approx 10^{6}$ tracers at three different epochs in a 
cubic sub-volume of the size of $10Mpc/h$.

Section \ref{sec:convergence} presents numerical tests to decide the optimal
setup of our tracer algorithm. In Sect.\ref{subsec:interp} we compare the use
of different interpolation scheme to assign tracers' velocities; 
in Sect.\ref{subsec:time} we discuss the effect of different injection
epochs; in Sect.\ref{subsec:sampling}
we investigate the role played by the spatial sampling strategy;
in Sect.\ref{subsec:number} we report on the suitable number of tracers
that should be adopted to avoid undersampling of the Eulerian
grid.


\begin{figure} 
\includegraphics[width=0.48\textwidth]{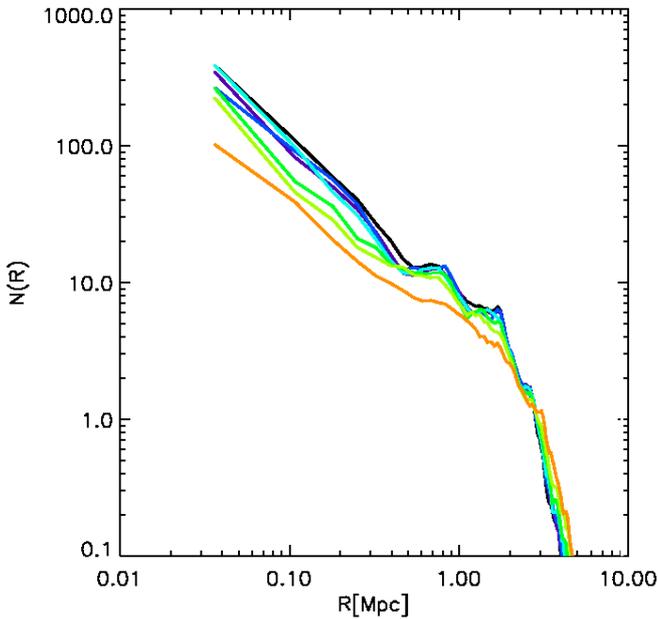}
\caption{Number density profile at z=0 for ten different generations of tracers
(one every $\approx 1 Gyr$), the black curve is for tracers generated
at $z=30$, while the orange one is for tracers generated at $z \approx 0.1$.} 
\label{fig:pro2}
\end{figure}

\begin{figure*} 
\includegraphics[width=0.95\textwidth,height=0.3\textwidth]{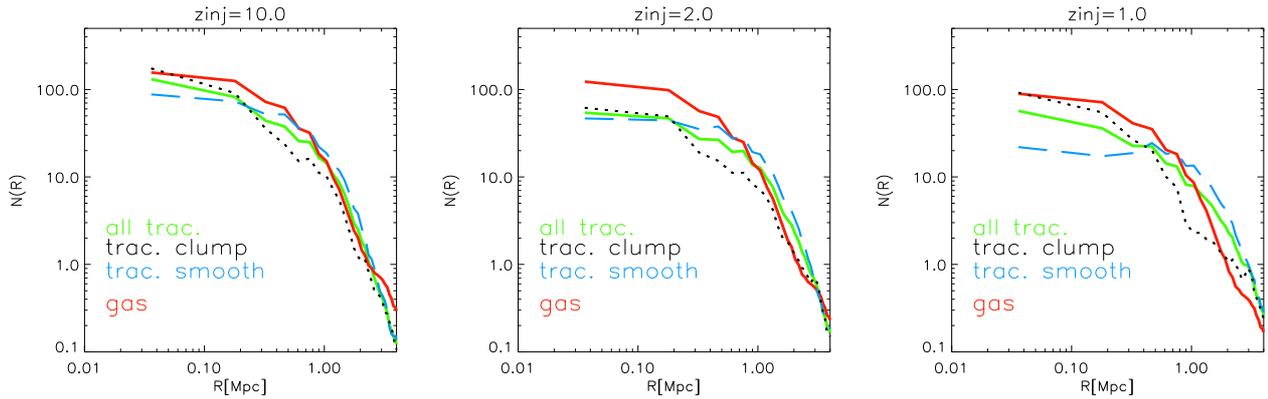}
\caption{Tracer number profiles at $z=0$ for three different epochs
of injection ($z_{inj}=10$, $z_{inj}=2$ and $z_{inj}=1$); the dotted black
lines show the distribution of tracers initially located in gas clumps,
the dashed blue lines show the distributions of tracers initially located in
a spatially uniform way, while the solid green lines show the average of
the two; the solid red lines show the real gas density profile of
the cluster at $z=0$ (as measured from the Eulerian cells). All runs have
the same number of tracers; the gas density profile was re-normalized to the 
total number of tracers inside $R_{vir}$.}
\label{fig:clump_smooth}
\end{figure*}

\section{Numerical tests}
\label{sec:convergence}

\subsection{What velocity for tracers?}
\label{subsec:interp}

The numerical convergence of our method was tested  
by investigating two approaches for the spatial interpolation to 
assign velocity to each tracer particle.

The Nearest Grid Cell (NGP) scheme 
allots a tracer the velocity 
of the gas on the closest grid point. 
With the the Cloud in Cell (CIC) interpolation scheme though, the
3--D velocity at the tracer position is calculated as

\begin{equation}
\vec{u} \equiv \sum_{ijk} \vec{v}_{ijk}\cdot (1-d_{ijk});
\end{equation}

 where $\vec{v}_{ijk}$ is the gas velocity and 
$d_{ijk}$ is the linear distance between the center of the
cells and the tracer, in units of the cell size 
(the sum is performed over the nearest eight cells). 

We tested the time interpolation procedure by implementing
an implicit second-order method with a NGP assignment for the 
tracer positions:

\begin{equation}
\label{eq:2order}
\vec{x}(t+\Delta t)=\vec{x}(t)+({\vec{u}+\vec{u'}}) \Delta t/2,
\end{equation}

where $\vec{u'}$ is the 3D velocity field at the time $t+\Delta t$.

First, we performed an idealized 2-D test which compares the above interpolation 
schemes for a rigid rotating disk
(Fig.\ref{fig:vortex}) for increasing resolutions of the 
underlying grid distribution ($64^{2}$, $128^{2}$ and $256^{2}$). 
Tracers are initially located at a fixed radius from
the center of the disk and
their motion is followed for 1500 time steps
(which is similar to the total number of time steps in typical clusters
runs) and the time step $\Delta t$ 
is timely computed with the same prescription from the Courant condition 
of our ENZO simulations.
For the  NGP and CIC schemes we also tested the possibility
of updating the tracer velocity every two or every four time steps, 
because this would reduce the computational cost in 
cluster simulations.
For a perfect interpolation 
procedure, tracers are expected to move in a perfectly circular way
at the initial radius of injection.
Figure \ref{fig:vortex} shows the time evolution of the radial position 
for a tracer in the rotating disk, placed in all cases at the initial
distance of $R/4$ ($R$ is the disk radius) from the rotation center.
In this simple setup, the CIC and NGP schemes produce identical results
and the different lines are superimposed for all choices of $\Delta t$.
At all resolutions, the adoption of a second order implicit method
for the time interpolation leads to convergent results, with a rapid 
oscillations of $<2$ cells around the radius of injection.
The CIC and the NGP methods led also to similar radial oscillations
on larger periods if the tracer position was updated every $\Delta t$ or $2\Delta t$ time steps, while they show an increasing
radial dispersion for the choice of $4\Delta t$ at all grid resolutions.
The above findings have been tested and confirmed even in the case of an angular
velocity that varies with time (which is reasonable as the Courant 
condition is designed to preserve time accuracy against any change on
the dynamical time of the system), and for different injection radii.
Since the typical grid 
size corresponding to the AMR volume of our
clusters is $N\sim 200-300$ cells (at the highest resolution), 
this test suggests that 
updating tracer positions every $\leq 2$ time steps either with a 
CIC or with a  NGP interpolation  may be a viable choice for
cluster simulation, where large rotation patterns may be present.
 
To investigate this issue further, 
we simulated with the CIC and the NGP scheme
the advection of tracers in a cluster run by using Eq.\ref{eq:one}.
Figure \ref{fig:map_path} shows the evolution of the projected
positions of four tracers in a cluster simulation from
$z=30$ to $z=0$.
The color coding shows the path according to the two integration
procedures: differences in the positions are usually $\leq 5$ cells after $\sim 150-200$ 
integration time steps. 

Figure \ref{fig:pro1} shows the radial number density profile
of tracers (i.e. mean number of tracers for unit of volume, for
shells around the cluster center) at $z=0$, adopting
$N \approx 10^{6}$ tracers that have been moved using both methods and Eq.\ref{eq:one}
for two different generation epochs, $z=30$ (blue lines) and for $z=2$ (red lines). 
We also show the distributions for an additional
run, which makes use of the NGP with 
positions updated every four time steps of
the simulation. Statistically, the three
methods provide consistent results, with no systematic
differences. 
Finally, in Fig.\ref{fig:pro_IIorder} we show the results or our test of our fiducial NGP interpolation with
updates every $2 \Delta t$ against the implicit second-order interpolation
scheme of Eq.\ref{eq:2order} for the advection of $N \approx 10^{6}$ tracers
injected at $z=30$ for the smallest cluster of the sample.
The agreement at $z=0$ is better than  $\sim 5$ percent at all radii, while
for $z=1.0$ there is a larger scatter of up to a factor $\sim 2$ within
$100kpc$ from the cluster center. 
Given the considerably larger amount of data that the implicit second
order method requires (since one must simultaneously consider 
the 3--D velocity field at two close time steps),  
and given the relatively small statistical difference compared to
the other more simple schemes from now on we will adopt the 
NGP method, which seems to provide a 
sharper reconstruction of the gas velocity field close to shocks, and we
will integrate the tracer positions every two  
time steps of the simulation.
As a cross check, we report in Sect.\ref{subsec:dispersion} an additional test to 
show that the results obtained with second
order integration and simple Euler step are very similar {\footnote{Similar
results on the convergence between Euler step and higher order time interpolation for 
ENZO simulations of the inter stellar medium were reported also in Federrath et al. (2008).}}.

\begin{figure} 
\includegraphics[width=0.45\textwidth]{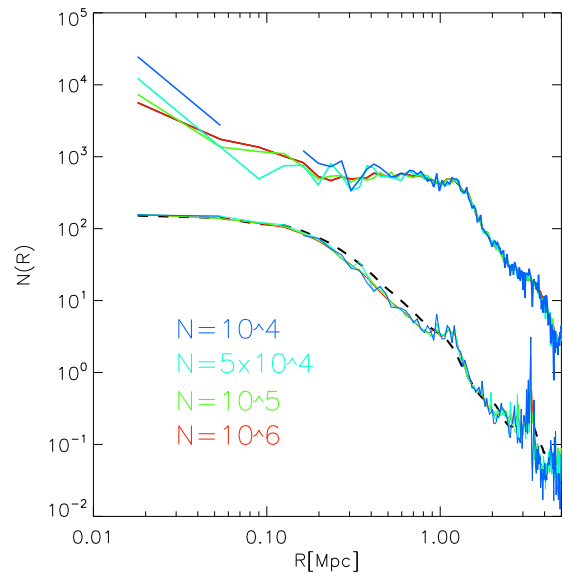}
\caption{Profiles of number distributions (bottom curves) and for
average density (in arbitrary units, top curves) for tracers in a cluster run as a function
of the total number of tracers. The black dashed line shows the volume weighted gas density profile measured in the Eulerian grid.}
\label{fig:pro3}
\end{figure}

\begin{figure*} 
\includegraphics[width=0.95\textwidth]{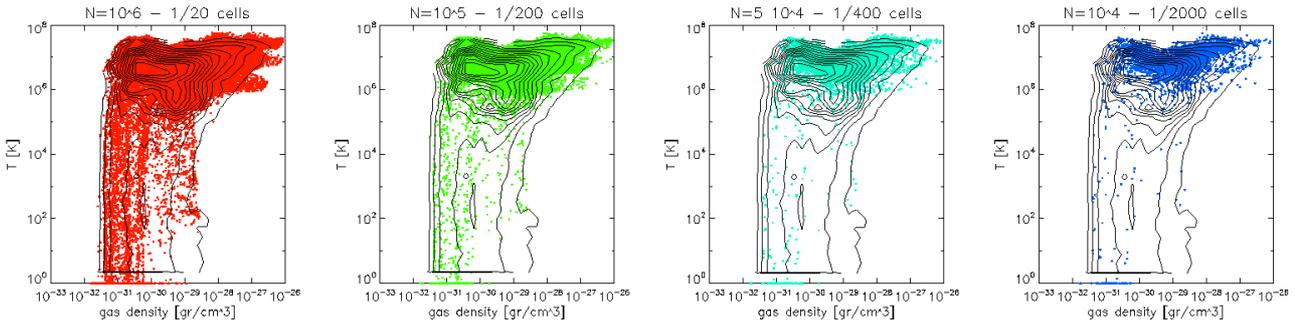}
\caption{Phase distribution for four tracers runs, with decreasing sampling of
the underlying Eulerian grid: from left to right, $N=10^{6}$, $N=10^{5}$, $N=5 \cdot 10^{4}$ and $N=10^{4}$ tracers; the number profiles were normalized
to the same total number for the sake of comparison.
 The isocontours show the phase diagram
for the underlying gas distribution, with sqrt-spacing.}
\label{fig:phase}
\end{figure*}

\subsection{The injection time.}
\label{subsec:time}

Matter accreted by clusters may
present different dynamical properties according to the cosmic
epoch of accretion. Consequently different generations of tracers will be 
spatially distributed depending on their injection epoch and on cluster
dynamical evolution. 

For example, Fig.\ref{fig:map2} shows the projected positions at $z=0$ of
three different generations of tracers in a cluster run.
Tracers were injected at $z=30$, $z=6$ and $z=0.1$),
with an initial spatial sampling of one tracer every four cells (in 1--D)
of the grid at the maximum refinement level.

Generations injected at earlier epochs are more concentrated because of
the cluster formation process. This is also evident from Fig.\ref{fig:pro2},
where we show the radial number density
profile of tracers at $z=0$ for the ten different generations in the
same cluster run. 
Differences at the level of $\sim 3-5$ in the number
density profile are found within $\leq 100kpc$ from the cluster
center, although all profiles have similar shapes,  and a trend to
present flatter distributions is found for the latest generations.

\subsection{The sampling strategy.}
\label{subsec:sampling}

Tracers basically represent a way to sample the Lagrangian evolution
of the physical quantities associated to a fluid described with an Eulerian
method, and 
the choice
of the sampling strategy to place tracers in the mesh cells
depends on the physical quantity of interest.
For example, if the evolution of gas matter accretion is the
quantity of interest, tracers should
sample the Eulerian space in a density-weighted way 
rather than in a spatial uniform manner. 
In fact, a regular volume sampling
of the grid would place most of tracers in smooth and under-dense cosmic regions,
while a density-weighted sampling would place most of them inside
the self-bound gas clumps.
For this
reason the adoption of the 
initial sampling strategy (volume-weighted, gas density-weighted, DM 
density-weighted, etc) must be defined according to the investigated physical process.

Figure.\ref{fig:clump_smooth} shows the results of a test where tracers were
injected at different epochs ($z=10$, $z=2$ and $z=1$) following two different approaches: a) tracers were
placed only inside the virial
volume of the 50 most massive halos in the AMR region at the three different
redshifts, with a
number profile equal to that of the gas in each clump
{\footnote{Halo positions and virial parameters were computed
according to a halo finder algorithm working with spherical gas+DM matter over-density,
specifically designed for Eulerian simulations (e.g. Gheller, Pantano \& 
Moscardini 1998; Vazza, Brunetti \& Gheller 2009)}; b) tracers were placed with a 
regular spatial sampling of the grid, as in the previous sections.
Although this represents just a crude test to study the environmental dependence
of tracers, the trends found are clear. 
Tracers injected in clumps produce more concentrated number density profiles
at $z=0$, while tracers injected in the smooth gas component 
have lower densities in the core at $z=0$ and flatter profiles
outside $\sim 0.5 R_{vir}$. This tendency is increasingly clear as the redshift of
injection decreases. 
For tracers injected at $z=10$, the agreement of the average profile with
the gas mass density distribution is generally within $\leq 50$ per cent. 
With a more accurate 
sampling of the gas density field at the epoch of generation, the
gas density profile at $z=0$ can be closely reproduced by tracers.

\subsection{How many tracers?}
\label{subsec:number}

The choice of a suitable number of tracers is important 
to find the best compromise between computational time and
accurate sampling of the underlying gas distribution.

We performed convergence tests on the number of tracers with
ten generations, with a number of tracers 
$N=10^6$, $N=10^{5}$, $N=5 \cdot 10^{4}$ and
$N=10^{4}$ for each generation. The time-step is kept the same
in all tests.

In Fig.\ref{fig:pro3}, we
plot the radial number density profile of tracers and the average
gas density profile at tracers positions.
Tracers provide a good statistical
sampling of the real gas density profile, with a scatter of less than $<10$ per 
cent inside the virial radius. 
Also the number profile of tracers (when normalized to the same total value)
shows convergent results, with the typical scatter
of less than a  factor $\sim 2$ within $500$kpc from the cluster centers 
when the cases $N=10^{4}$ and $N=10^{6}$ are compared. 

Figure \ref{fig:phase} shows the 
density vs temperature phase diagrams for the four samplings discussed
above(as colored points), 
compared with the phase diagram from the underlying distribution
of gas within the AMR region. The comparison shows that even
if the tracer distributions agree well within the
main halos, larger differences due to poor sampling in low density regions
emerge in runs with a sampling worse than one tracer every $\sim 100$ gas
cells. We therefore consider that, in order to have a fairly good representation
of the cluster regions and also of the accretion pattern outside clusters, an 
initial sampling with at least 1 tracer every $\sim 70-80$kpc/h in 1--D (i.e. $N \sim 10^{6}$ tracers in the AMR region)
for every generation is necessary.

\begin{figure*} 
\includegraphics[width=0.95\textwidth]{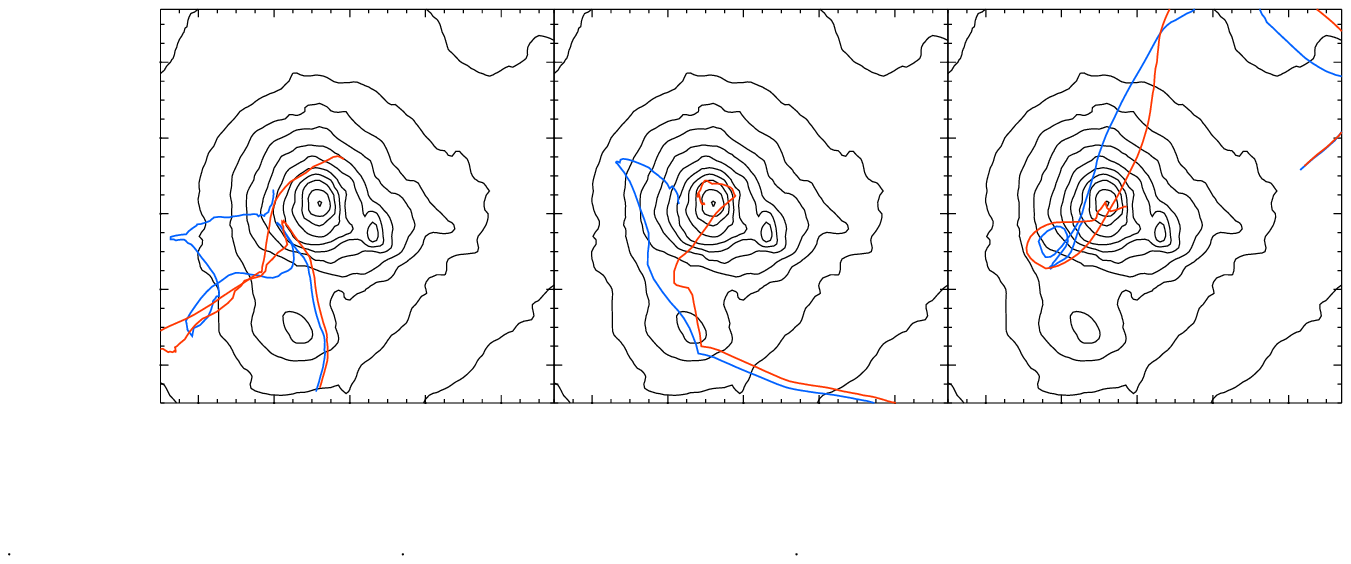}
\caption{Projected paths of three couples of tracers evolved from $z=30$
o $z=0$ from an initial separation of 1 cell=36kpc, for run H7. Isocontours
show the projected gas density at $z=0$. The side
of the image is 3.8Mpc}
\label{fig:disp_pathz}
\end{figure*}

\begin{figure} 
\includegraphics[width=0.45\textwidth]{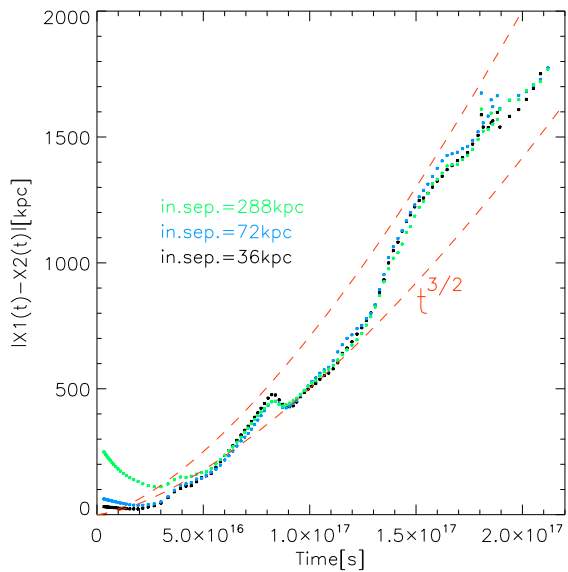}
\caption{Time evolution of the pair tracers dispersion for $N=10^{6}$ tracers in run H7. Couples with three initial separations
are considered: $|\vec{X_{0}}=\vec{X(0)}|=36kpc$ (black dots),  $|\vec{X_{0}}=\vec{X(0)}|=72kpc$ (blue dots) and  $|\vec{X_{0}}=\vec{X(0)}|=288kpc$ (green dots). The additional dashed lines show two possible $\propto t^{3/2}$ trends
to guide the eye.}
\label{fig:disp_veg}
\end{figure}

\begin{figure*} 
\includegraphics[width=0.45\textwidth]{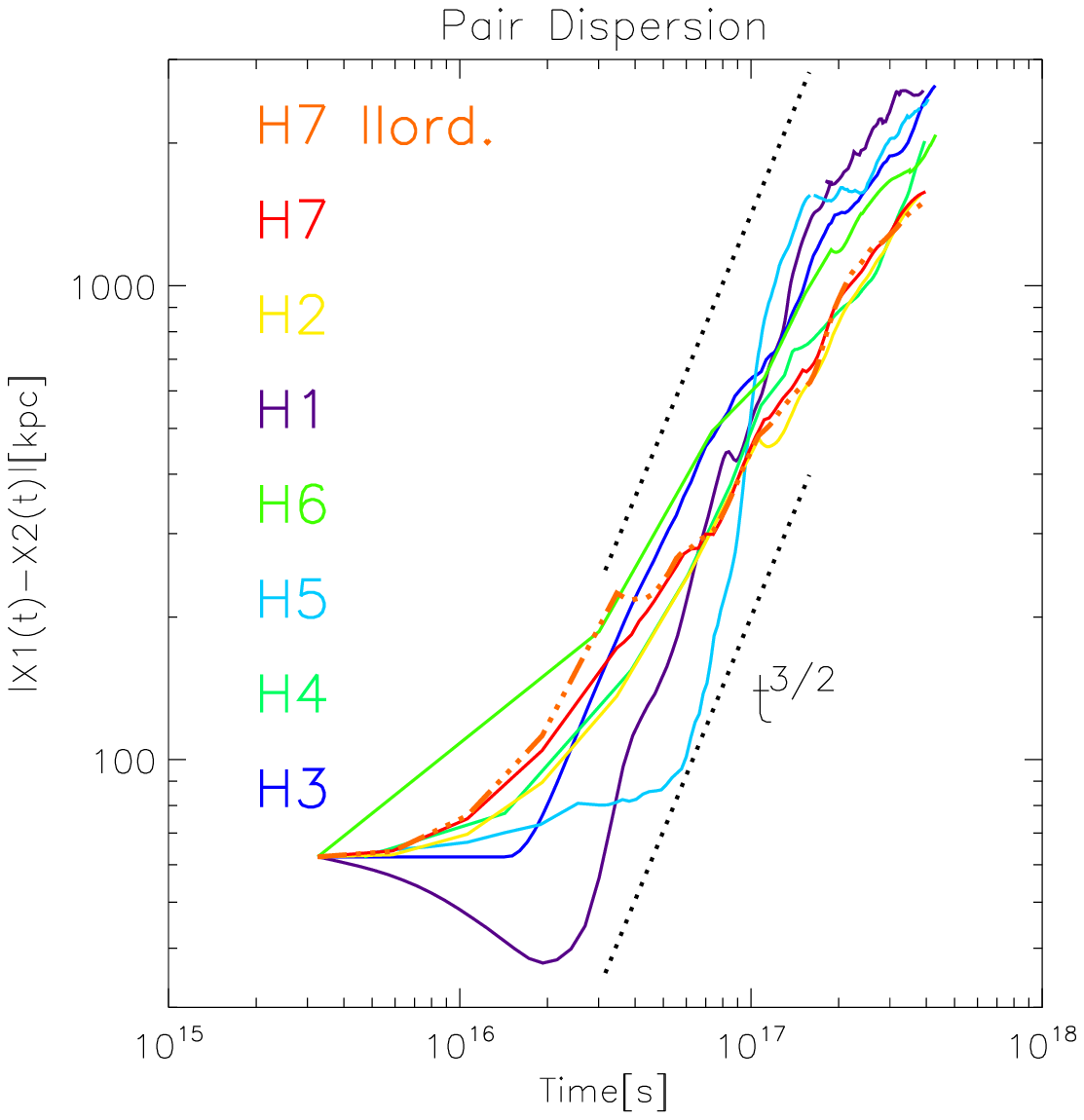}
\includegraphics[width=0.45\textwidth]{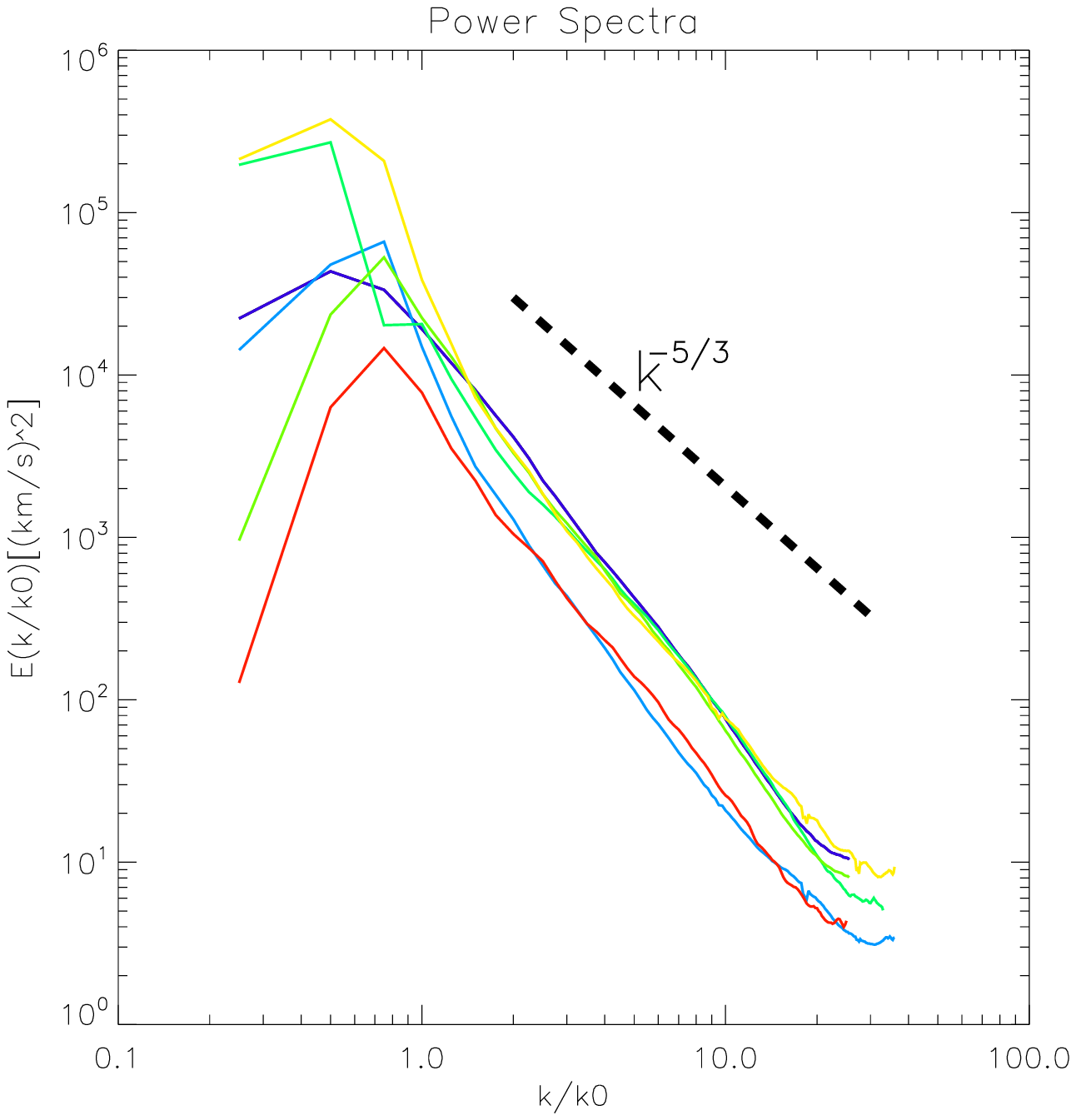}
\caption{{\it Left:}Time evolution of the pair of tracers dispersion for $N=2 \cdot 10^{6}$ tracers in all runs for initial separations of 72kpc (two cells). The additional thick dashed line
shows the slope for the Richardson law, $R(t) \sim t^{3/2}$. {\it Right:} 3-D velocity power spectrum of the galaxy clusters in the sample. The spatial scale $k$ was normalized to the scale
corresponding to the virial radius of every cluster $k0\approx 2 \pi/R_{vir}$. The additional dashed lines show the $-5/3$ scaling to guide the eye. Colors are the same as in the left panel.}
\label{fig:disp_L1}
\end{figure*}

\section{Results}
\label{sec:results}

\begin{figure*} 
\begin{center}

\includegraphics[width=0.24\textwidth]{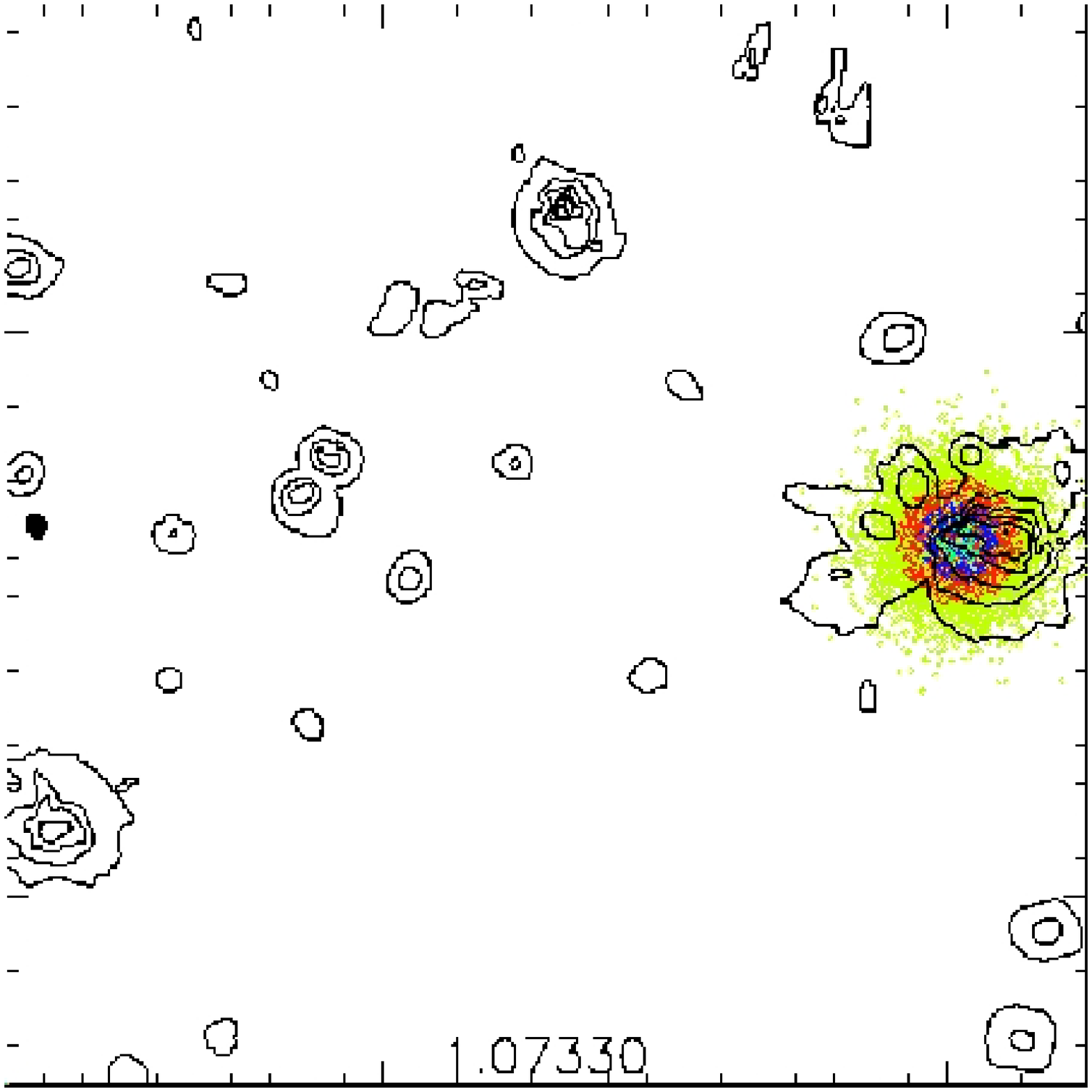}
\includegraphics[width=0.24\textwidth]{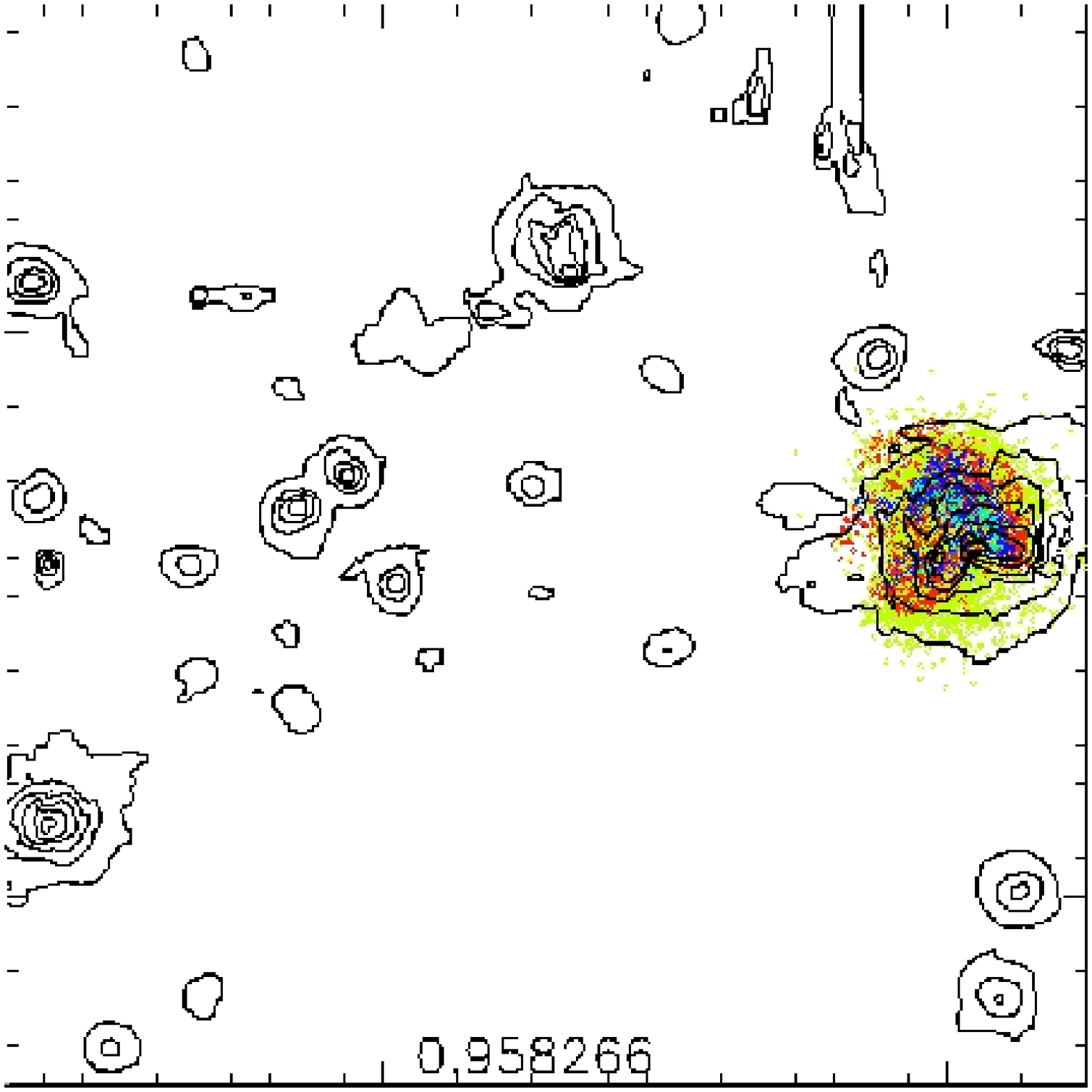}
\includegraphics[width=0.24\textwidth]{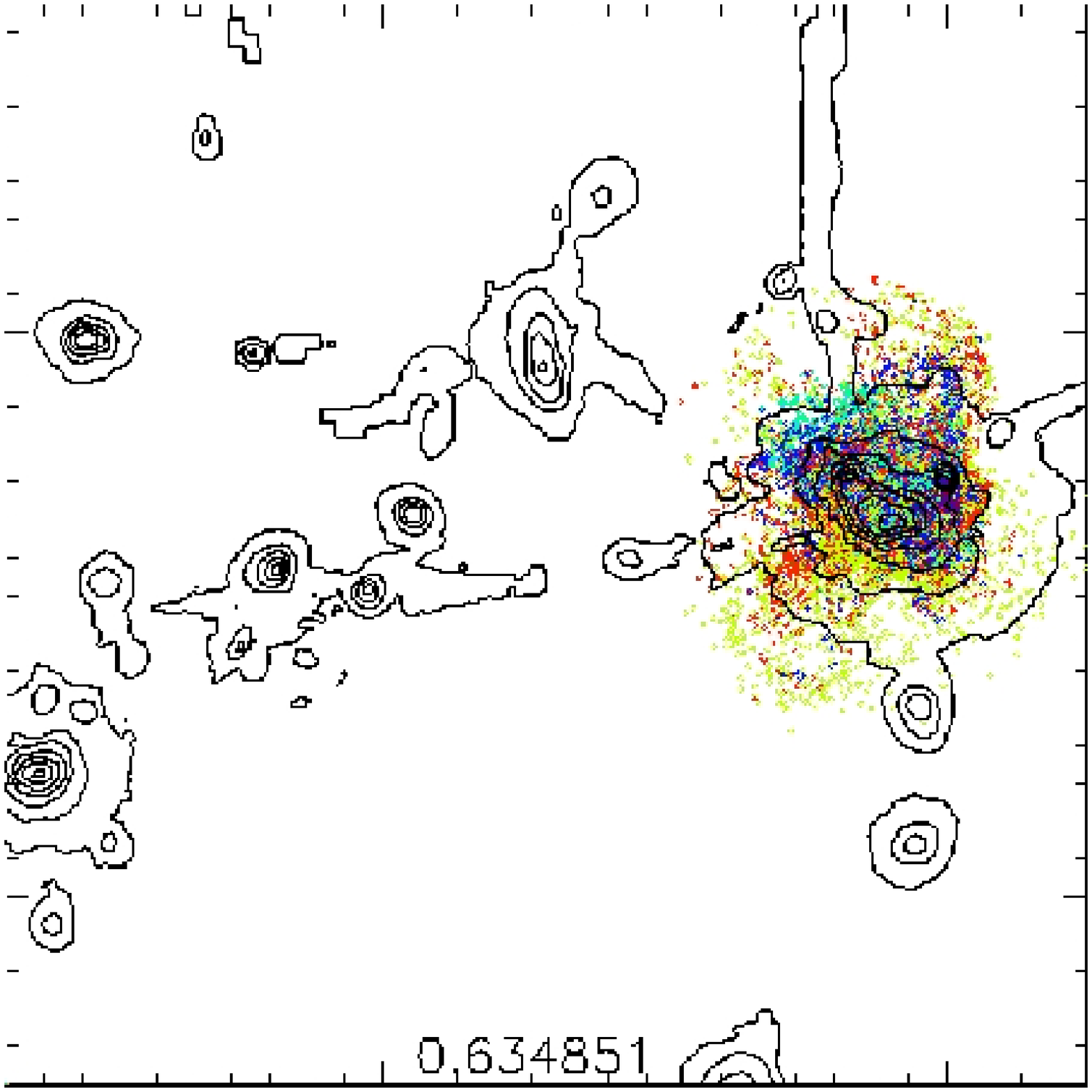}
\includegraphics[width=0.24\textwidth]{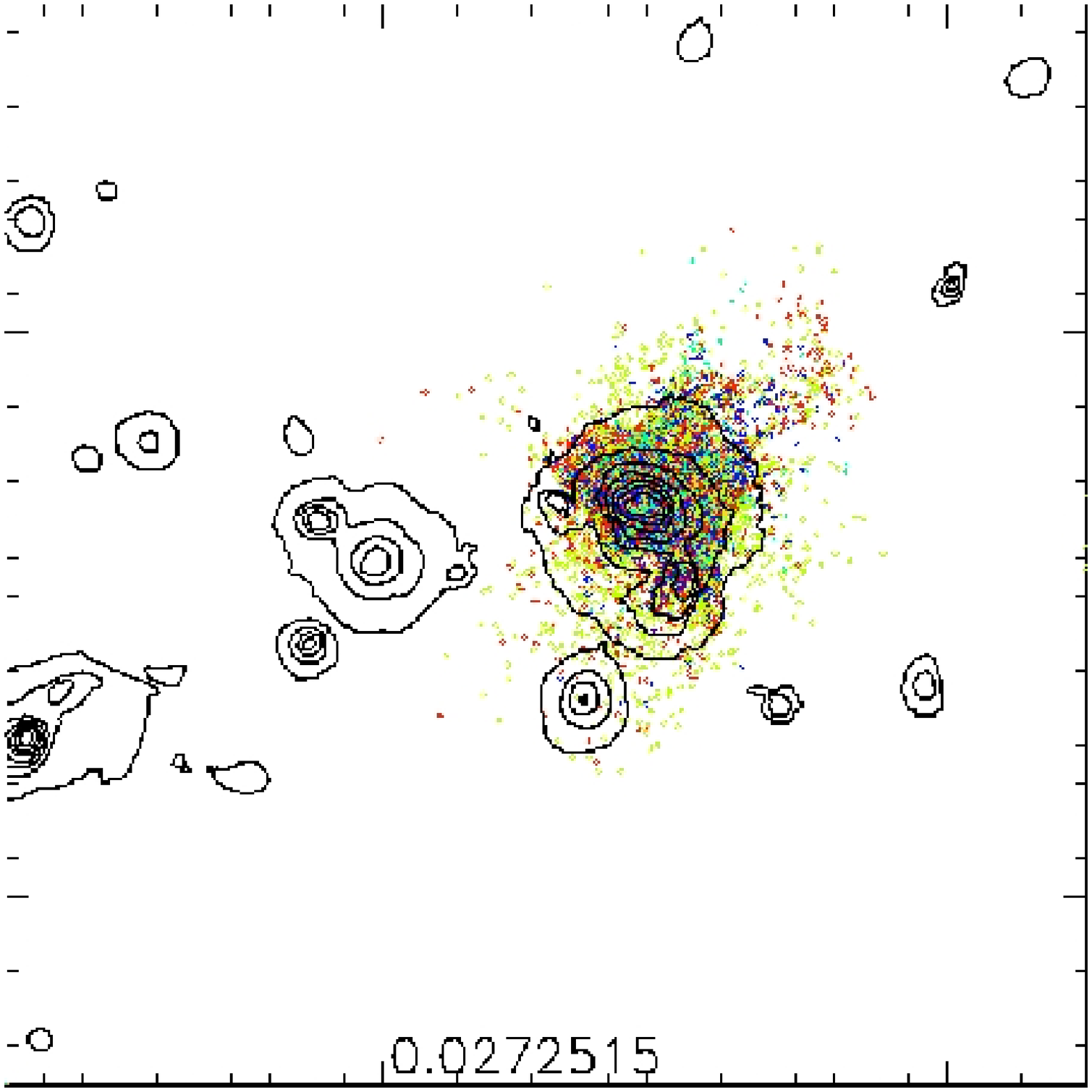}
\includegraphics[width=0.24\textwidth]{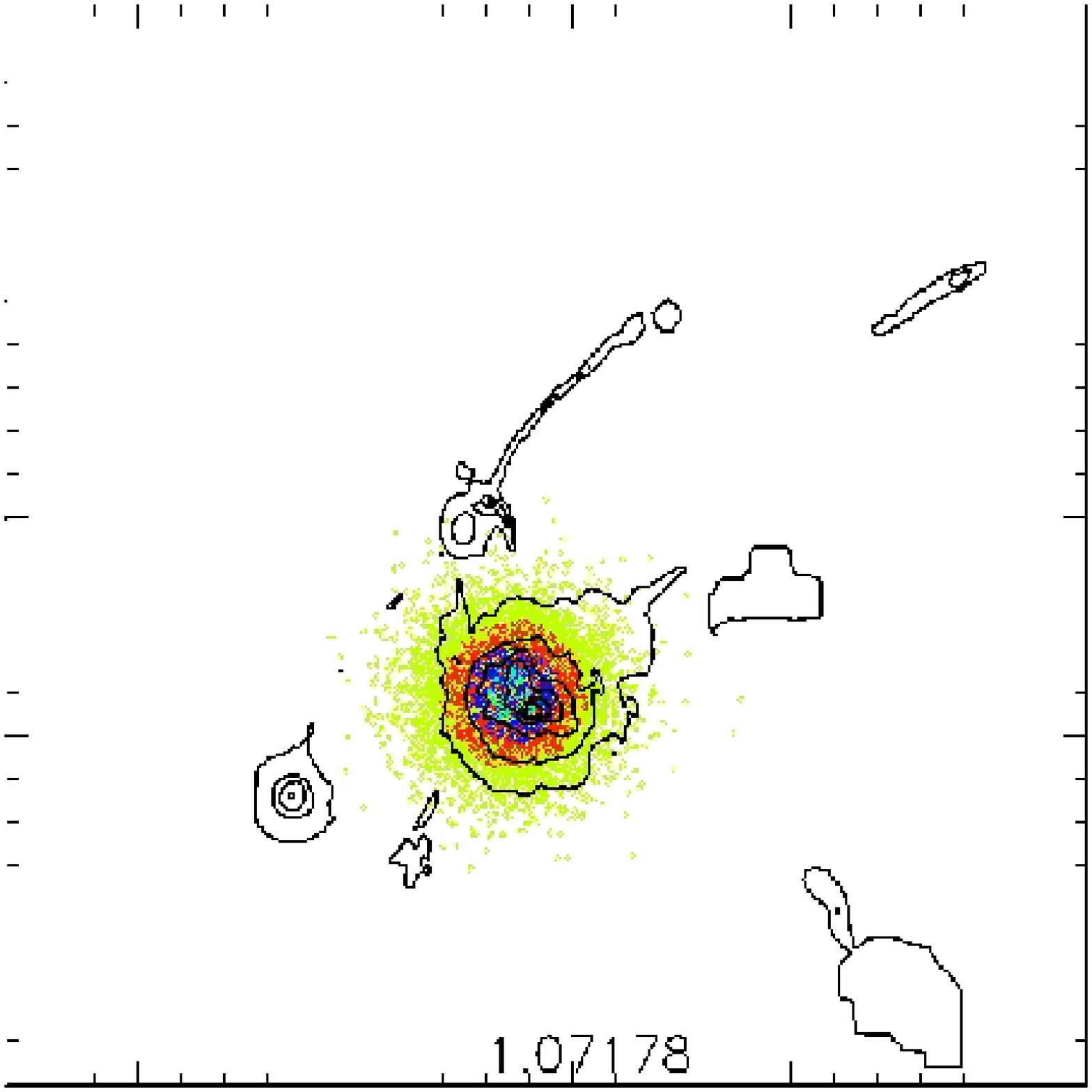}
\includegraphics[width=0.24\textwidth]{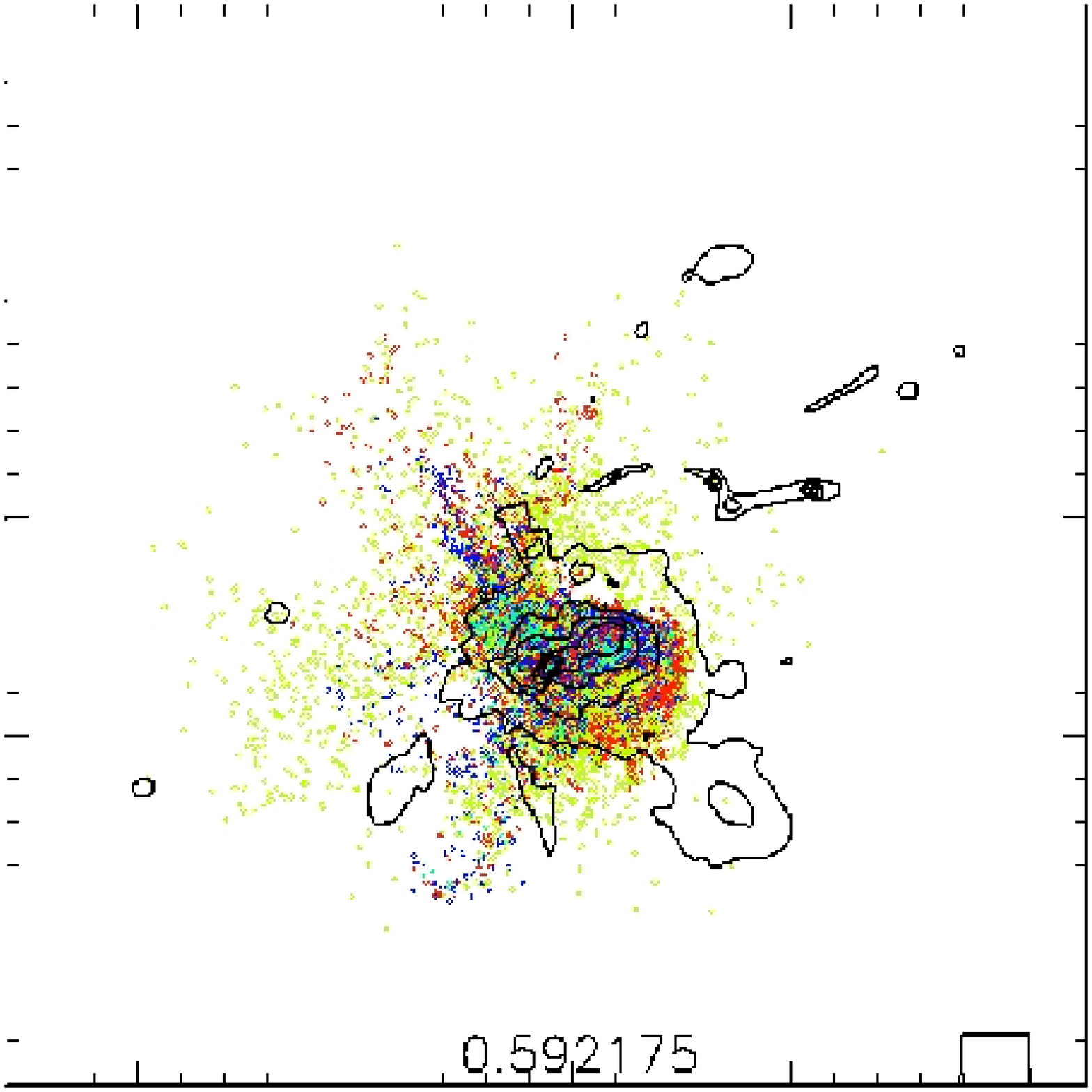}
\includegraphics[width=0.24\textwidth]{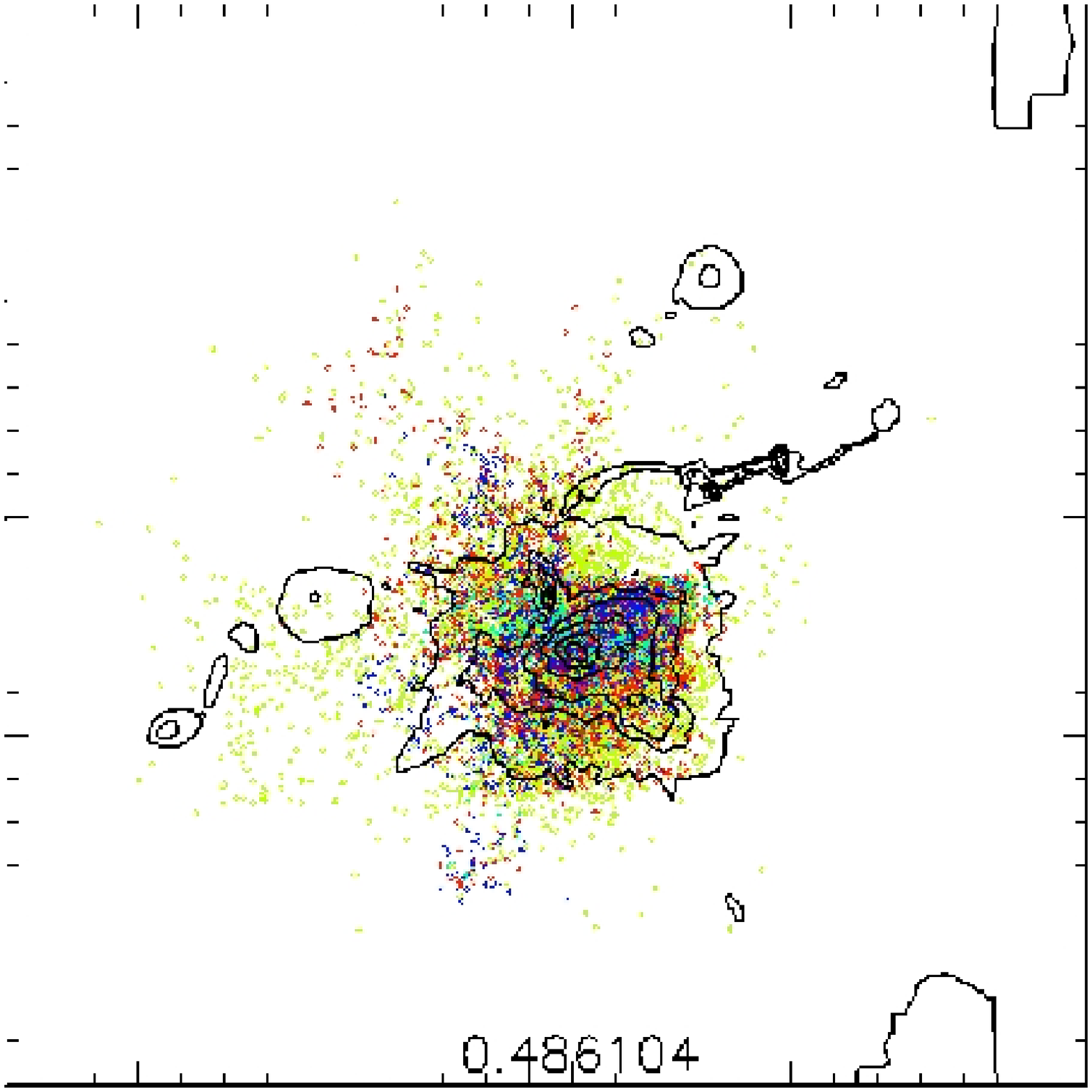}
\includegraphics[width=0.24\textwidth]{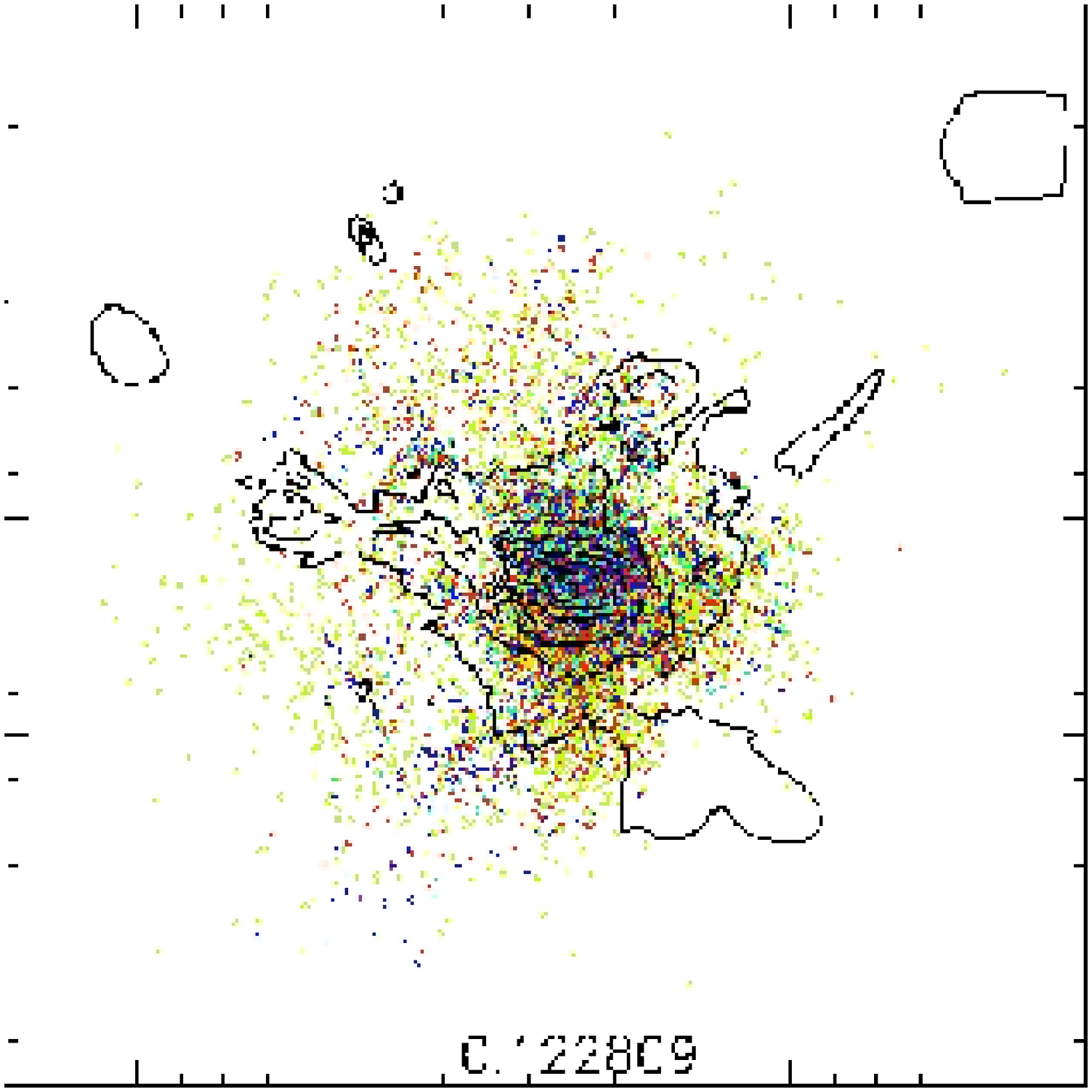}
\includegraphics[width=0.24\textwidth]{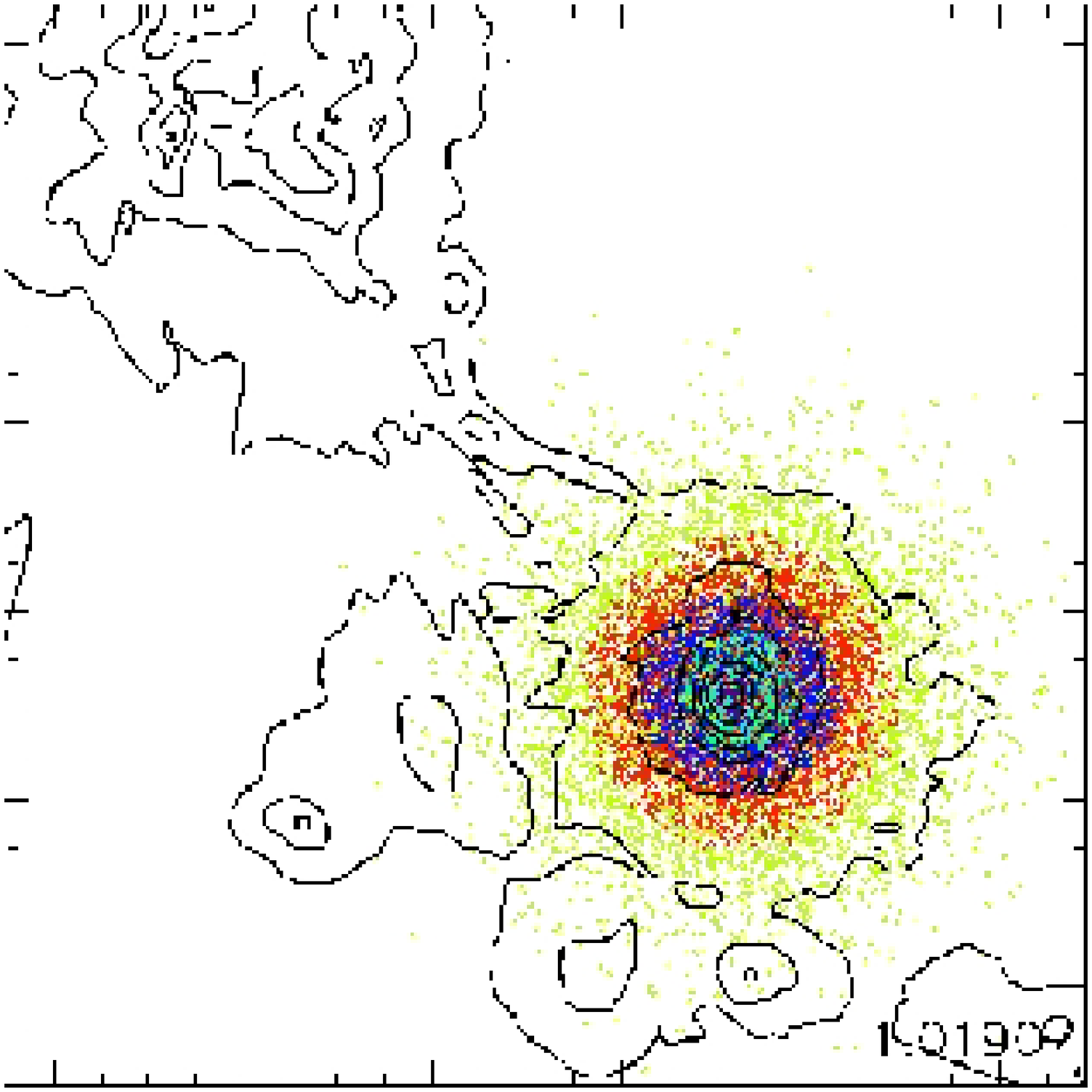}
\includegraphics[width=0.24\textwidth]{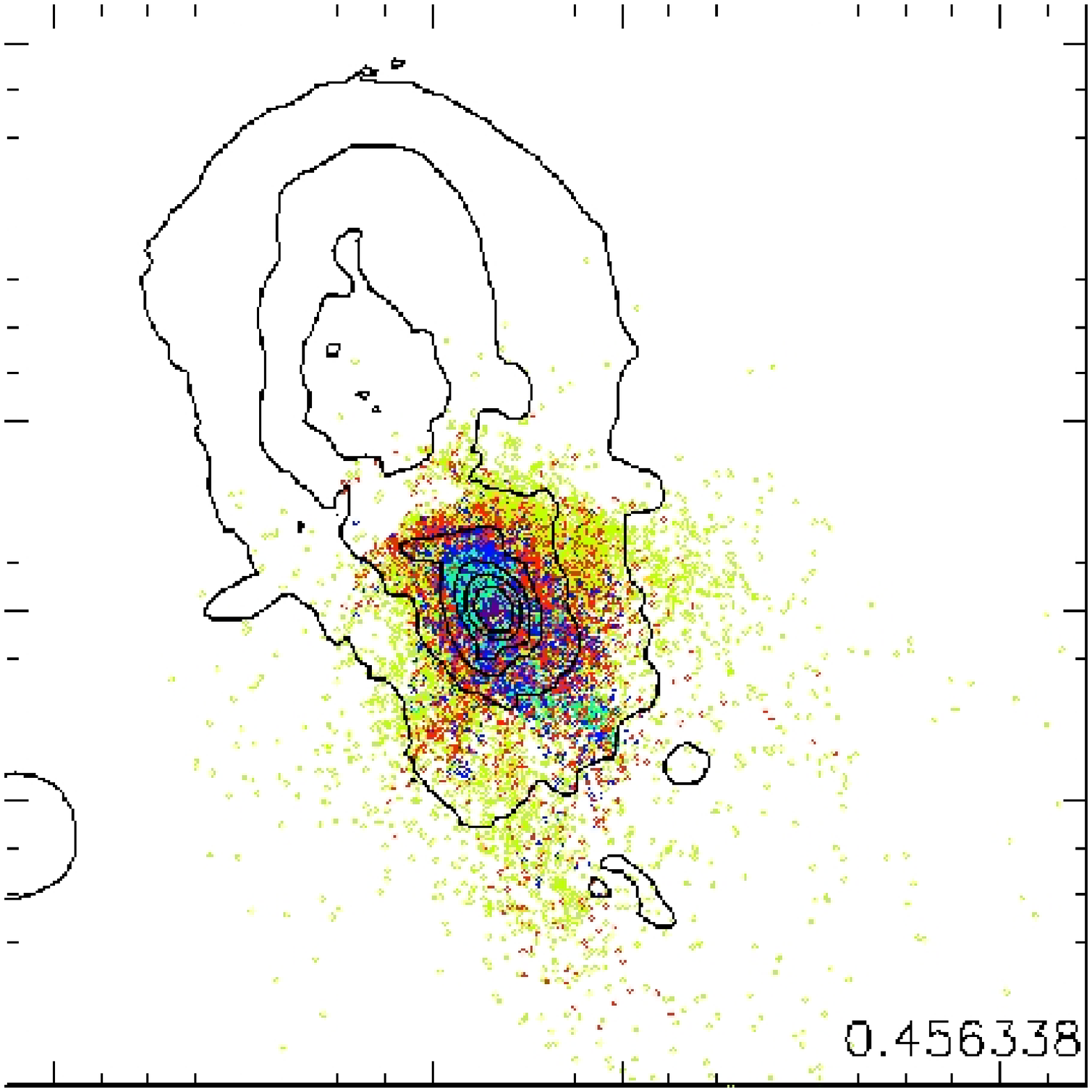}
\includegraphics[width=0.24\textwidth]{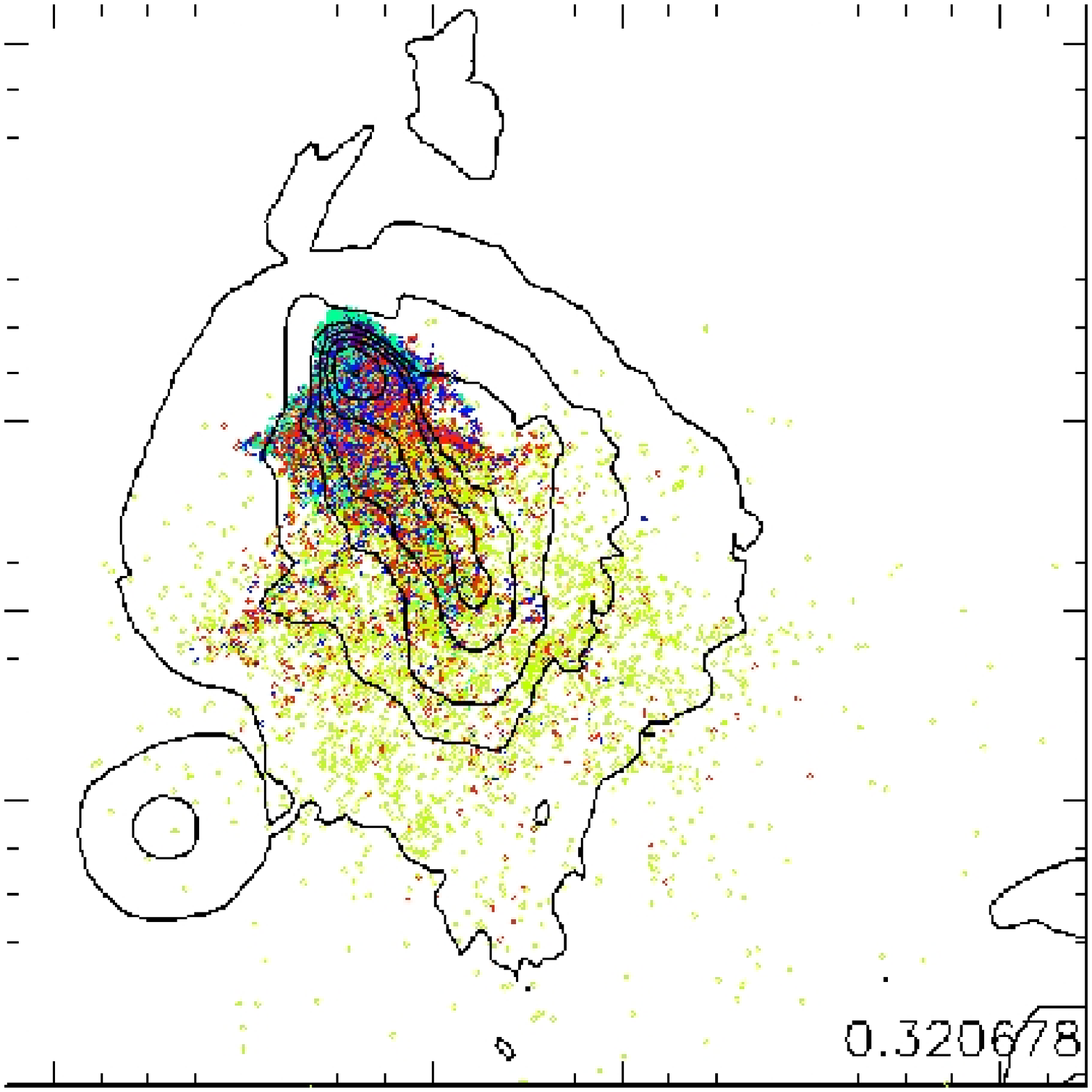}
\includegraphics[width=0.24\textwidth]{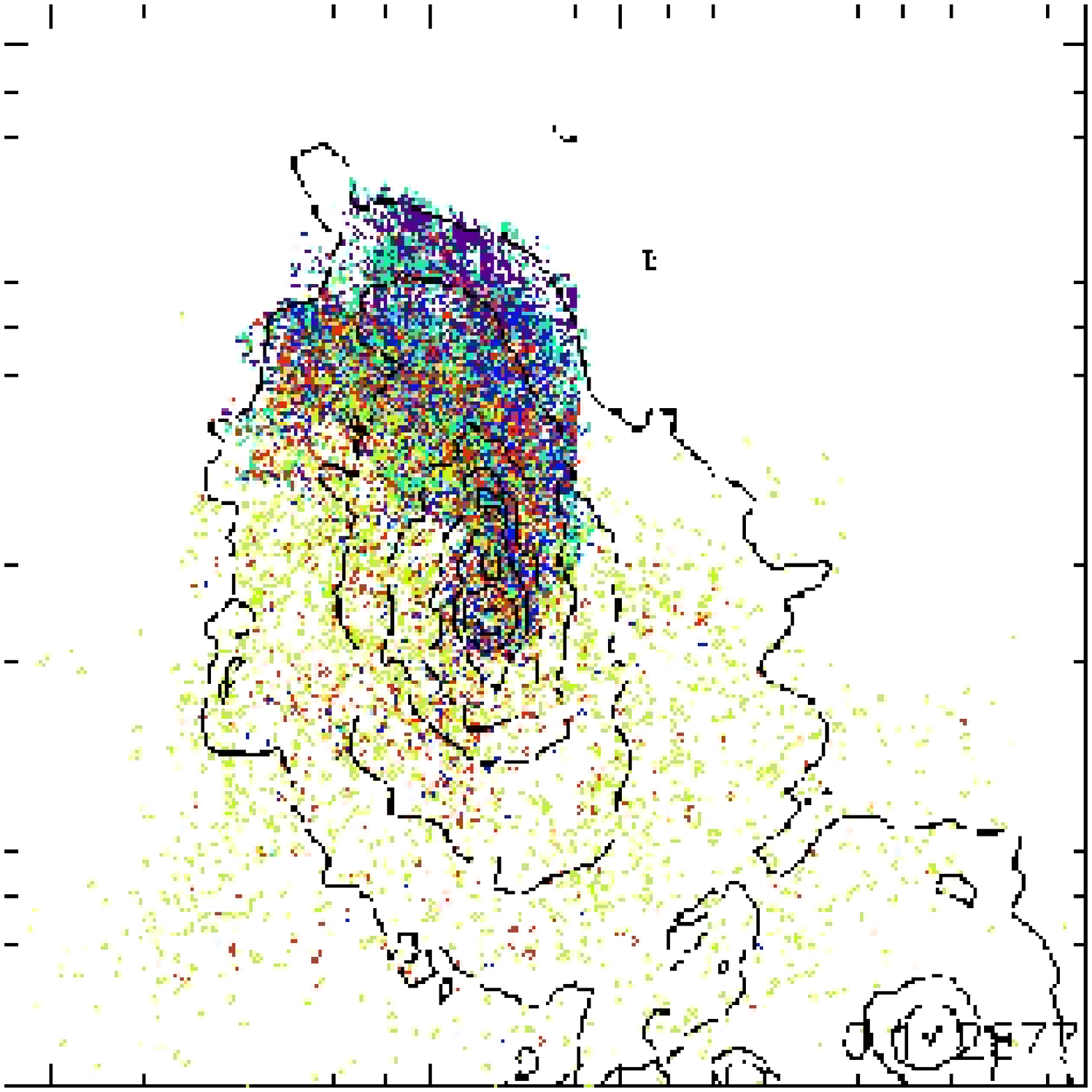}
\caption{Evolution of projected positions of  tracer 
families for run H3 (top panels),
H1 (central panels) and H5 (lower panels). Every color refers to a different
shell of origin.
Overlaid are the contours of the projected
gas density (the first contour is for $\rho_{gas}=50 \rho_{cr}/f_{b}$, where
  $\rho_{cr}$ is the critical density of the Universe and $f_{b}$ is the 
baryon fraction; contours are spaced with $\sqrt{2}$ intervals), the side of the image and the LOS are $15Mpc$ (top and central
panels) and $12Mpc$ (lower panels), respectively.}
\label{fig:map_mixing_fami}
\end{center}
\end{figure*}

\subsection{Tracers dispersion statistics}
\label{subsec:dispersion}

The presence of complex and turbulent velocity fluctuations in the simulated ICM
has been established in a number of works which were performed with very different
numerical techniques (e.g. Norman \& Bryan 1999;
Dolag et al.2005; Vazza et al.2006; Iapichino \& Niemeyer 2008; Lau et al.2008;
Vazza et al.2009; Xu et al.2009).
Using the same AMR technique applied in this work, Vazza et al.(2009) were
able to measure the 3--D velocity power spectrum of a simulated
galaxy cluster over a range of $\sim 100$ in spatial scales. 
Although cosmological simulations provide a simplified view of the ICM, 
the presence of subsonic turbulent velocity fields in clusters is 
in line with existing observations (e.g. Schuecker et al.2004; Henry
et al.2004; Sanders et al. 2009).

Complex motions in the ICM affect the mixing and transport processes in 
the baryon gas. 
In Fig.\ref{fig:disp_pathz} we show the evolution of the projected position
of three random couples of tracers, injected at $z=30$ with an initial
separation of 36kpc.  
The tracers follow a rather laminar path in the first stage of their evolution,
while their paths become tangled and fairly different when they enter the collapsing region
of the forming cluster.

\bigskip

As a guide reference to understand the motions
of tracers within the complex gas/DM velocity field in galaxy clusters one
may use studies customarily performed for isotropic and incompressible
turbulence (e.g. Bec et al.2006; Bec et al.2009 and references therein). 
In this case the {\it tracer pair dispersion} statistics
(i.e. $P(t) \equiv  <|\vec{X_{1}(t)}-\vec{X_{2}(t)}|>$,
where $\vec{X_{1}(t)}$ and $\vec{X_{2}(t)}$ are the position of 
couples of tracers at any time) are expected to show a simple behavior 
with time with well-known regimes
(e.g. Elhmaidi et al.1993; Zouari \& Babiano 1994; Schumacher \&
Eckhardt 2002). 
The dispersion has an initial ballistic regime $\sim t$ as long as
the particles lie within the viscous sub-range. 
For times larger than the Lagrangian integral scale 
(i.e. the time associated with the auto-correlation function of the Lagrangian
velocity) the relative distance between tracers increases diffusively
as in an uncorrelated Brownian motion, i.e. $\sim t^{1/2}$.
For intermediate time scales  the 
dispersion follows the Richardson law, $\sim t^{3/2}$ (Richardson 1926).

Figure \ref{fig:disp_veg} shows the evolution between $z=30$ and $z=0.5$ of the pair
dispersion statistics derived with 
$N=10^{6}$ couples of tracers in the same cluster run. 
Tracers are initially 
placed with a random
sampling of the grid at $z=30$ and for three different initial separations 
between the pairs: $|\vec{X_{0}}=\vec{X(t=0)}|=36$, $72$ and $288$ kpc. 
After $\approx 1.5 Gyr$ the trend of the
pair dispersion loses any dependence on the initial separation, and the time
evolution of the pair dispersion becomes consistent with 
a $\propto t^{3/2}$ scaling.

This a general behavior found in our clusters, as shown in the left
panel of Fig.\ref{fig:disp_L1}.
As soon as the cluster formation starts (i.e. $z\leq 2$),
the pair dispersion increases following a behavior consistent with
a $\sim t^{3/2}$ scaling, although significant scatter and episodic
variations across this scaling are found. 
In the same panel, we additionally show the evolution of the 
pair dispersion statistic for cluster H7, by using the implicit 
second order interpolation scheme introduced in Sect.\ref{subsec:interp} (Eq.\ref{eq:2order});
the reported trend is independent of the time interpolation
scheme, and the scatter between the two methods is much smaller than
the pair dispersion at all times.

Even if the mass accretion
histories of clusters and their velocity field may be different,
the emergence of a common behavior suggests that the transport of 
tracers due to gas motions in the ICM is similar for most of the lifetime 
of simulated clusters. 
 In the right panel of Fig.\ref{fig:disp_L1} we show the 
3--D power spectrum, $E(k)$, of the velocity field at $z=0$ for all clusters 
of the data sample. 
The curves of $E(k)$ are calculated with a
standard FFT-based approach, 
adopting a zero-padding technique to account for the non-periodicity
of the considered volume (see also Vazza et al., 2009). 
 We also verified tat the use of some apodizing function, which that would help
to avoid spurious effects due to the sharp cut at the boundaries of the AMR region,
provides only negligible changes in the estimate of $E(k)$ (and only at scales
close to the Nyquist frequency of the spectra), since the average velocity
fields at the boundaries of the AMR region are much smaller compared to the
velocity field inside clusters.
In order to emphasize the similarity between our clusters, the spatial
frequency $k$ has been rescaled to that of the the virial radius of each cluster, $k_{0} \approx 2\pi/R_{vir}$. 
$E(k)$ shows a maximum scale 
in the range corresponding to $\sim 1-2 R_{vir}$ and a well-defined 
power-law behavior for nearly two orders of magnitude in spatial scale; 
the flattening found from scales corresponding to $\sim 1/(8 \Delta)$
is due to the numerical dissipation in the PPM scheme (e.g.
Porter \& Woodward 1994; Kitsionas et al.2009). 

The power spectrum in the velocity field is due to the combination
of the turbulent motions inside the cluster volume with the pattern of 
matter accretions that surrounds the cluster volume, and with
the laminar infall motions that penetrate inwards. 
The maximum coherence length of this complex velocity field is constrained
by the size of the clusters themselves and this easily explains why the power spectrum
peaks at 1-2 $R_{vir}$. 
Tracer particles are frozen inside this ``correlated'' velocity
field, and the distance between two tracers is always smaller than
the maximum scale of $E(k)$: this explains the Richardson-like
behavior of the pair dispersion reported in Figs.\ref{fig:disp_veg}-\ref{fig:disp_L1}.
The complex velocity field in the volume of simulated clusters
allows for a fairly fast transport of particles, because large scale
motions are strong, with particles separating by $\approx 100-200$ kpc
distances in about 1Gyr.
We are therefore lead to the conclusion that in the analyzed set of simulated
galaxy clusters the efficiency of particle transport
is much larger than that due to thermal diffusion
(e.g.Shtykovskiy \& Gilfanov 2009), and that the typical scale
of transport is larger than the scale of turbulent diffusion induced
by central AGN activity (e.g. Rebusco
et al.2005). Therefore the effect of large scale turbulent
and mixing motions in the cluster volume is likely a key
ingredient in any practical modeling of gas 
mixing processes in the ICM.

\begin{figure} 
\includegraphics[width=0.48\textwidth]{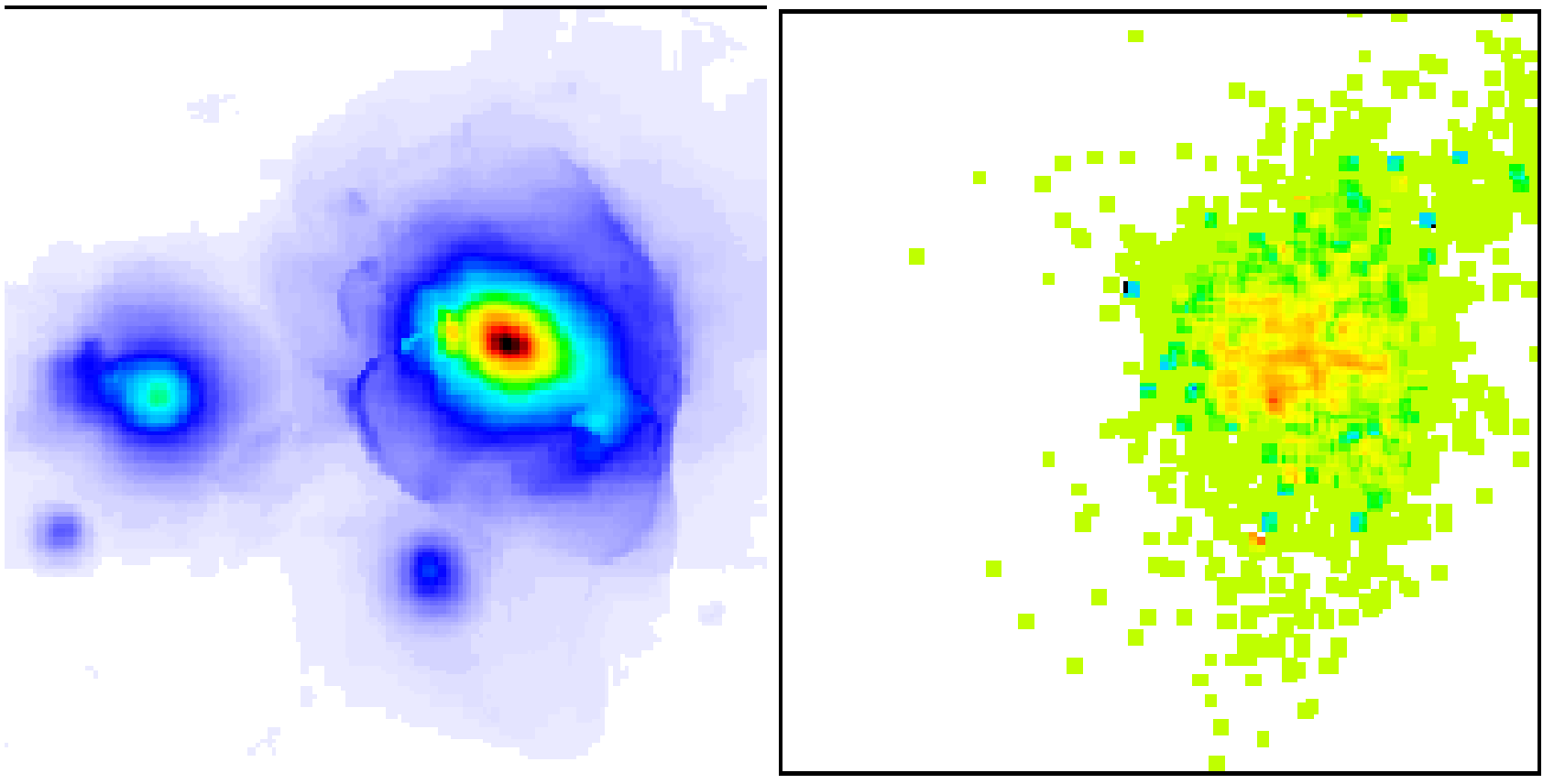}
\includegraphics[width=0.48\textwidth]{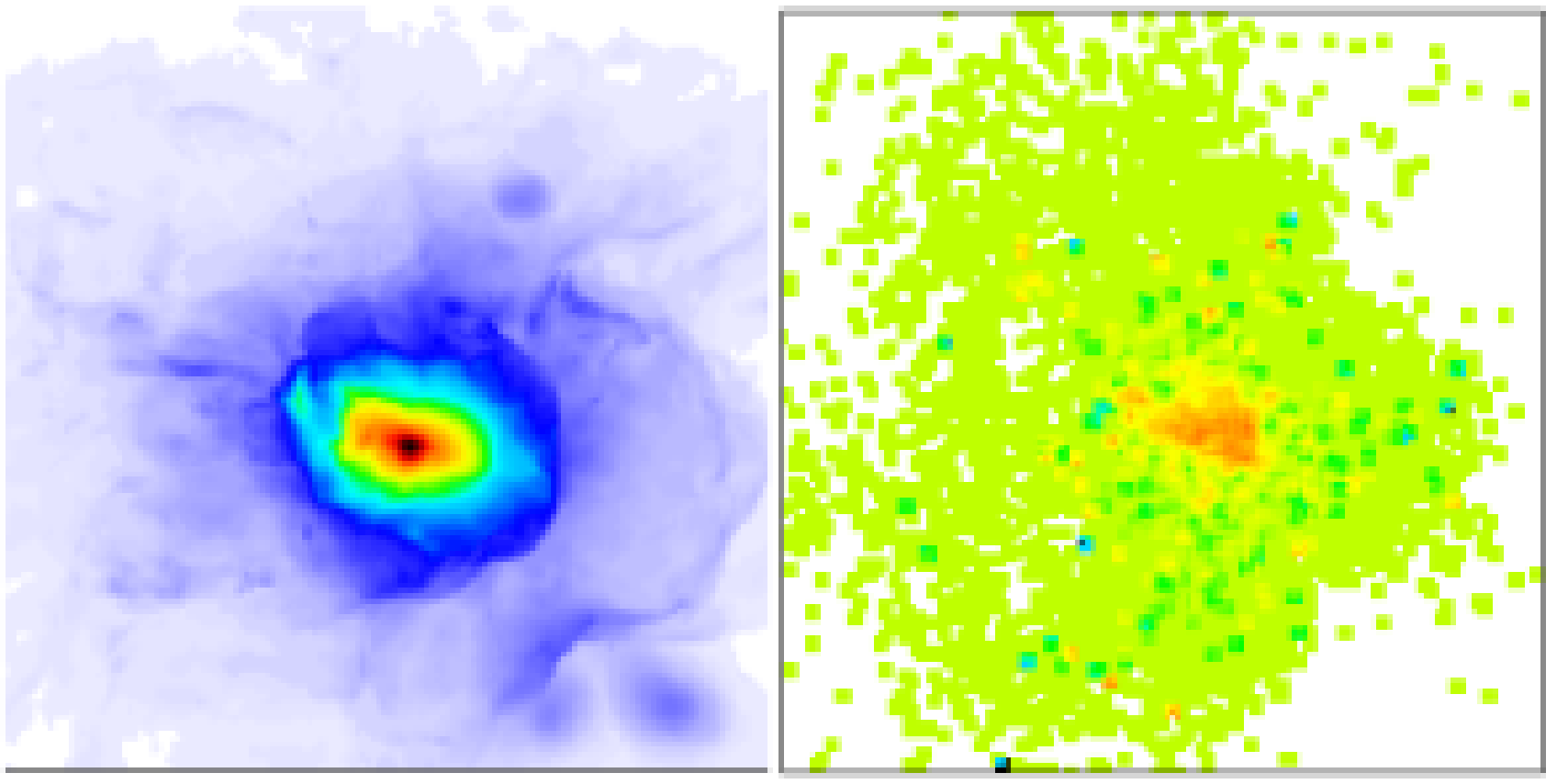}
\includegraphics[width=0.48\textwidth]{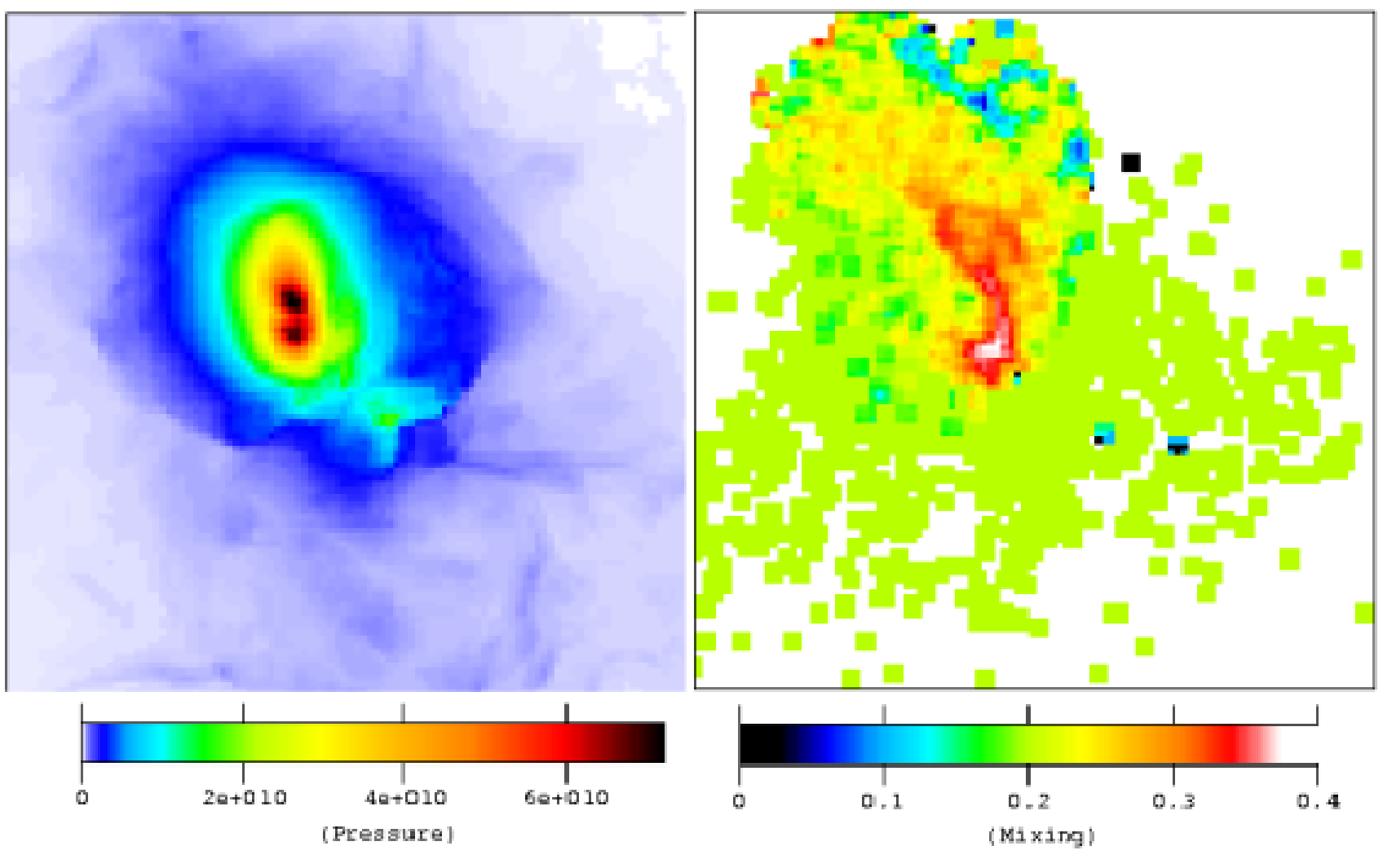}
\caption{Map of projected gas pressure (left panels) and of projected
tracers mixing (right) panels, for run H3 (top), H1 (center) and H5 (bottom)
at $z=0$. The side of the images and the LOS are 5Mpc;
the mixing maps are produced by considering only cells containing more than one
tracer.}
\label{fig:map_mixing}
\end{figure}


\begin{figure} 
\includegraphics[width=0.24\textwidth]{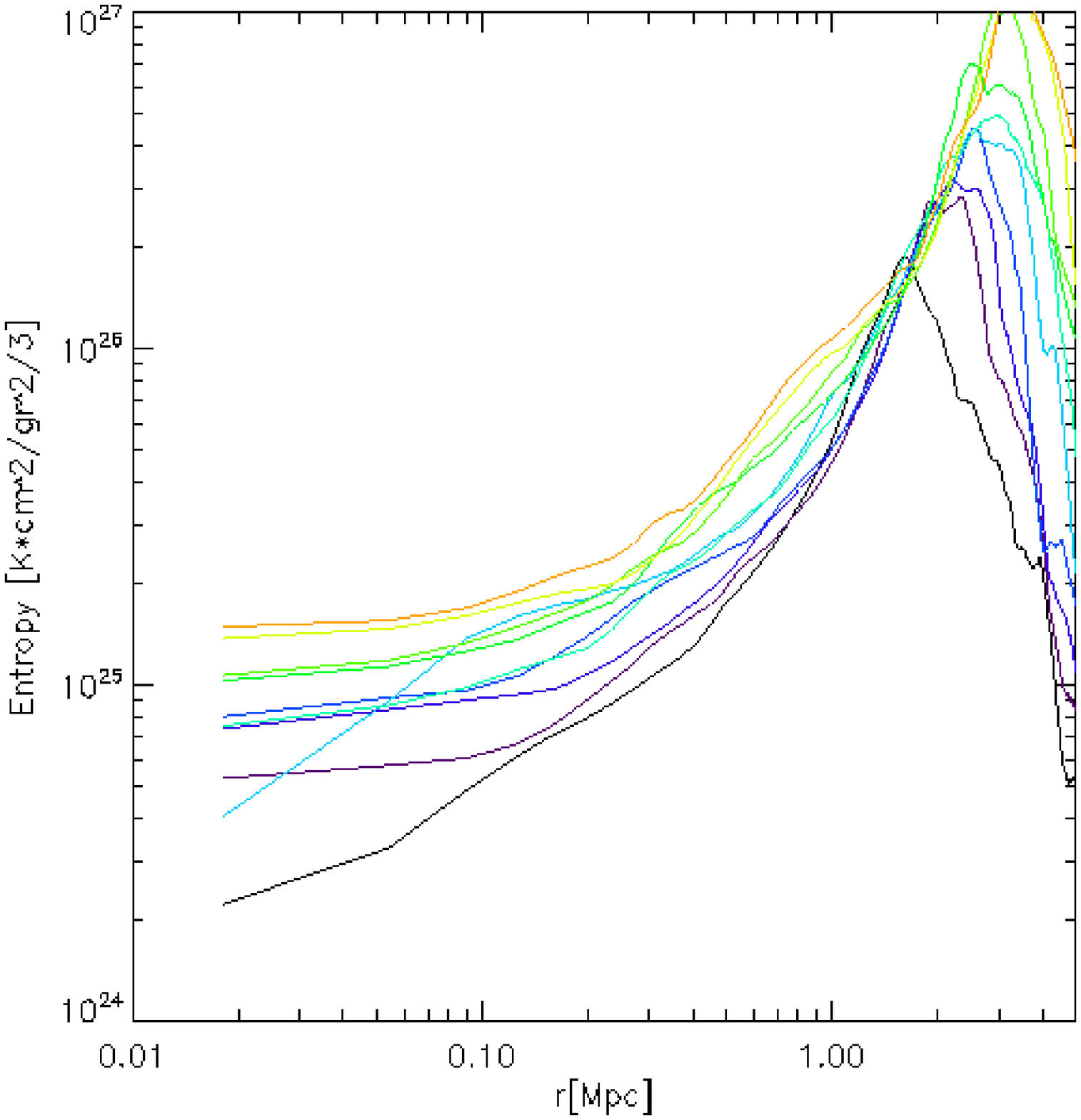}
\includegraphics[width=0.24\textwidth]{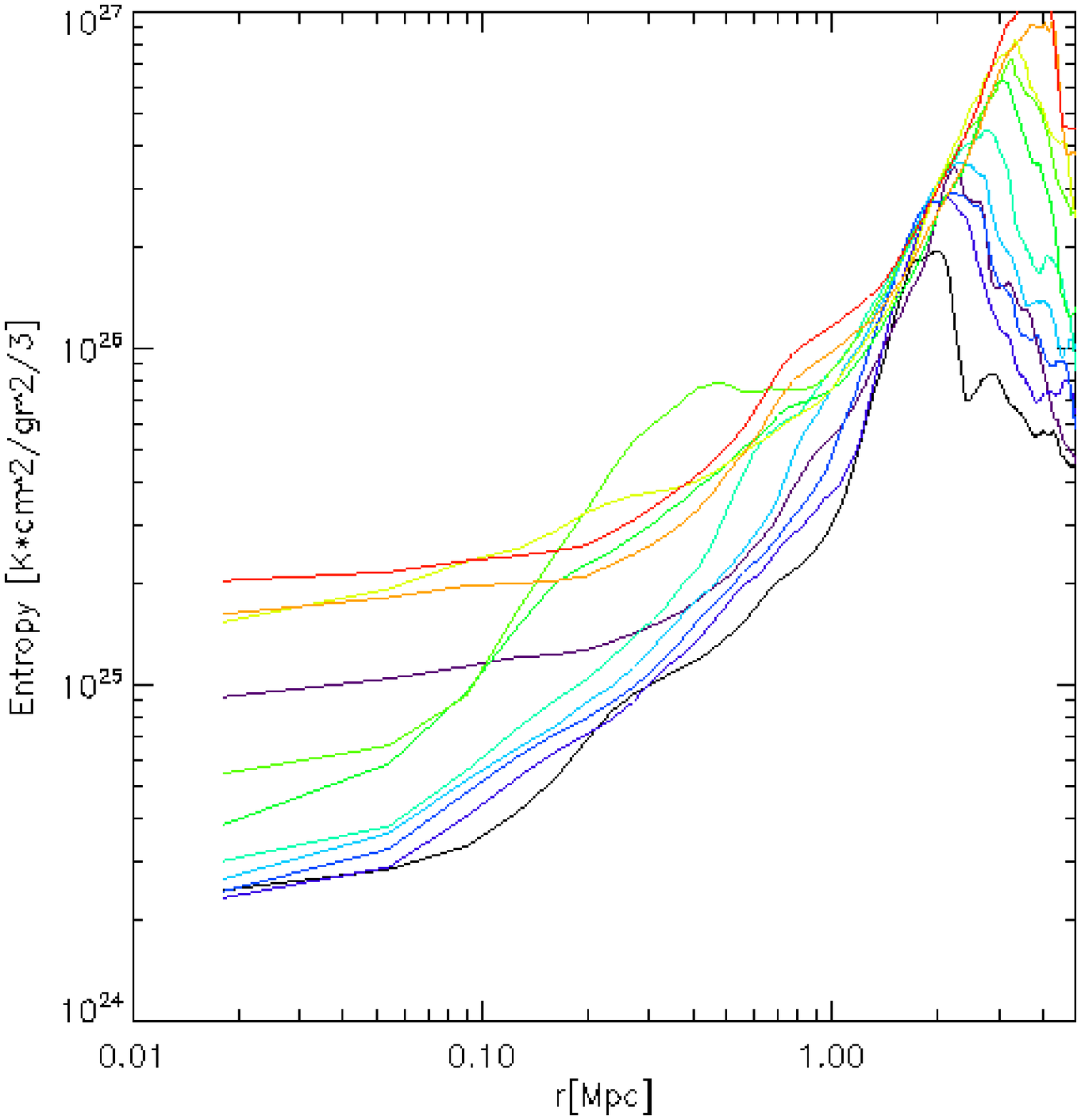}
\includegraphics[width=0.24\textwidth]{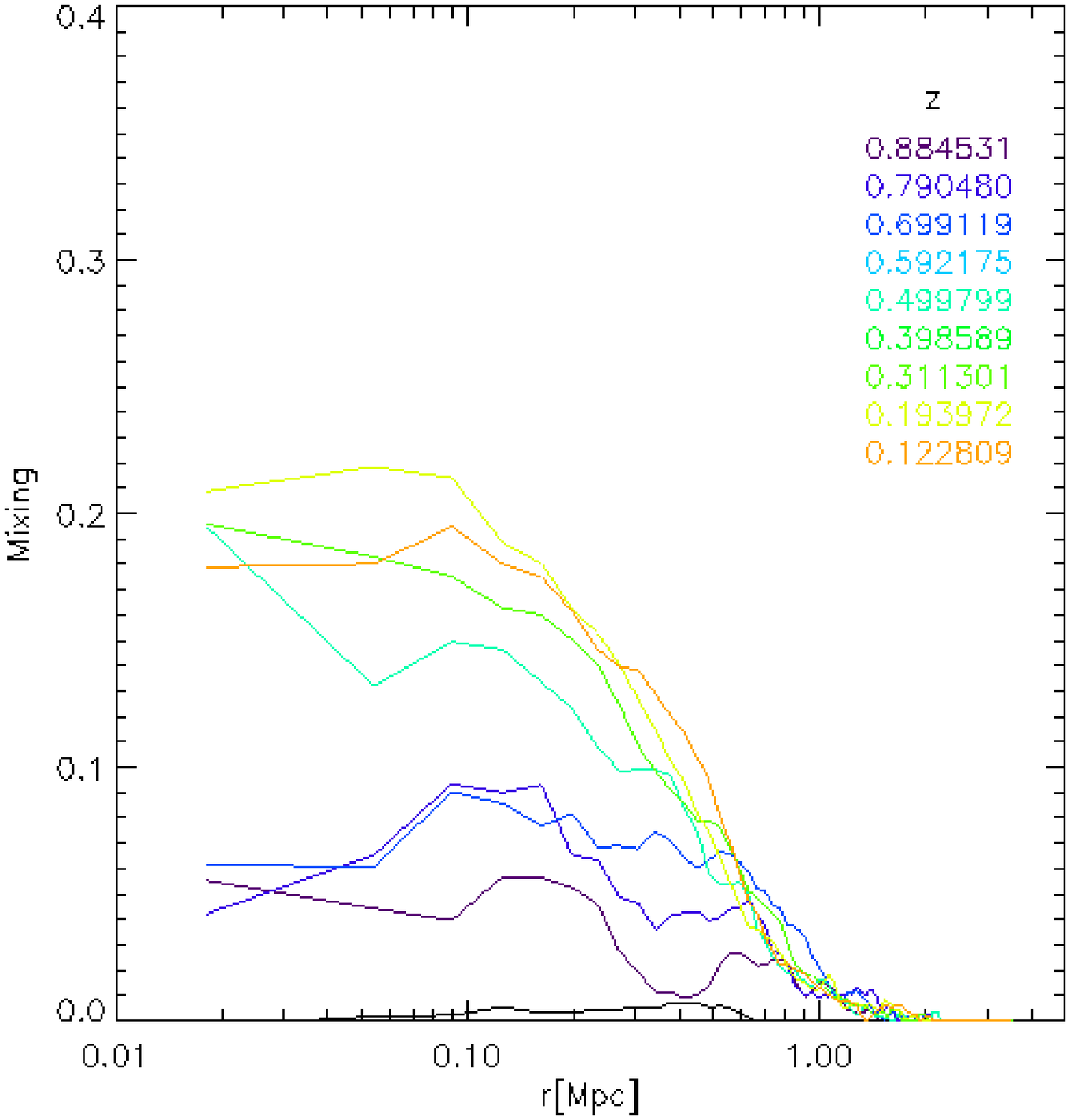}
\includegraphics[width=0.24\textwidth]{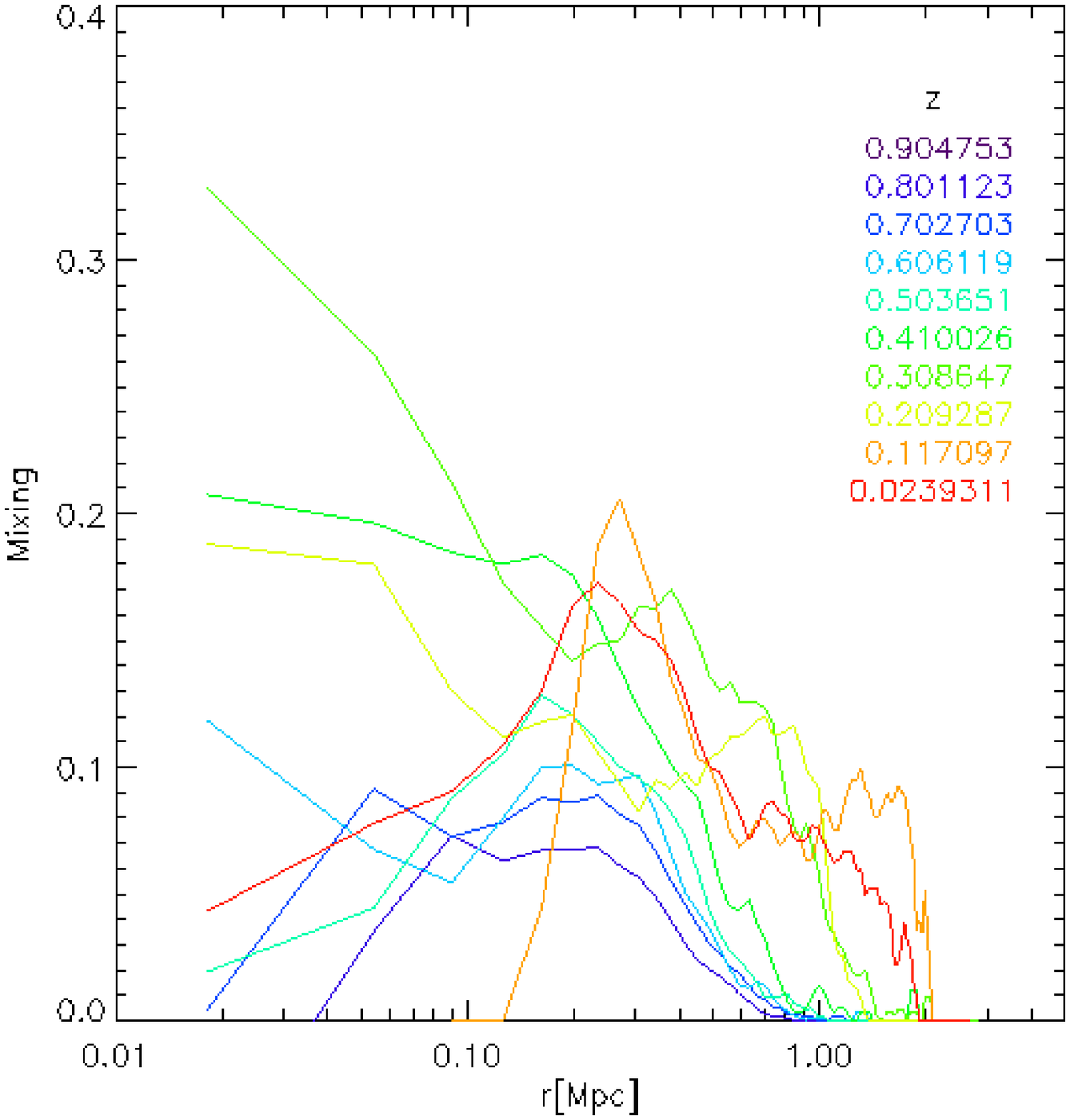}
\caption{Evolution of gas entropy profile ({\it top panels}) and of
mean mixing ({\it bottom panels}) for 
 H1 (left) and H5 (right).}
\label{fig:map_mixing3}
\end{figure}

\begin{figure} 
\includegraphics[width=0.45\textwidth]{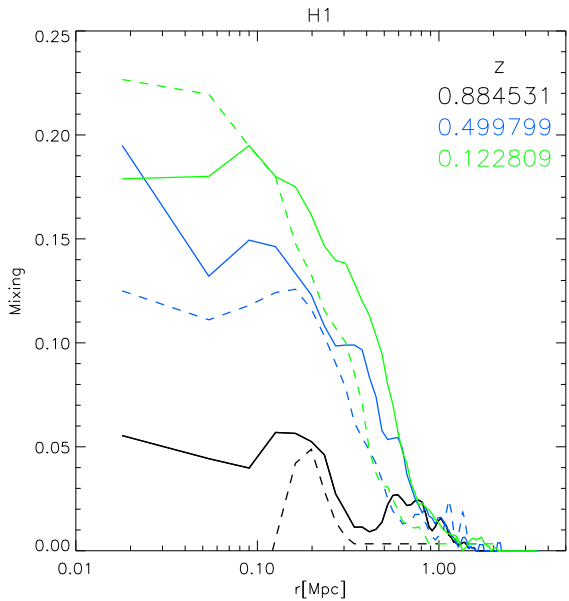}
\caption{Evolution of mean mixing profile for cluster H1 at three
redshifts, for the standard mesh refinement strategy ({\it dashed lines})
and for our implemented mesh refinement strategy ({\it solid lines}).}
\label{fig:map_mixing4}
\end{figure}

\subsection{Mixing}
\label{subsec:mix}

The "volume-averaged" statistics examined in the previous 
sections 
show that the transport of gas particles in the ICM 
is 
similar from cluster to cluster.
We might however question if
the final spatial distributions of tracers show any
dependence on the dynamical state of the host cluster. 
We explored this issue by applying a simple initial
tracers setup in a subset of representative objects of our sample
and focused on the 
evolution of radial mixing (e.g. mixing of tracers originated at different
distance from the center of clusters) as a function of their dynamical
evolution.

A number of $\approx 10^5$ tracers was randomly placed within the cluster
with a number density that follows the gas density
profile of the clusters at $z=1$.
We divided every tracer generation into 
ten families, by dividing the initial distribution to ten concentric
shells with an equal number of tracers and we followed the evolution
of the system by keeping the information of the shell of origin
for each tracer.

In Fig.\ref{fig:map_mixing_fami} we show maps of the tracer evolution
in clusters
H3, H1 and H5. 
These objects are good representatives of
typical evolutive histories in our simulations: H3 does not experience any
significant accretion of gas/DM satellites after $z \leq 0.2$, and is 
fairly relaxed at $z=0$;  
H1 is interested by continuous accretions of gas/DM satellites 
and presents an elongated structure at the final epoch of observation; 
H5 experiences a major merger at $z \sim 0.4$, and retains
a very perturbed dynamical state even at $z=0$.

We quantified the 
the degree of mixing 
in every cell between tracers of "s" families and the other tracers 
through

\begin{equation}
{\it M_{s}} = 1-(|n_{s}-\sum{n_i}(i \neq s)|/\sum n_{i}),
\end{equation}

where $n_{s}$ is the number density of the tracers within a cell (at the highest resolution
level)
and the sum refers to all the species of tracers. 
This formula generalizes the more simple case of mixing between two species
(e.g. Ritchie \& Thomas 2002).

The total mixing in each cell, {\it M}, is the volume average between all
species, ${\it M} = \sum_{s} {\it M_{s}}/N_{s}$, where $N_{s}$ is the number
of families considered. The interpretation of this parameter is simple: for a cell where $n_{1} \approx
n_{2} \approx n_{3} ...\approx n_{10}$ the different families are well mixed 
and we have ${\it M} \rightarrow 1$, while
${\it M} \rightarrow 0$ implies no mixing within the cell.

Figure \ref{fig:map_mixing} shows the projected distributions of the mixing
parameter ${\it M}$ at $z=0$ for 
the three clusters considered above together
with projected maps of gas pressure along the line of sight.
We found that the differences in the dynamical history of the three
objects produce different spatial distribution of mixing at $z=0$. 
H3 shows a uniform pattern of mixing, with ${\it M} \sim 0.2-0.3$ across the
whole cluster core. On the other hand  
H5 presents the most patchy mixing map with a large
plume of the size of $\sim 2$Mpc in the direction of the major merger
and a maximum mixing of ${\it M} \sim 0.4$ between the cores of merging
clusters.


In Fig.\ref{fig:map_mixing3} we compare the evolution of the gas 
entropy profiles for H1 and H5 (top panels) with that of the mixing
profiles (bottom panels).
We found that efficient mixing is always confined in central low entropy regions. 
The very regular evolution of mixing in H1
closely reflects the regular evolution of the gas entropy
profile. At the final epoch of observation, a flat mixing
profile with ${\it M} \approx 0.2$ was found within the 
entropy core at $r<100kpc$, and outside this core
${\it M}$ falls quickly to very low values.
This is not surprising: in the idealized ICM represented by
these simulations, convective stability to radial displacement
is provided
by the classic Schwarzschild criterion, 
$dlog(S)/dr>0$ (Schwarzschild 1959). A very flat entropy
core is just marginally stable to radial perturbations, whereas
the steep rise of the entropy profile at $r>100kpc$
provides a very stable configuration to radial perturbations.
The situation is more complex for the major merger system
H5. The merger event at $z\sim 0.4-0.5$ affects the entropy profile and
also drives a more extended pattern of mixing. At 
the final epoch a peak of mixing is found 
at $r \sim 300kpc$, and significant mixing is found even out of $r=1-2Mpc$.

This large mixing pattern in extended merger systems 
confirms previous works that analyzed idealized 2-body encounters with the SPH
techniques (e.g. Ritchie \& Thomas 2002; Ascasibar
\& Markevitch 2006), and thus
our exploratory study extends their results to fully
cosmological Eulerian simulations. In addition, as 
shown by Mitchell et al.(2009), major mergers in Eulerian
AMR codes are more efficient  
in mixing the ICM compared to mergers in SPH
codes, since in SPH mixing during the central stages of the merger
is significantly prevented by the suppression of 
instabilities by the artificial
particles viscosity; our findings seem to support the presence of large
eddies that provide efficient mixing during the central stages of cluster 
mergers.
It is important to compare results for the evolution of mixing obtained
through our refinement
strategy (Sect.\ref{subsec:amr}) with those obtained with
a standard mesh refinement strategy. For this purpose we re-simulated
cluster H1, allowing only for mesh refinement according
to the gas/over-density criterion and applied the same initial setup
of spherical shells of tracers to measure the evolution of mixing
at all times (Fig.\ref{fig:map_mixing4}).
Even if the overall behavior is similar, the refinement criterion
using velocity jumps and gas/DM over-density produces a larger mixing
especially at large radii, and the progressive accretion of this gas also 
induces large mixing in the innermost cluster regions at later redshifts.
This
clearly shows that the artificial suppression of chaotic motions 
due to coarse numerical resolution may reduce
in a sizable way the mixing in the simulated ICM, even within the 
virial volume of clusters.

\begin{figure} 
\includegraphics[width=0.24\textwidth]{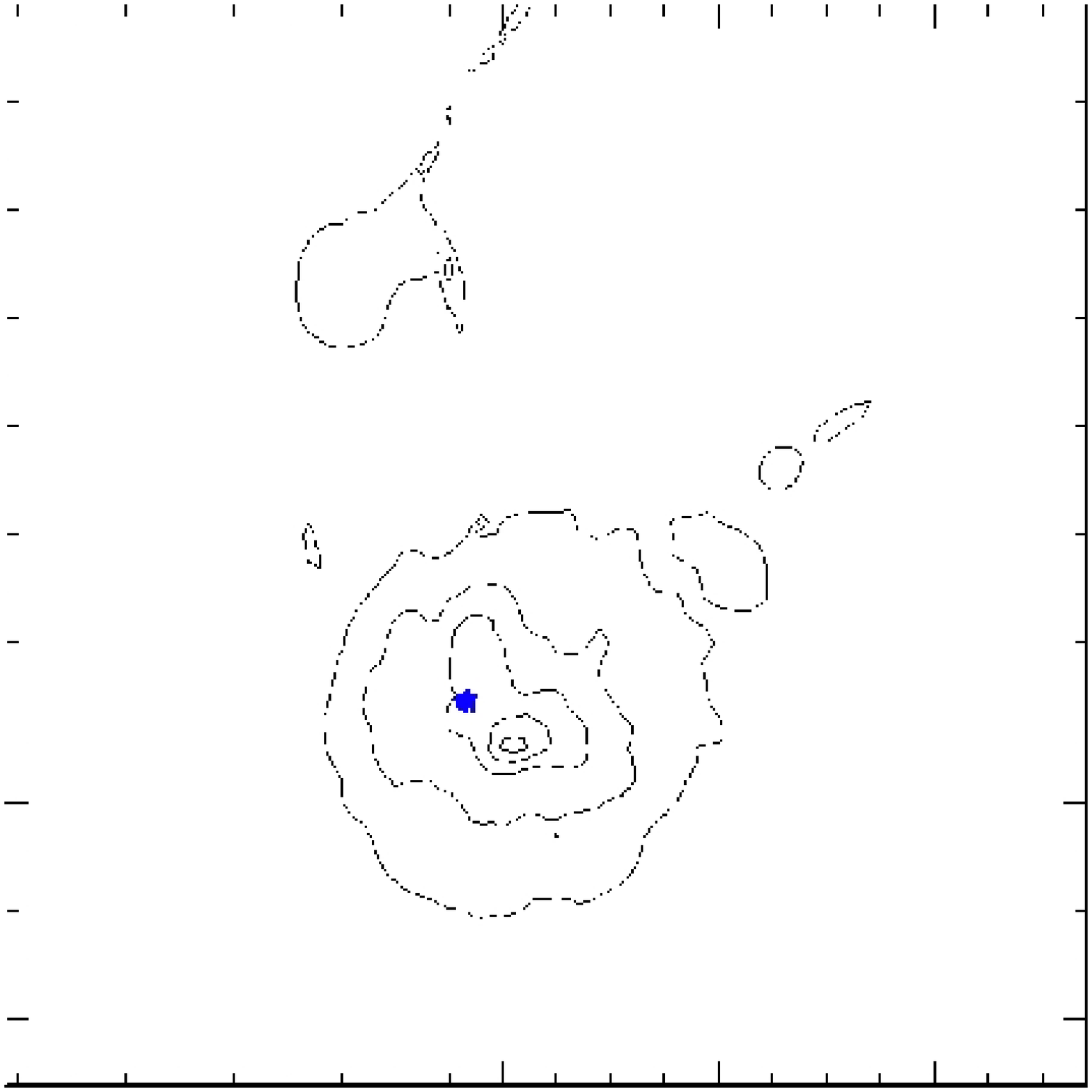}
\includegraphics[width=0.24\textwidth]{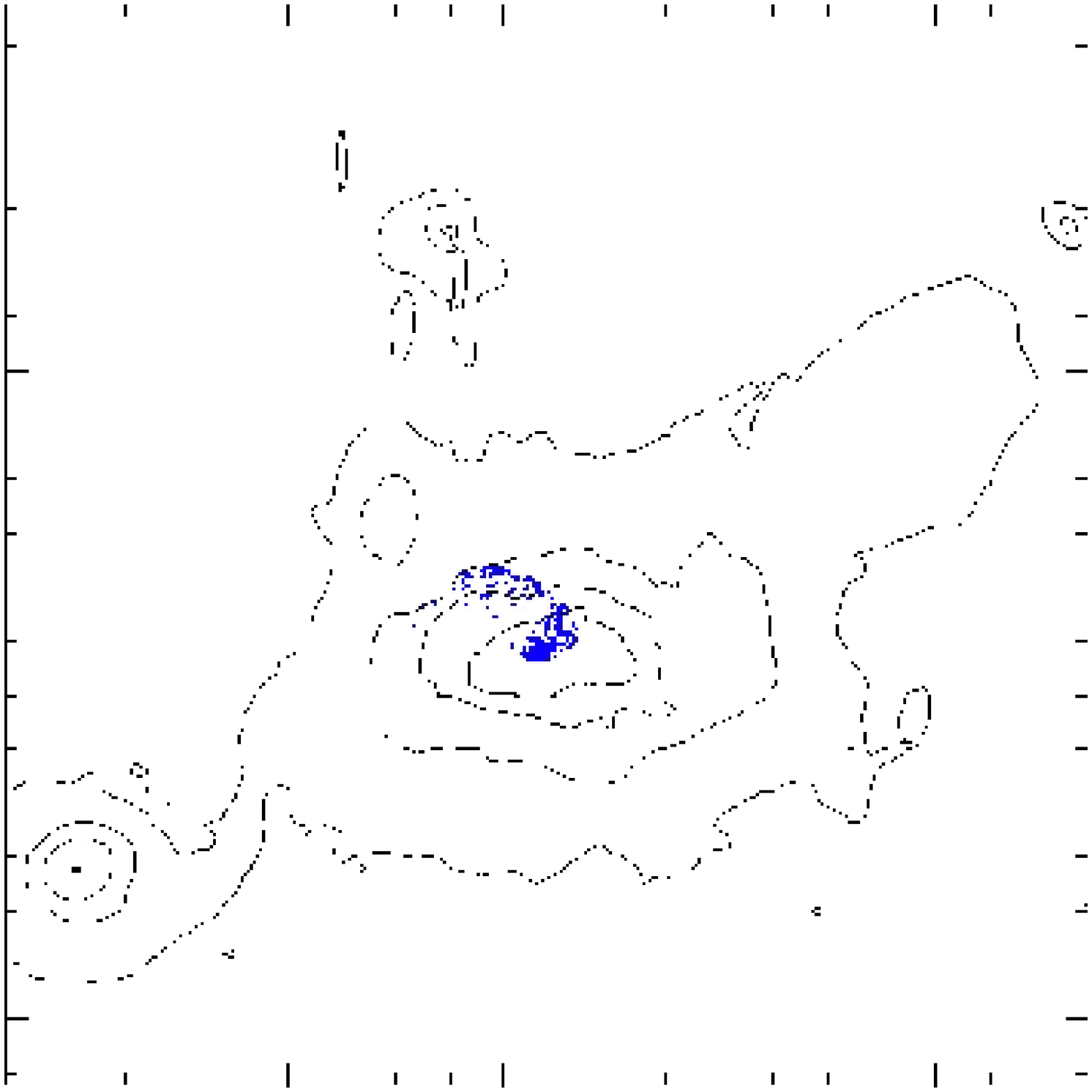}
\includegraphics[width=0.24\textwidth]{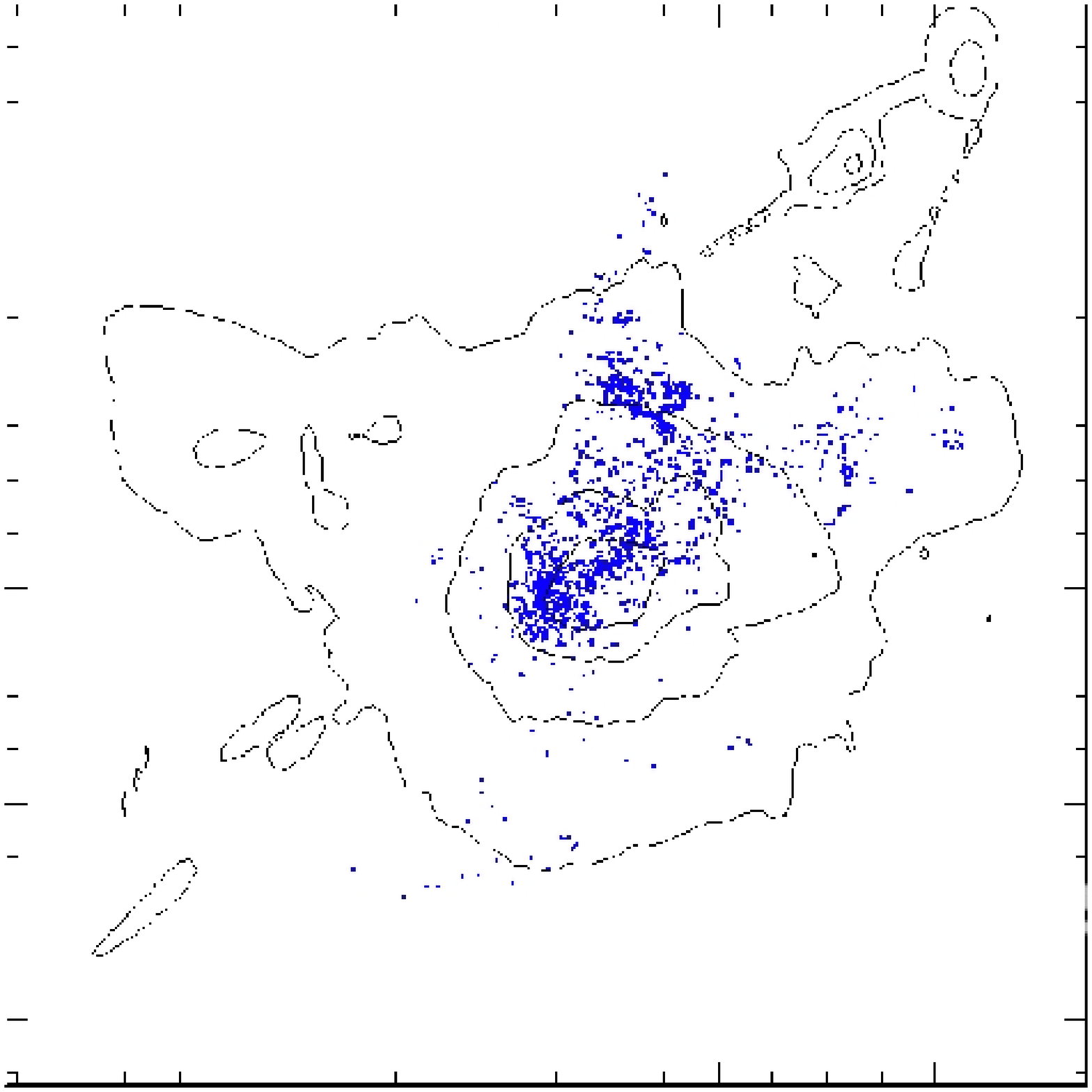}
\includegraphics[width=0.24\textwidth]{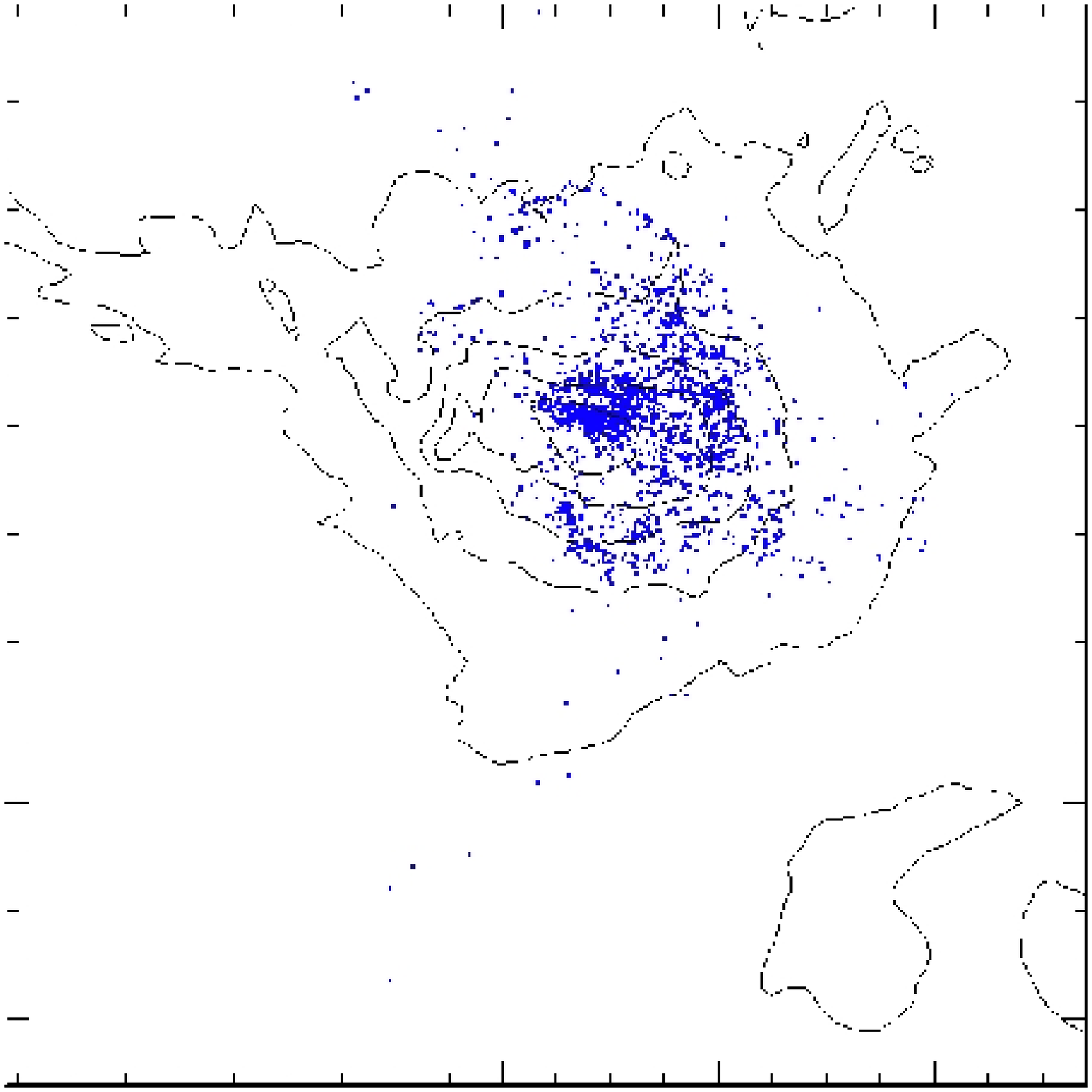}
\caption{Time evolution of the projected positions for metal tracers
injected at $z=1$ from a single cD galaxy in the center of H1. Gas density
contours are generated as in Fig.\ref{fig:map_mixing_fami}, the side
of the images is $\sim 5Mpc$.}
\label{fig:galaxy_single}
\end{figure}



\subsection{Application to the metal mixing in the ICM}
\label{subsec:metals}

X-ray observations proved
that the ICM hosts $\sim 0.6$ per cent of heavy elements
in mass, corresponding to a mean metallicity  $Z \sim 0.3 Z_{\odot}$,
where $Z_{\odot}$ is the solar metallicity (e.g. Werner et al.2008 for a up-to-date review).
The distribution of metals in galaxy clusters is characterized by profiles that are
peaked
toward the core of cooling flow clusters and rather flat in all the others
(Vikhlinin et al.2005; Pratt et al.2007).
Two dimensional metallicity maps from nearby clusters provided evidence for
complex and non symmetric distributions of heavy elements (e.g. Sanders \& Fabian 2006; Finoguenov et al.2006).
Furthermore, recent observations shed some light on the significant 
evolution with decreasing redshift
of the cluster metallicity (Balestra et al.2007; Maughan et al.2007).

From the theoretical point of view, 
several processes can contribute to the observed patterns
and evolution of metals in the ICM: galactic winds, ram pressure stripping, 
galaxy-galaxy interactions, intra-cluster supernovae, intra-cluster stellar
populations etc (e.g. Schindler \& Diaferio 2008 and references therein).
Dealing with all these multi-physics and multi-scale processes is a challenging
task for any cosmological simulation. As a consequence, although the overall scenario of metals evolution can
be reproduced by simulations with an acceptable agreement with observations 
(e.g. Cora 2006; Kapferer et al.2007; Tornatore et al.2007; Wiersma et al.2009), the details of the
process are still largely uncertain.

\bigskip

Although  our simulations do not account for any treatment of chemical
elements production and evolution, our tracer method can provide new
hints about the ways injected metals are spread and mixed
in the ICM during cluster evolution.

We present here a simple set of exploratory studies, where initial
injection profiles of metal tracers are assumed to reproduce the metal
release from galactic winds, and where the advection of metal tracers is
followed as a function of the evolution of the underlying
gas distribution.
The DM mass resolution in our runs is high enough to
detect {\it single} galaxy-like structures in the DM distribution
by means of the same halo-finder algorithm mentioned in
Sect.\ref{subsec:mix}. Therefore below 
we will use the term {\it galaxy} for
massive DM clumps with $N \sim 10^{2} - 10^{3}$ DM particles.
We explore three (idealized) scenarios describing the initial
injection of ``metal tracers'': a) a single
injection at $z=1$ from the central cluster galaxy; b)
ten injections which are spaced in time (in the range $1<z<0$) from
the central cluster galaxy; c) ten injections as in case 'b' (in the range $1<z<0$) from the ten
most massive galaxies  within the AMR volume.

Following Rebusco et al.(2006), we modeled the profile of injected
metals from a galaxy with a $\beta$-profile

\begin{equation}
a(r)=a_{0}\{1+(r/r_{a})^{2}\}^{-\beta},
\end{equation}

where $a_{0}$ is the solar abundance, $r_{a}$ is the
scale radius for each galaxy, and $\beta=0.5$. With a very simple
prescription we adopted a constant value $r_{a}=36kpc$ (one cell)
for all galaxies and a 
normalization $a_{0}$ that depended on the virial DM mass
of each cluster (with $a_{0}=1$ for the main cluster galaxy and 
$a_{0,gal} \propto M_{dm,gal}/M_{dm,main}$ for the other
galaxies).

Our choice of parameters was just 
a trial, designed to fall within the range of the parameters for
galaxies studied in Rebusco et al.(2006).
With this simple setup, we explored the 
effects played by spatial transport on the
mixing of metals deposited in the evolving ICM.

In Fig.\ref{fig:galaxy_single} we show the evolution of the projected
positions of $N \approx 4 \cdot 10^{4}$ metal tracers in the cluster
H1 for model 'a'.

Even if the dynamical history of this cluster
at $z=0$ is fairly regular, the amount of mergers and
accretions of matter experienced
at $z \leq 1$ is 
enough to 
spread the metals out to $\sim 500kpc$ {\footnote {We also notice that significant
offsets ($\sim 100kpc$) between the position of the DM and gas peaks are
common; since X-ray observations are usually referred to the luminosity
peak to derive profiles, we stress that this may have an effect because 
the bulk of metals is released by galaxies.}.

The evolution of the number density profile of metal tracers  from 
a single injection at $z=1$ (model 'a') is shown
in the left panel in Fig.\ref{fig:metal_prof_ev} for
the same cluster. This injection scenario produces a very flat distribution
of metals at $z\approx 0.1$. Even if the qualitative 
behavior is similar to that reported in 
Rebusco et al.(2005), we notice that in our case the
distance where metal pollution is efficient in the ICM is 
larger, which is
consistent with the larger transport pattern presented in Sect.\ref{subsec:mix}.

The result of ten injection epochs (one every $\approx 700Myr$) 
from the central galaxy (model 'b')
is shown in the 
central
panel in Fig.\ref{fig:metal_prof_ev}. This 
obviously increases by $\sim 10$ the final metal content in the innermost cluster region
and is found to produce a monotonically decreasing tracer distribution at all
radii.

A case of more astrophysical relevance is perhaps our model 'c', which
assumes multiple injections from the ten most massive galaxies in the AMR region
(see right panel in Fig.\ref{fig:metal_prof_ev}).
The corresponding spatial evolution of metal tracers in cluster H1
with overlaid gas density contours, is shown in 
Fig.\ref{fig:galaxies}. In this case the metal distribution is quite regular
at all redshifts compared to the case of injection from 
the central cluster galaxy only.

\bigskip

The role played by mergers in the disruption of cool cores 
is still an object of debate (e.g. Santos et al.2009 and references therein),
although  observations support 
the destruction
of cool-cores by cluster mergers (Allen et al.2001; Sanderson
et al.2006; Rossetti \& Molendi 2009). Indeed, observations suggest a dichotomy between
the metallicity profiles in the cool-core and non-cool core
clusters (e.g. De Grandi et al.2004).
Our simulations neglect any treatment of cooling and of energy feedback
mechanisms from central AGN, and consequently we cannot address this 
in a self-consistent way.
We can speculate though that the most relaxed clusters in our sample
are similar to cool core clusters, while clusters with mergers are more
similar to non-cool core clusters.

We calculated the metallicity profiles for four clusters in our 
sample (two with a fairly relaxed dynamical
evolution and two with a major merger in the range $0 \leq z \leq 1$)
to investigate the effect of cluster dynamics on the shape of 
metallicity profiles.
The total iron mass in the ICM is normalized so that
the metallicity at the position of the central galaxy in the
cluster was $Z=1$; the gas mass
was directly taken from the cells in the simulation.
In Fig.\ref{fig:prof_iron} we show the profiles of the four clusters
at $z \approx 0.1$. The profiles are 
computed for spherical shells of a 
width of 36kpc and they were centered on the peak of the X-ray
bolometric cluster emission; the contribute of freshly (un-evolved)
metal tracers from the central cluster galaxy at the redshift
of observation was removed in post-processing.

We found a remarkable trend:
``relaxed'' clusters showed a very {\it peaked}
metallicity profile, while 
``merger'' clusters showed a more complex behavior with
{\it flat} profiles up to the outer regions.
Since the
injection of metal tracers in all runs was done in the same
way and with the same duty cycles and no physical
mechanism other than hierarchical clustering was at work here,
this test suggests that {\it very different} metallicity
profiles in the ICM may result from the effect of different
matter accretion histories.

Future
works accounting for a realistic chemical enrichment model and 
cooling/feedback processes may produce {\it quantitative} and self-consistent
predictions, yet our results {\it qualitatively}
suggest that the observed dichotomy of metallicity profiles might be simply
explained in terms of different cluster dynamics; tracers offer a valuable
method to investigate this important issue.

\begin{figure*} 
\includegraphics[width=0.33\textwidth]{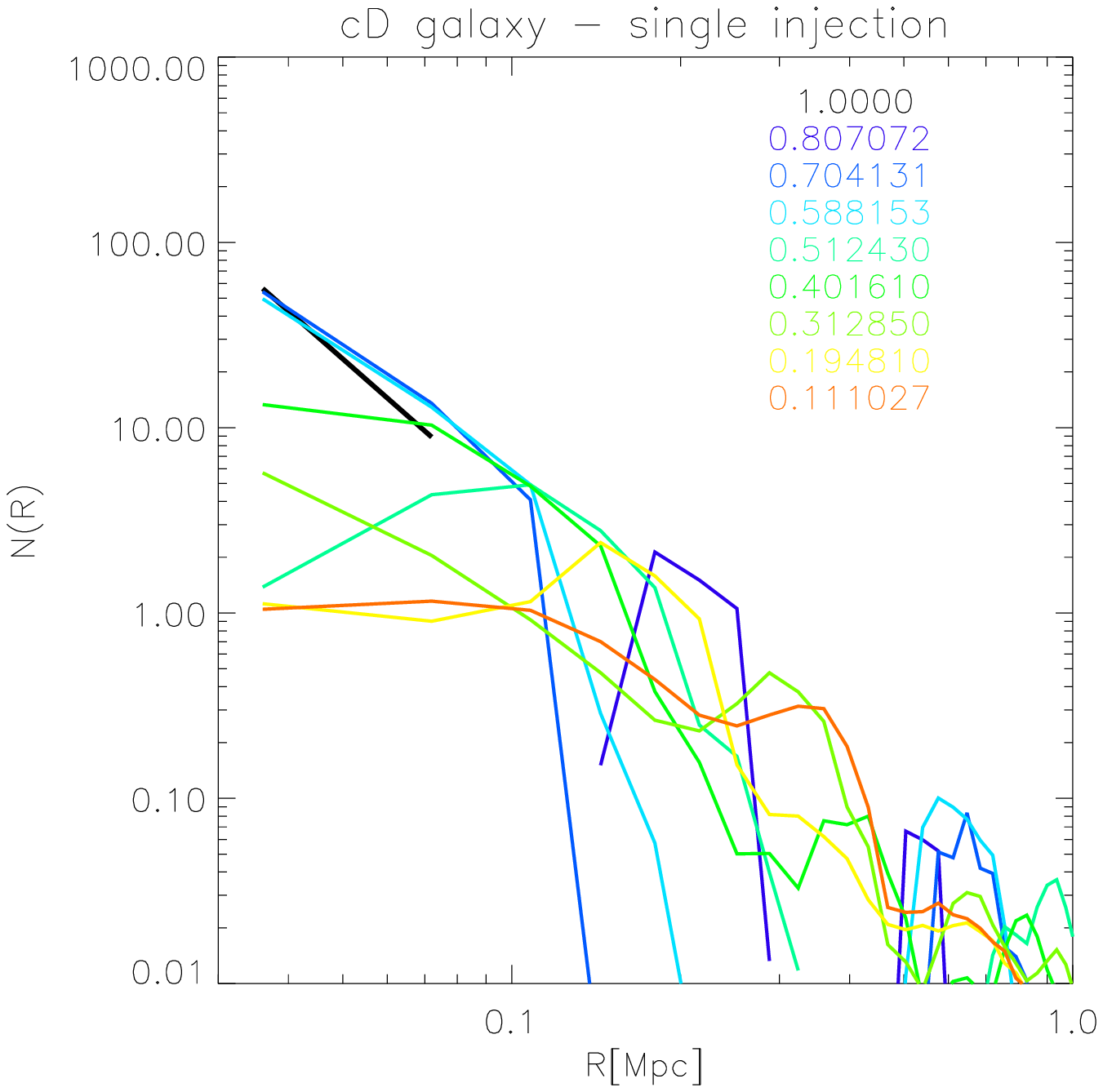}
\includegraphics[width=0.33\textwidth]{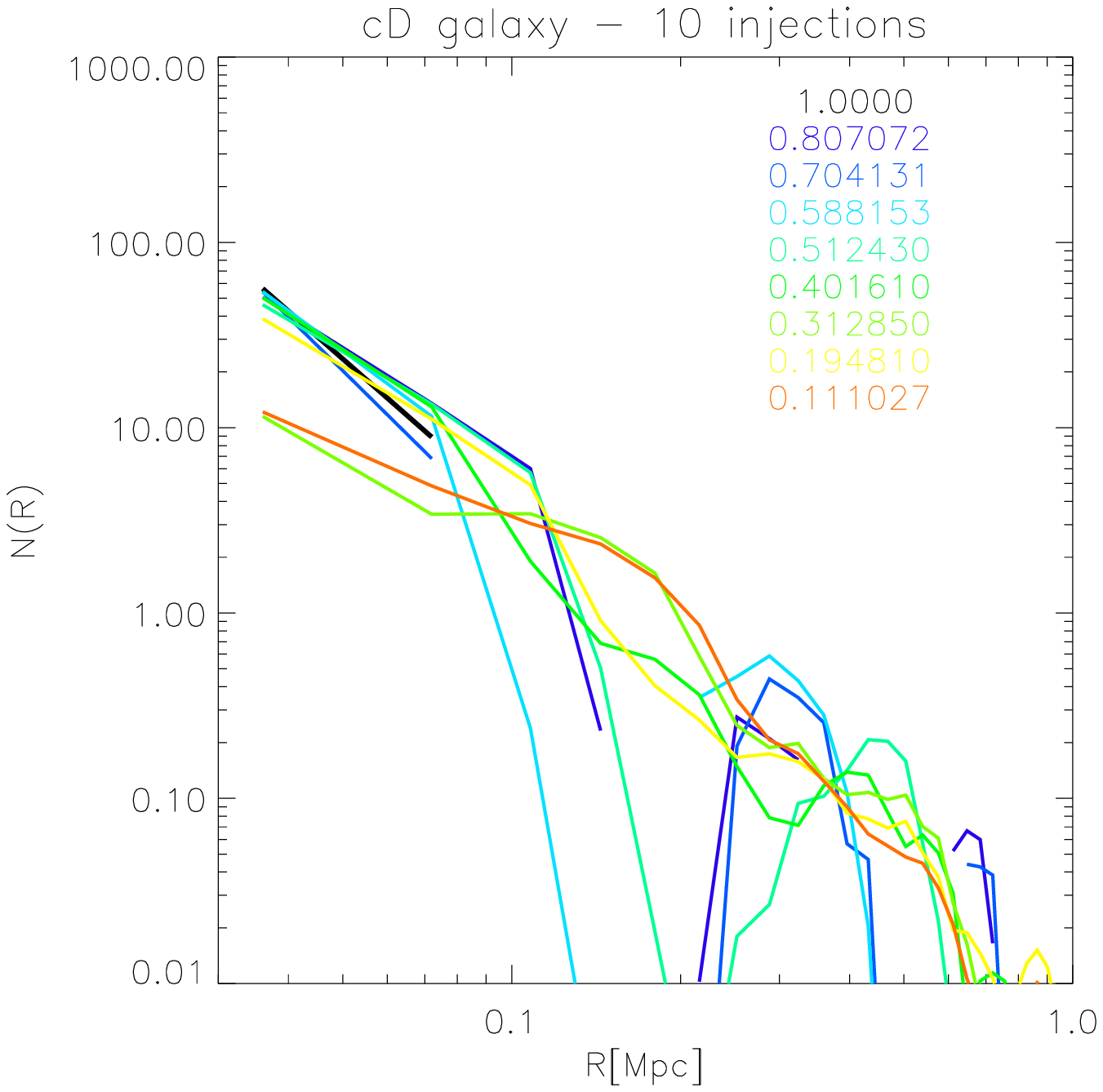}
\includegraphics[width=0.33\textwidth]{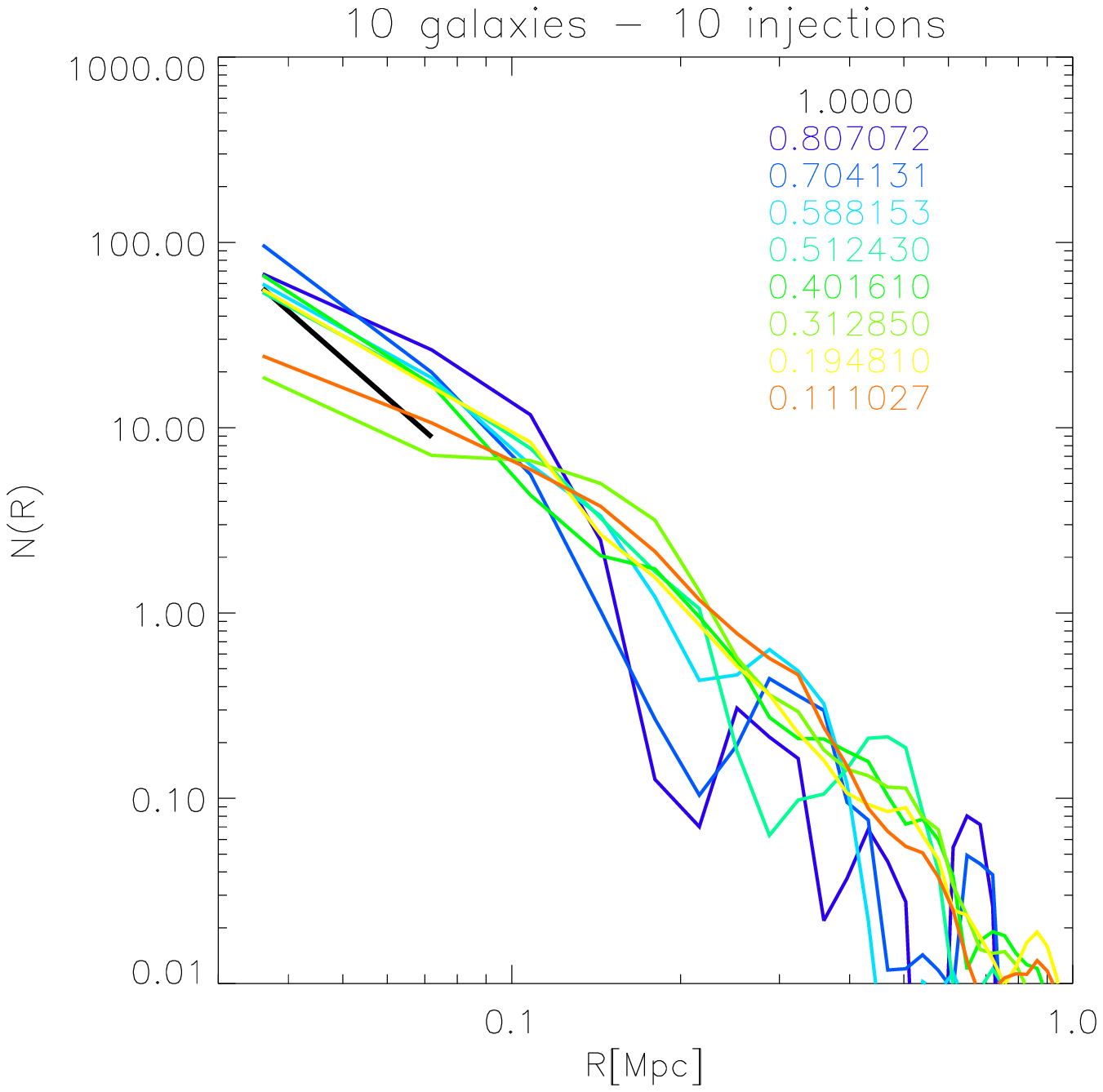}
\caption{Evolution of the number profile of metal tracers for cluster H1, for 
the 3 injection recipes introduced in Sect.\ref{subsec:metals}: single injection
at $z=1$ from the central cD galaxy ({\it left}), several injections from 
the central cD galaxy ({\it center}) and several injections from 10 galaxies
({\it right}). The contribution from each galaxy in the 'c' model 
has been normalized as 
explained in Sect.\ref{subsec:metals}.}
\label{fig:metal_prof_ev}
\end{figure*}

\begin{figure} 
\includegraphics[width=0.24\textwidth]{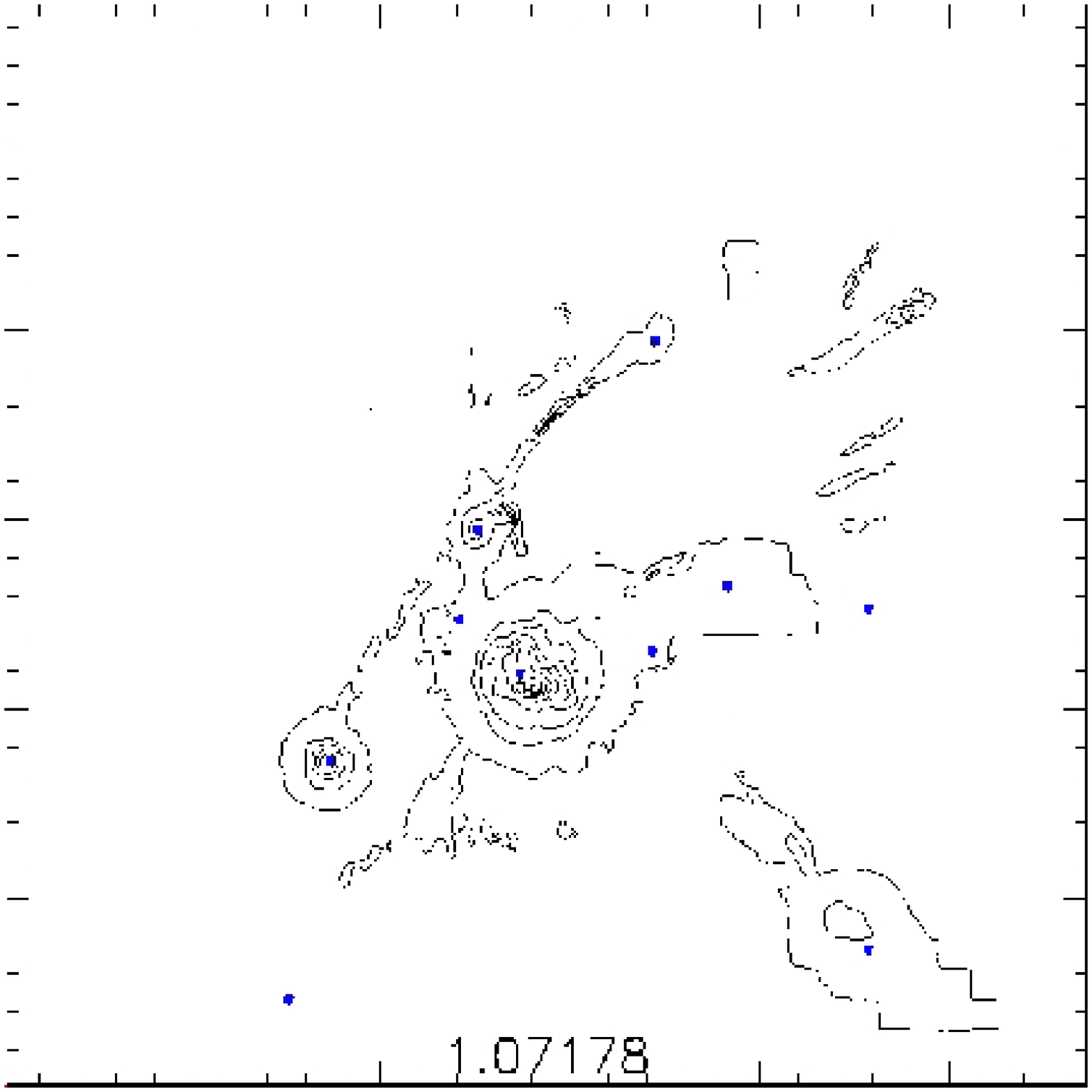}
\includegraphics[width=0.24\textwidth]{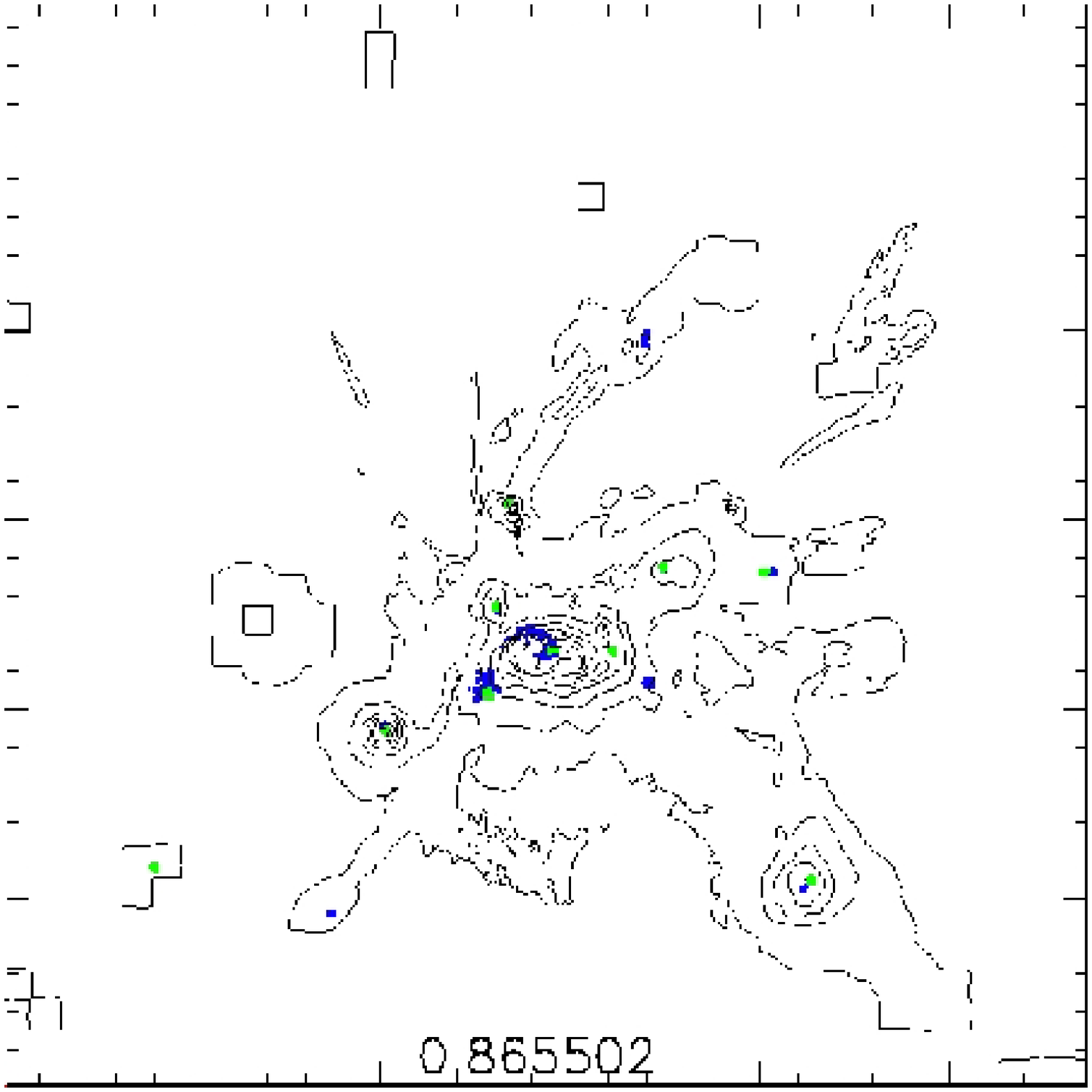}
\includegraphics[width=0.24\textwidth]{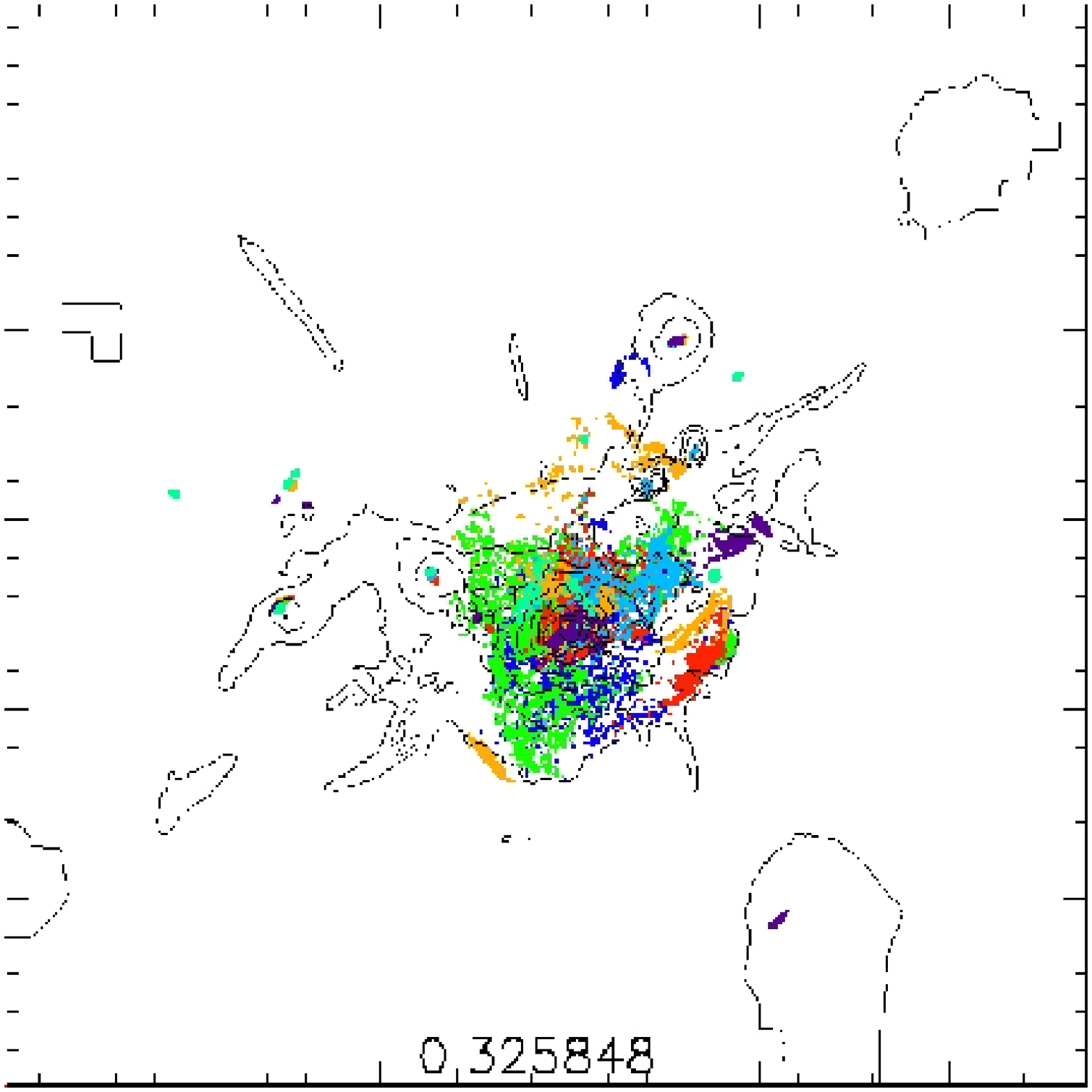}
\includegraphics[width=0.24\textwidth]{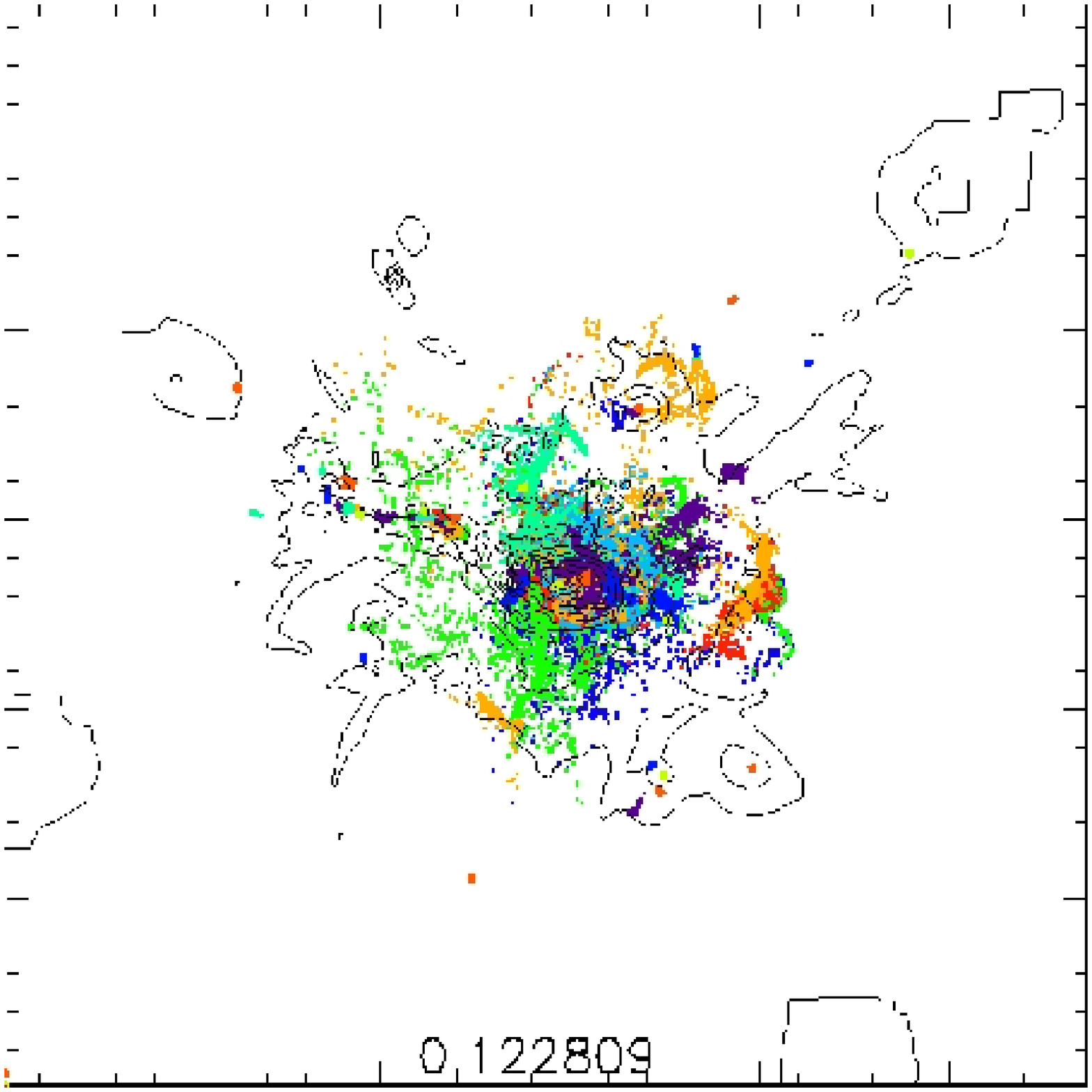}
\caption{Time evolution of the projected positions for metal tracers
injected every $\approx 700Myr$ from different galaxies in the AMR
region of cluster H1. The different colors refer to the different
epochs of injection; gas density
contours are generated as in Fig.\ref{fig:map_mixing_fami}. The side
of the images is $\sim 15Mpc$.}
\label{fig:galaxies}
\end{figure}

\begin{figure} 
\includegraphics[width=0.45\textwidth]{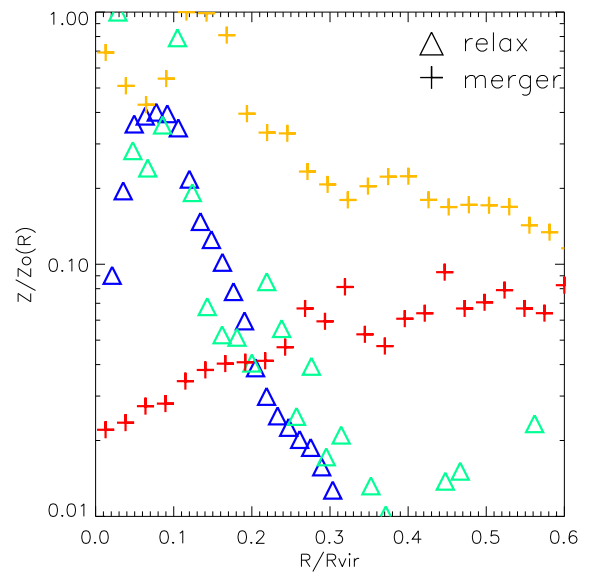}
\caption{Mean metallicity profile for the relaxed clusters H1 and H3 (triangles) and for 
the perturbed clusters H5 and H6 (crosses) at $z=0.1$.}
\label{fig:prof_iron}
\end{figure}

\begin{figure}
\includegraphics[width=0.24\textwidth]{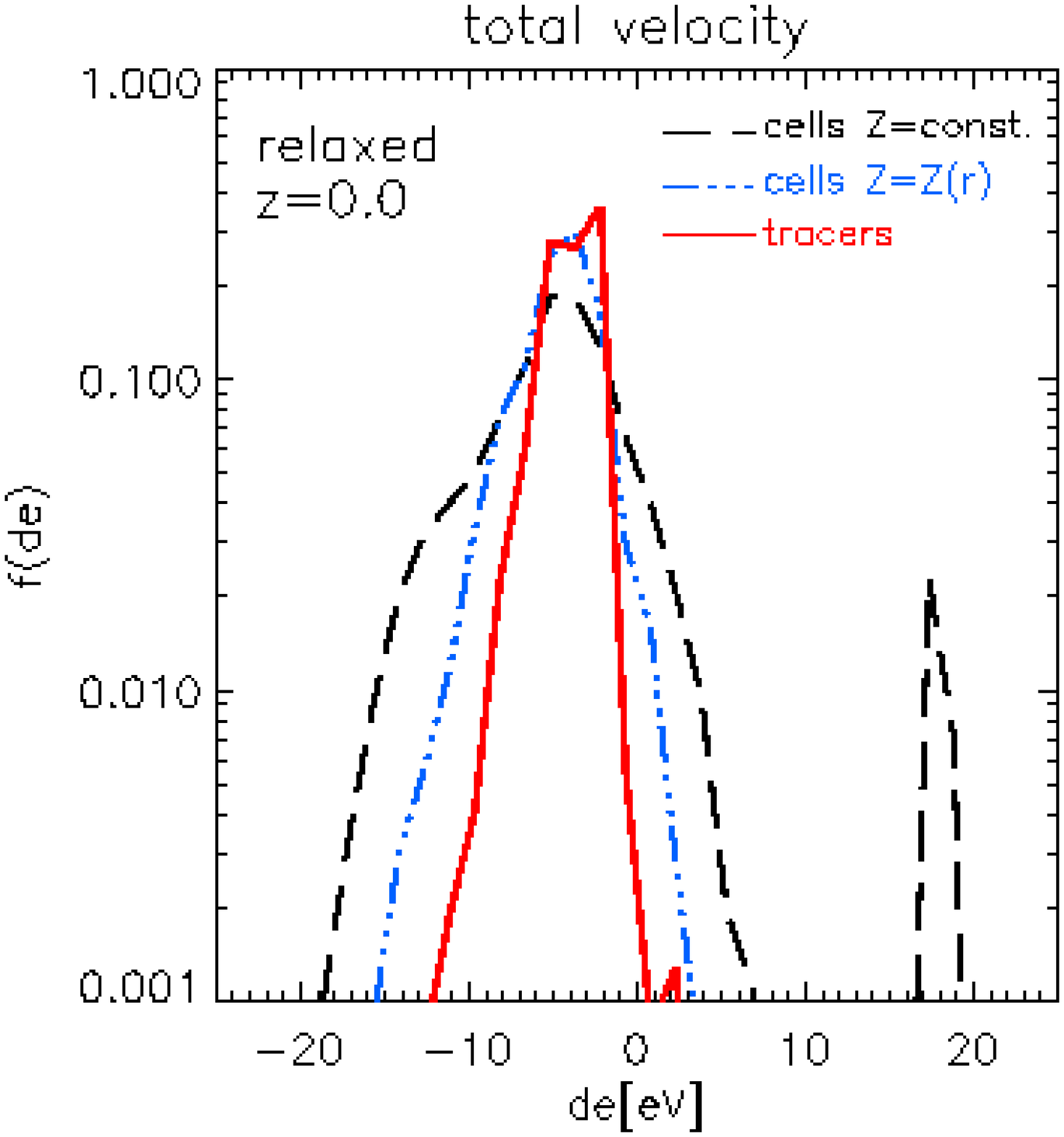}
\includegraphics[width=0.24\textwidth]{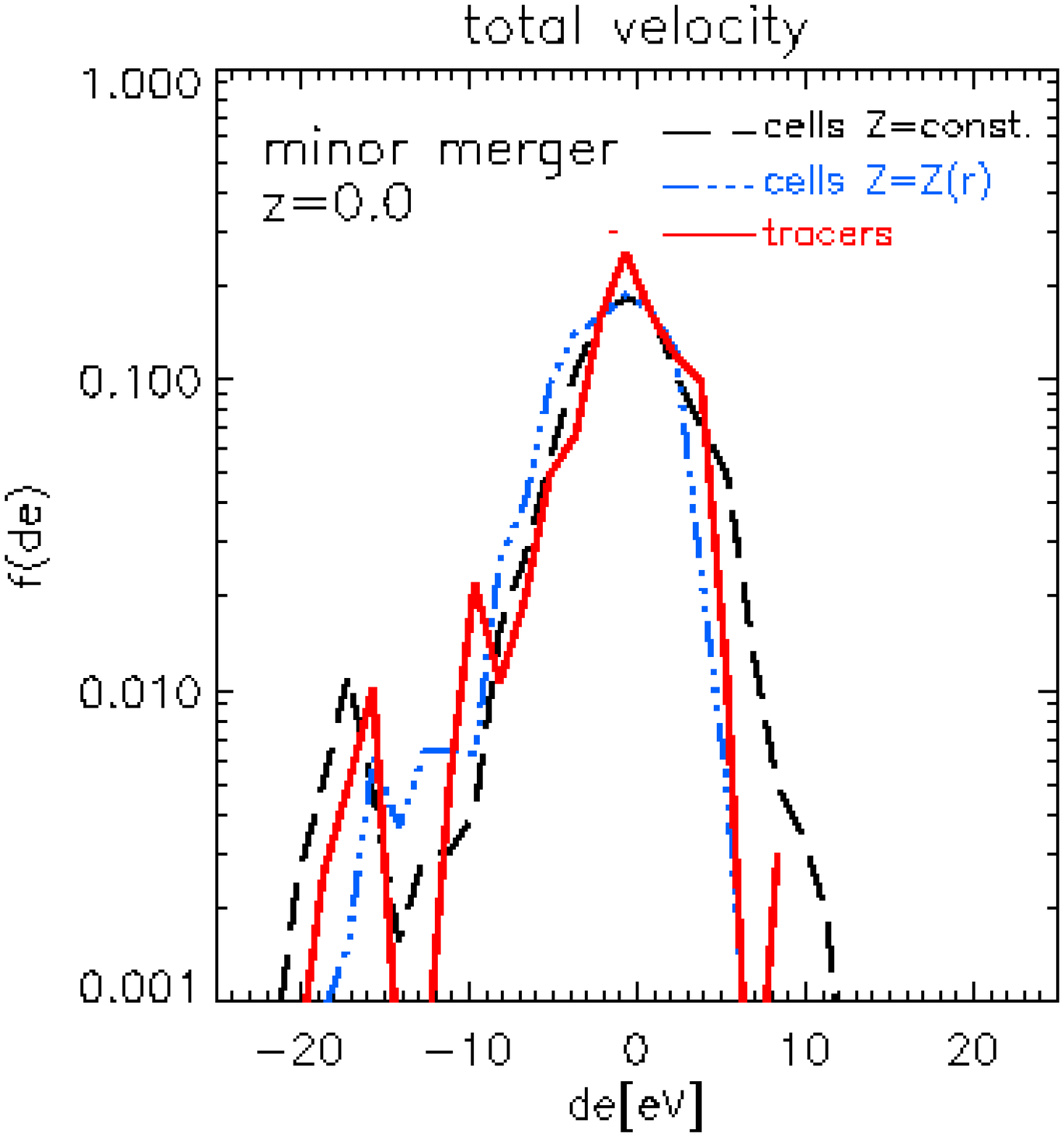}
\includegraphics[width=0.24\textwidth]{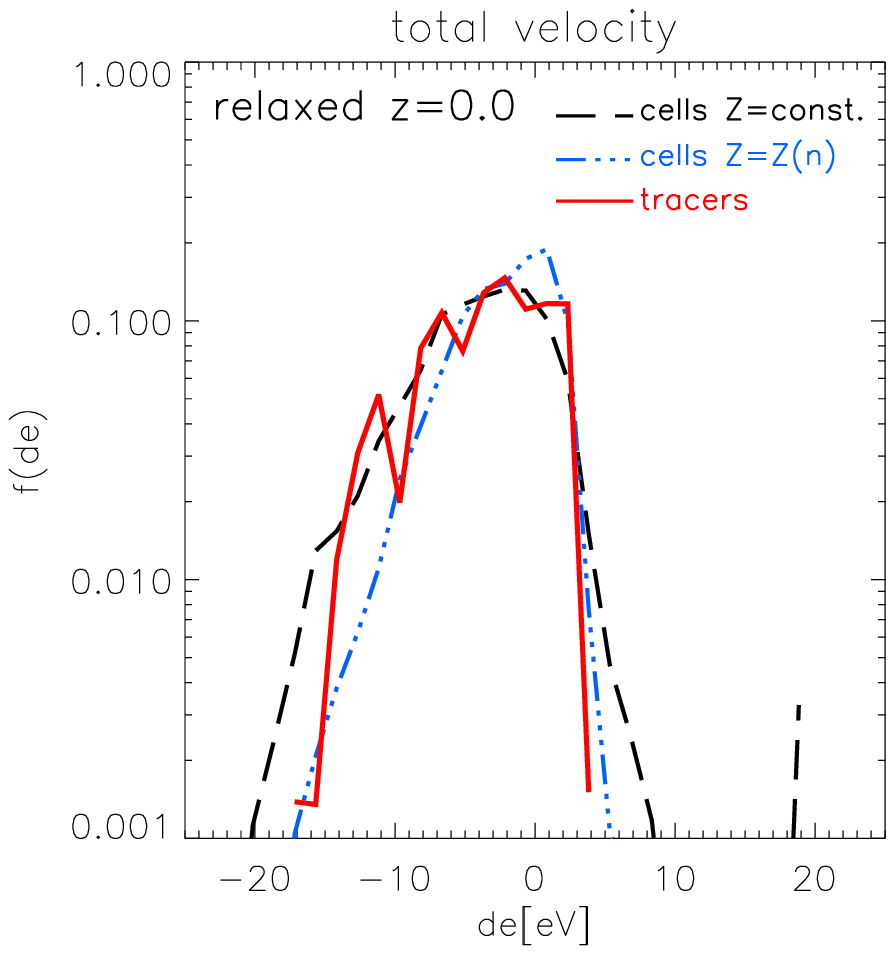}
\includegraphics[width=0.24\textwidth]{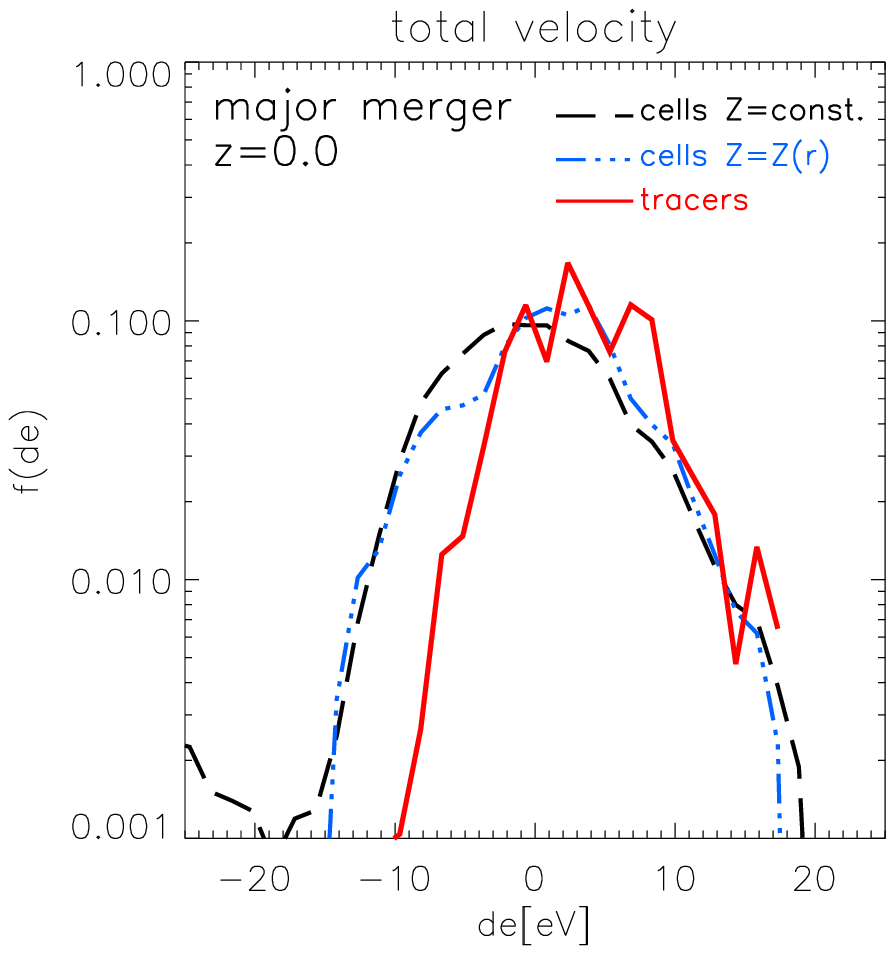}
\caption{Simulated iron line profile for clusters H1, H3, H4 and H5. The red lines show the line emissivity simulated assuming the tracer metallicity, the black lines are for a constant metallicty and the blue lines are for an average metallicity which scales with radius.}
\label{fig:metals1}
\end{figure}

\subsection{Iron line profile.}
\label{subsec:iron_line}

Turbulent gas motions can lead to a Doppler shift broadening 
of X-ray emitting metal lines, in excess to their 
intrinsic width. Measuring the broadening of 
metal lines in the future will likely provide the most direct
evidence of subsonic turbulent motions in the ICM {\footnote{Very 
recently Sander et al. 
(2009) for the first time placed direct limits on the turbulent broadening
on the emission lines in a cool core galaxy cluster with XMM-Newton. 
The ratio between
the turbulent to thermal energy density in the cluster core is found to
be less than 13 percent within 30kpc from the cluster center, which is
consistent with the typical turbulent to thermal energy ratios found in the
centres of our simulated clusters, e.g. Vazza et al.(2009).}}
In a number of works (Sunyaev et al.2003; Inogamov \& Sunyaev 2003; Dolag et al.2005)
it was proved that if turbulent motions are at the level of what
is produced in cosmological numerical simulations, the shape of the 
Fe XXIII line would be enough modified to allow
studies of turbulence in the ICM.
Here we extend these works by taking into account the 3--D metallicity
distribution that comes out in our simulations adopting the 
'c' model as fiducial injection model for metal tracers.

We thus estimated the emission along the line of sight of the 6.702 keV iron
line with columns through the center of the four galaxy clusters in our sample.
For every tracer we assumed a line emissivity proportional to the second
power of its metal content (which was derived as in Sect.\ref{subsec:metals}),
and for simplicity we neglected any dependence on the gas temperature. The effect of thermal broadening and the blending with other iron lines is also
neglected.
As a control check, we also computed two additional iron emissivities directly from 
the cells of the simulations: in the first case we simply assumed a fixed metallicity
for all cells, and in the second case we assumed a uniform metallicity radial
profile in the form of $Z/Z_{\odot} \propto (r/R_{vir})^{-1/2}$ (e.g. Vikhlinin et al.2005).
In all cases, the velocity of the center of mass of every cluster
has been removed.

Figure\ref{fig:metals1} shows the simulated iron line profile, for clusters
H1, H3, H4 and H5 at $z=0$. For every cluster the signal was extracted along
columns with a section of $300kpc/h$ and an energy resolution of $\Delta E=2eV$
was assumed.
All emissions were normalized to have the same total 
luminosity=1 within the column.

For clusters without a major merger, the centroid of the iron 
line and the general shape of the line from the tracers agreed 
reasonably with those from the cells.
In the case of the major merger system (H5)
the centroids and the shapes of the lines were quite different.
Consistently with what was found 
in the previous sections, this is because mergers
create complex and patchy distributions of metal tracers, which cannot
be fully described by a simple radial profile.
This is further highlighted in Fig.\ref{fig:metals2}, which is obtained at the epoch
of the major merger in H5.

We conclude that 
the realistic modeling of the iron line broadening requires an appropriate
modeling of the distribution of metals in the cluster volume.

\bigskip

Tracers also allow for the decomposition of the convoluted iron line profile in the 
contributions from the different generations of metal tracers.
Figure \ref{fig:metals3} shows the comparison of the emissivity profile for
cluster H1, H3 and H5 (in this case the section of the column is 
set to $600kpc/h$ to also highlight the contribution from tracers distributed at larger
distances from the cluster center) and the 
relative contributions from different 
generations of tracers. 
The case of the the major merger system highlights that
the signal from the total iron line is the convolution of very different
emissions coming from different generations of metal tracers and that
the earlier generations are responsible for much broader and asymmetric
components compared to the latest generations that were injected after
the major merger.

\begin{figure}
\includegraphics[width=0.45\textwidth]{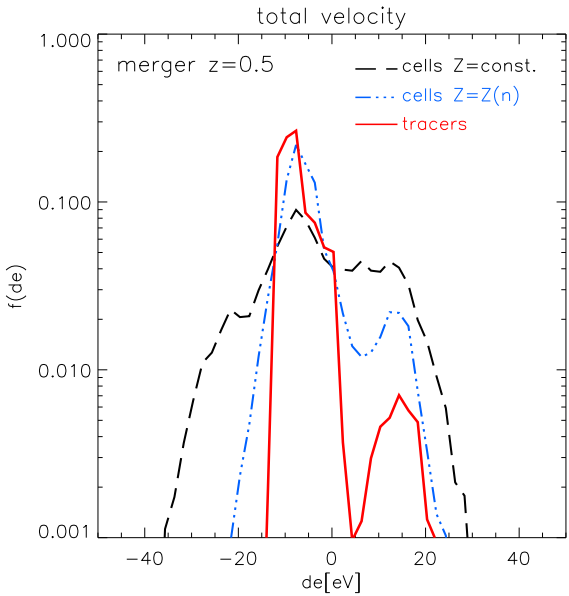}
\caption{Simulated iron line profile for cluster H5 at $z=0.5$. The
meaning of lines is the same of Fig.\ref{fig:metals1}.}
\label{fig:metals2}
\end{figure}

\begin{figure*}
\includegraphics[width=0.3\textwidth]{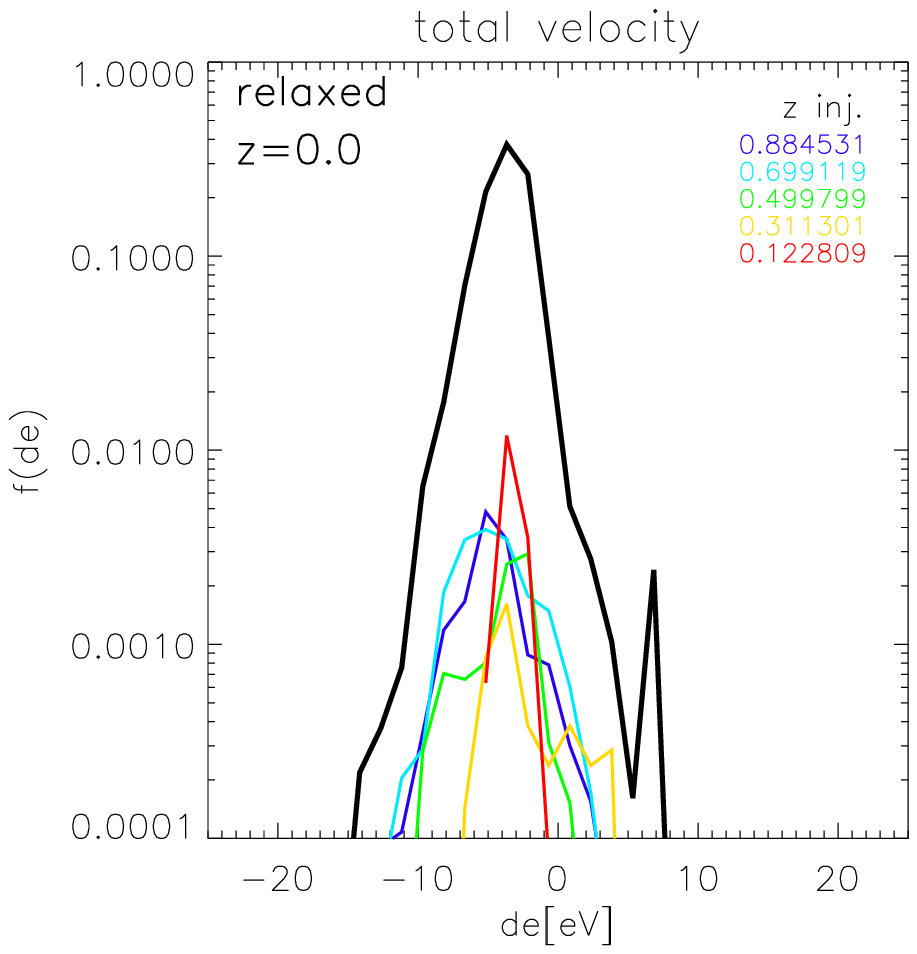}
\includegraphics[width=0.3\textwidth]{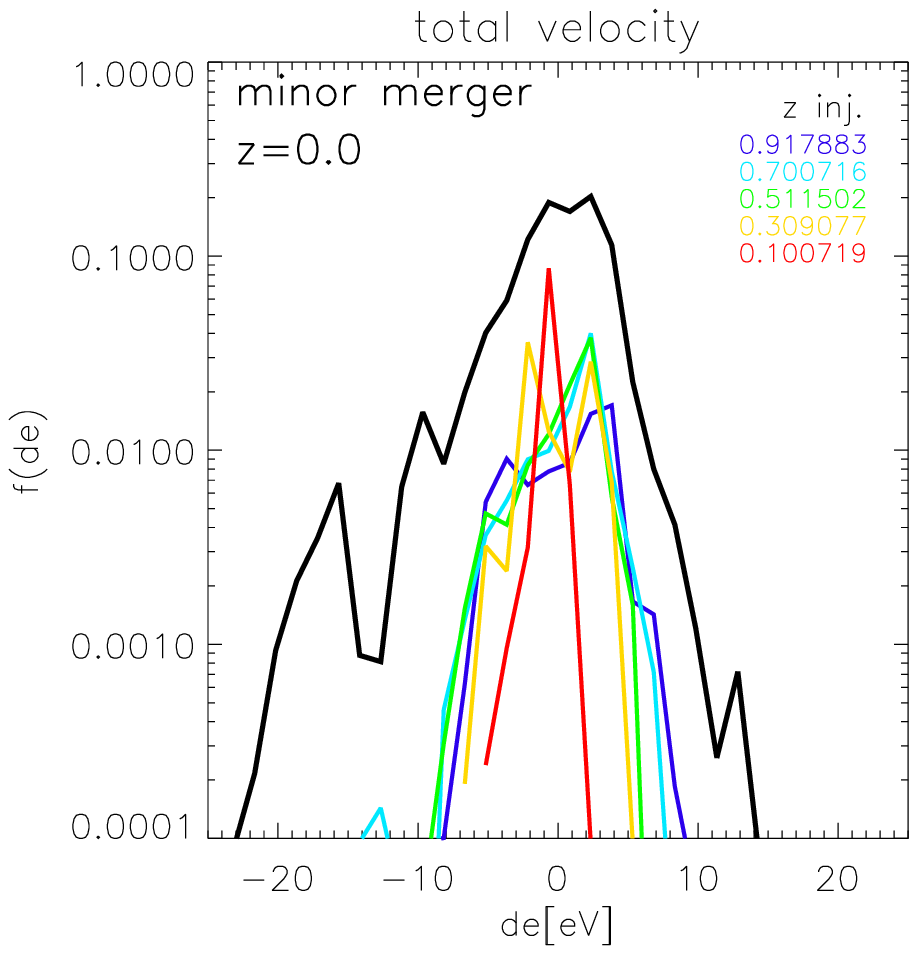}
\includegraphics[width=0.3\textwidth]{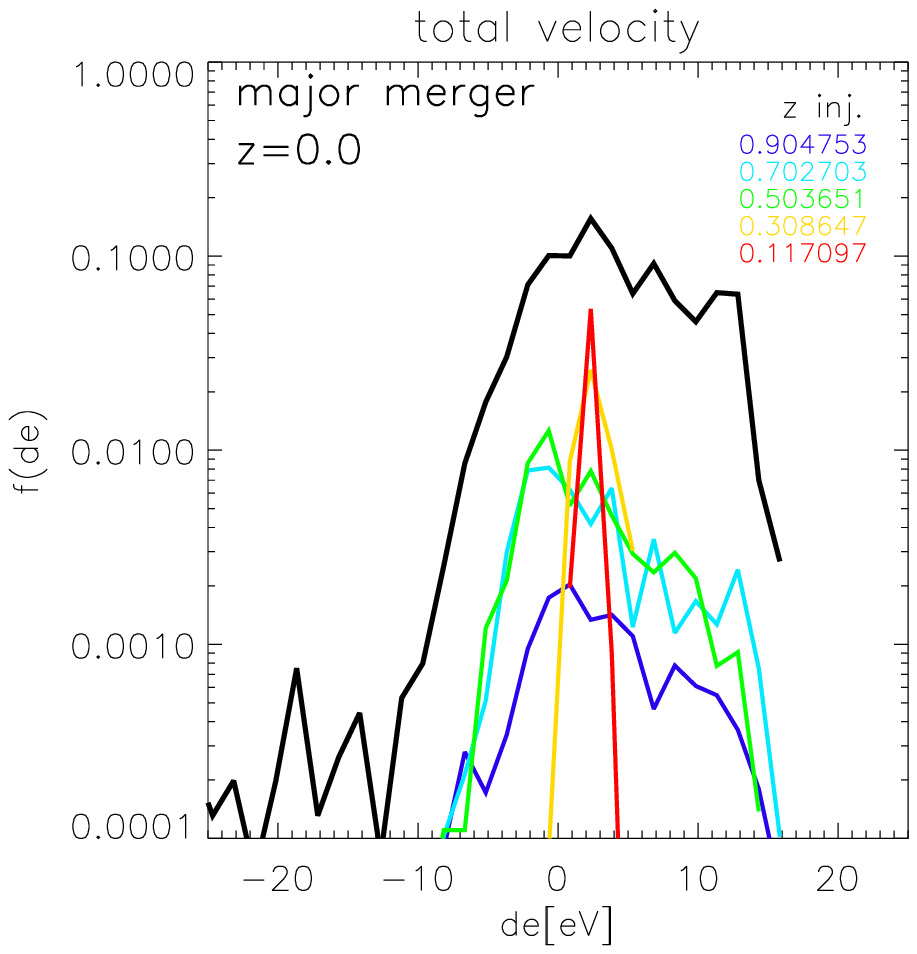}
\caption{Simulated iron line profile for clusters H1, H3 and H5 at $z=0$. The
black lines show the total line emissivity, while the colored lines show
the contribution for a sub-sample of tracer generations at different epochs.}
\label{fig:metals3}
\end{figure*}

\section{Conclusions}
\label{sec:conclusions}

In this work we studied the mixing properties
of the ICM in simulated galaxy clusters, using 
tracer particles that are advected by the gas flow
during the evolution of cosmic structures.
This approach allowed us to obtain valuable information on 
the Lagrangian properties of the ICM, and  allowed progress 
in our understanding of the 
transport processes in Eulerian simulations of galaxy clusters.
Seven galaxy clusters with total masses in the range $M \sim 2 - 3 \cdot 10^{14}M_{\odot}/h$ and different 
dynamical states were simulated with an updated version
of the ENZO code (Norman et al.2007), where an additional
refinement criterion was designed to reach good
spatial accuracy at shock waves up to $\sim 2 R_{vir}$ from
the center of galaxy clusters (Vazza et al.2009).
In a post-processing, we injected and followed 
passive tracers, and tracked their spatial evolution 
with high time resolution.
Numerical tests were presented and discussed to find the
optimal strategies to inject tracers, update their positions
in time and sample the galaxy cluster volume (Sect.\ref{sec:convergence}).
We then applied this tracer technique to some interesting
problems concerning the mixing of the ICM.

First, we showed that the pair dispersion statistics has a 
well constrained behavior in time: 
for most of the cosmic time the separation between close couples of 
tracers shows a very regular evolution in all clusters,
consistently with the basic $\propto t^{3/2}$
scaling expected for a simple turbulent model (Sect.\ref{subsec:dispersion}).
This finding is well consistent with the independent measurement of
3-D velocity power spectrum for the same clusters, and confirms that a sizable amount of complex subsonic motions 
correlated on cluster
scales account for 10-30 per cent of the thermal energy
inside $R_{vir}$ of our simulated clusters.

Second, we quantified the degree of radial mixing of the ICM 
by following the evolution of tracers injected in regular shells 
within the cluster volume at $z \approx 1$.
The radial mixing morphology was found to be strongly dependent on the dynamical
history of every cluster.
Large plumes (with the size of $\sim Mpc$) where gas matter is efficiently mixed
 were found
for a major merger cluster. On the other hand more regular
and concentrated mixing profiles were measured for more 
relaxed clusters.

Third, we assumed a simple model of metal injection from galaxies 
identified in the simulations around the clusters and 
studied the metal enrichment
of the ICM by tracking the advection of ``metal'' tracers (tracers with
a different iron content as a function of their galaxy of origin) deposited
in the cluster volume.
In this case also, the dynamical state of galaxy clusters and its
matter accretion history were found to affect the final distribution 
of metal tracers at evolved epochs. In particular,
the simulated clusters showed a hint of dichotomy in metallicity 
distributions
between
clusters experiencing only minor mergers and clusters with major mergers, with
the latter showing a broader distribution of metals.

Finally, we simulated the emissivity of the Fe XXIII line from the center
of our clusters, using the 3--D distribution of metals produced through
our metal tracer injection (model 'c').
We reported significant departures from the simplified assumption of a constant
or a radially symmetric metallicity profile, particularly for
merger systems. This stresses the importance of studying in detail
the connection between the metal enrichment history and the matter
accretion history of the ICM.

\bigskip

In conclusions, we believe that the coupling of high resolution Eulerian
numerical simulations and a tracer technique represents a very powerful tool to
study crucial open problems concerning the mixing properties
of the diffuse gas in galaxy clusters, and the injection and the advection
of cosmic rays particles in galaxy clusters, which will be 
subject of future works.

 \section{acknowledgments}
We thank the anonymous referee, whose suggestions helped us to improve
the presentation of this paper.
 The authors thank R. Brunino for valuable support at 
CINECA Supercomputing Center,  M. Nanni, F. Tinarelli and 
M.Tugnoli for valuable technical support at the Radio Astronomy Institute (Bologna), and G.Tormen for useful discussions. 
We acknowledge partial 
support through grant ASI-INAF I/088/06/0 and PRIN INAF 2007/2008, and the 
use of computational resources under the CINECA-INAF 2008-2010 agreement.

 \end{document}